\documentclass{article} 

\usepackage[utf8]{inputenc} 
\usepackage[T1]{fontenc}    
\usepackage{ifthen}

\newboolean{arxiv}
\setboolean{arxiv}{true}  

\ifthenelse{\boolean{arxiv}}{\usepackage{lmodern}

\newcommand{\IEEEoverridecommandlockouts}{}

\newcommand{\IEEEauthorblockN}[1]{#1}
\newcommand{\IEEEauthorblockA}[1]{#1}
\newcommand{\IEEEauthorrefmark}[1]{%
	\ifthenelse{\equal{#1}{1}}%
		{${}^*$}%
		{\ifthenelse{\equal{#1}{2}}%
				{${}^\dagger$}%
				{${}^\ddagger$}%
		}%
}

\newcommand{\IEEEQED}{$\square$}
\newcommand{\IEEEQEDopen}{$\square$}

\newenvironment{IEEEproof}{\begin{proof}}{%
    \end{proof}\ignorespacesafterend
}

\newcommand{\appendices}{\appendix}
}{}
\usepackage[english]{babel}
\usepackage{cite}
\usepackage[pdftex]{graphicx}
\usepackage{amsmath}
\usepackage{amsfonts}
\usepackage{amssymb}
\ifthenelse{\boolean{arxiv}}{\usepackage{amsthm}}{}
\usepackage{mathtools}
\usepackage[cal=boondoxo]{mathalfa}
\usepackage{algpseudocode}
\usepackage{algorithm}
\usepackage[inline]{enumitem}
\usepackage{booktabs}
\usepackage{tabu}
\usepackage{url}
\usepackage[colorlinks=true,allcolors=black,urlcolor=blue]{hyperref}
\usepackage[normalem]{ulem}
\usepackage{caption}
\usepackage{subcaption}

\usepackage{xcolor}
\definecolor{red}{HTML}{d62728}
\definecolor{green}{HTML}{2ca02c}
\definecolor{blue}{HTML}{1f77bf}
\definecolor{purple}{HTML}{9467bd}
\definecolor{orange}{HTML}{ff7f0e}

\definecolor{red}{RGB}{204, 0, 0}
\definecolor{blue}{RGB}{0, 51, 204}

\graphicspath{{figures/}}
\DeclareGraphicsExtensions{.pdf,.jpeg,.png}
\DeclarePairedDelimiter\abs{\lvert}{\rvert}

\newtheorem{theorem}{Theorem}
\newtheorem{lemma}[theorem]{Lemma}   
\newtheorem{proposition}[theorem]{Proposition}
\newtheorem{corollary}[theorem]{Corollary}
\newtheorem{definition}{Definition}
\newtheorem{candidate}{Candidate Definition}
\newtheorem{property}{Property}
\ifthenelse{\boolean{arxiv}}{\theoremstyle{definition}}{}
\newtheorem{counterexample}{Counterexample}

\usepackage[colorinlistoftodos,backgroundcolor=lightgray!50,linecolor=black,textsize=scriptsize]{todonotes}
\usepackage[margin=1in,marginparwidth=0.7in]{geometry}
\usepackage{marginfix}  
\usepackage{sidenotes}  
\makeatletter
\RenewDocumentCommand\sidenotetext{oo+m}{
	\IfNoValueOrEmptyTF{#1}{%
		\@sidenotes@placemarginal{#2}{\textsuperscript{\{\thesidenote\}}{}~\scriptsize{#3}}%
		\refstepcounter{sidenote}%
	}{%
		\@sidenotes@placemarginal{#2}{\textsuperscript{#1}~#3}%
	}%
}
\RenewDocumentCommand\sidenotemark{o}{
	\@sidenotes@multichecker%
	\IfNoValueOrEmptyTF{#1}{%
		\@sidenotes@thesidenotemark{\{\thesidenote\}}%
	}{%
		\@sidenotes@thesidenotemark{#1}%
	}%
	\@sidenotes@multimarker%
}
\makeatother

\ifthenelse{\boolean{arxiv}}{\linespread{1.1}}{\linespread{1.3}}

\newcommand{\Gstar}{\mathcal{G}^*}
\newcommand{\Vstar}{\mathcal{V}^*}
\newcommand{\Estar}{\mathcal{E}^*}
\newcommand{\G}{\mathcal{G}}
\renewcommand{\H}{\mathcal{H}}
\newcommand{\V}{\mathcal{V}}
\newcommand{\Vip}{\V_{\text{ip}}}
\newcommand{\E}{\mathcal{E}}
\newcommand{\T}{\mathcal{T}}
\newcommand{\C}{\mathcal{C}}
\newcommand{\A}{\mathcal{A}}
\newcommand{\B}{\mathcal{B}}
\newcommand{\Et}{\mathcal{E}_t}

\newcommand{\F}{\mathcal{F}}
\newcommand{\Q}{\mathcal{Q}}
\renewcommand{\P}{\mathcal{P}}
\newcommand{\R}{\mathcal{R}}
\renewcommand{\S}{\mathcal S}

\newcommand{\given}{\,\big\vert\,}
\newcommand{\xor}{\oplus}

\newcommand\independent{\protect\mathpalette{\protect\independenT}{\perp}}
\def\independenT#1#2{\mathrel{\rlap{$#1#2$}\mkern2mu{#1#2}}}

\newcommand{\what}[1]{\widehat{#1}}

\IEEEoverridecommandlockouts

\title{Information Flow in Computational Systems}

\author{
	\IEEEauthorblockN{
		Praveen Venkatesh\IEEEauthorrefmark{1},
		Sanghamitra Dutta\IEEEauthorrefmark{2}
		and Pulkit Grover\IEEEauthorrefmark{3}
	}\\
	\IEEEauthorblockA{
		Electrical \& Computer Engineering,
		and the Center for the Neural Basis of Cognition\\
		Carnegie Mellon University
		\\
		\IEEEauthorrefmark{1}\href{mailto:vpraveen@cmu.edu}{\texttt{vpraveen@cmu.edu}}
		\IEEEauthorrefmark{2}\href{mailto:sanghamd@andrew.cmu.edu}{\texttt{sanghamd@andrew.cmu.edu}}
		\IEEEauthorrefmark{3}\href{mailto:pulkit@cmu.edu}{\texttt{pulkit@cmu.edu}}
	}
}

\begin{document}

\maketitle
\thispagestyle{plain}
\pagestyle{plain}

\begin{abstract} 
	We develop a theoretical framework for defining and identifying flows of information in computational systems. Here, a computational system is assumed to be a directed graph, with ``clocked'' nodes that send transmissions to each other along the edges of the graph at discrete points in time. We are interested in a definition that captures the dynamic flow of information about a specific \emph{message}, and which guarantees an unbroken ``information path'' between appropriately defined inputs and outputs in the directed graph. Prior measures, including those based on Granger Causality and Directed Information, fail to provide clear assumptions and guarantees about when they correctly reflect information flow about a message. We take a systematic approach---iterating through candidate definitions and counterexamples---to arrive at a definition for information flow that is based on conditional mutual information, and which satisfies desirable properties, including the existence of information paths. Finally, we describe how information flow might be detected in a noiseless setting, and provide an algorithm to identify information paths on the time-unrolled graph of a computational system.
\end{abstract} 

\section{Introduction} 

\subsection{Motivation}

Neuroscientists\footnote{A short version of this paper has appeared in the 2019 IEEE International Symposium on Information Theory~\cite{Venkatesh2019How}.} often seek an understanding of how information flows in the brain while it performs a particular task~\cite{Almeida2013Tool,Brovelli2004Beta,Bar2006Topdown,Greenberg2012Visuotopic}. As a concrete example, consider the experiment performed by Almeida et al.~\cite{Almeida2013Tool}, where they examine how images of common handheld tools are processed in the brain. In simple terms, the question they investigate is this: when attempting to identify a handheld tool, does one make use of knowledge of how to manipulate it? Two hypotheses present themselves: \begin{enumerate*}[label=(\roman*)]
	\item the answer to the above question is \emph{yes}, so we should expect that information about a tool's identity \emph{first} flows from visual cortex to motor cortex (the area responsible for processing manipulation), \emph{before} synthesis of visual and motor information occurs at the area of the brain responsible for object recognition.
	\item The answer to the aforementioned question is \emph{no}, so we should expect that the information about tools' identities \emph{first} flows from visual cortex to the area responsible for object recognition, \emph{after} which this information arrives at motor cortex.
\end{enumerate*} Thus, distinguishing between these hypotheses is equivalent to determining the \emph{path} along which information about a tool's identity flows in the brain. What methods can neuroscientists use to gain such an understanding? What formal theory underlies such an analysis? How does one mathematically define colloquially-used terms such as ``information flow''? These are the fundamental questions we try to answer in this paper.

Information flow is a concept that appears in several contexts, across fields ranging from communication systems, control theory and neuroscience to security, algorithmic transparency, and deep learning. While our primary motivation comes from neuroscience, the theory that we develop is broadly applicable to any system which can be modeled in the form of a directed graph, with nodes that communicate functions of their inputs to other nodes, and where transmissions are observable. For example, several kinds of social networks readily fit this bill, and one might wish to analyze how information spreads in such networks. Our framework is also general enough to analyze information flow in various kinds of Artificial Neural Networks: this could be useful for identifying specific paths that carry information distinguishing two or more classes, or for intelligently pruning an Artificial Neural Network post-training.

In the field of neuroscience, studying normal and diseased brain function involves gaining insight into how information is processed in the brain. Attaining such insight, in turn, may require determining how information flows between various parts of the brain. Thus, a nuanced understanding of information flow in the brain could help with diagnosing and treating brain diseases: a subject that is currently of immense interest with numerous efforts around the world~\cite{Hammond2007Pathological,Smith1998Microcircuitry,Grace2000Gating,Lalo2008Patterns}. More generally, an understanding of information flow is essential when considering how one might intervene to affect the output of a computational system, be it modulating how information spreads in a social network, or complementing dysfunctional components of the nervous system through stimulation (such as in retinal and cochlear implants).

\subsection{Our Goal and Approach}

Our overarching goal in this paper is develop a formal theory for understanding information flow in neuroscientific experiments. In order to properly scope our task, we choose to restrict our attention to ``\emph{event related experimental paradigms}''~\cite{Samuels2018Event}, a set of standard neuroscientific experimental design protocols where a \emph{stimulus} is shown to an animal subject or human participant, whose brain signals are being recorded. This restriction also allows us to decide precisely what \emph{kind} of information flow we are interested in, since in general, the phrase ``information flow'' can refer to more than one notion in neuroscience. We identify two dominant interpretations of ``information flow'': \begin{enumerate*}[label=(\roman*)]
	\item the first refers to information about a specific quantity or variable that is of interest to the experimentalist, which in this paper we refer to as the ``message'';
	\item the second refers to information in the abstract, and is usually used to describe the fact that one area of the brain ``drives'' or ``influences'' another area through the transmission of some information: in this interpretation, one is not interested in \emph{what} is being communicated, only that the communication is \emph{occurring}.
\end{enumerate*}
In this paper, we focus only on the first interpretation of the phrase, where we are interested in information about a \emph{specific message}, and we wish to track how information about this message flows within the brain. All references to ``information flow'', henceforth, refer only to the first interpretation. This is particularly common in event-driven paradigms, where the neuroscientist investigates how the brain responds to a carefully chosen set of stimuli, and examines how information contained in these stimuli (or alternatively, information contained in the \emph{response}) flows through the brain.

Given that we are interested in information about a specific message, what are we in pursuit of when we say information \emph{flow}? Broadly speaking, we want to develop a measure that will allow us to examine \emph{how} and \emph{at what times} information about a specific message flows from one area of the brain to another. In particular, we think of the brain as a \emph{computational system} executing an algorithm, and we want to capture how information about different variables might flow between different computational nodes of this system. A given ``message'' variable may be stored at a particular node for some time, a function may be computed using this message, and then the result may be passed on to a different node. A node that transmits information about the message at one time instant may not do so at a later time instant; thus information flow is a dynamic or time-dependent quantity. The computational system should allow for all of these possibilities, and our ensuing measure of information flow should enable us to \emph{track the path traversed by the message\footnote{or information derived thereof} through this system, over time}. This is the principal goal of our theoretical development, and will guide many of our decisions in model design.

We approach this goal by formally defining a computational system model: one based on nodes that represent distinct computational areas of the brain. These nodes can potentially represent the brain at any scale: single neurons, groups of neurons, or even whole brain regions, depending on the measurement modality and the kind of experiment being performed. The computational model we develop borrows various ideas from across several fields. The basic ingredients of the computational model, based on a \emph{graph with computational nodes}, derives from Thompson's work on VLSI complexity theory~\cite{Thompson1980Complexity}. In order to attain a \emph{dynamic} picture of information flow on the edges of the computational graph, and to deal with cycles in the flow of information, we use the idea of time-unrolling a graph, taking inspiration from Network Information Theory~\cite{Ahlswede2000Network}. Finally, to describe how the computational nodes can compute stochastic functions of variables based on their current inputs, we use the idea of Structural Causal Models from the field of Causality~\cite{Pearl2009Causality,Peters2017Elements}.

Within this computational model, we define a new measure for information flow about a \emph{specific message} that captures the \emph{dynamic} nature of information transmission. Ultimately, the measure should allow us to track how information about the message flows through the system, in the form of an \emph{unbroken information path}---this will be a key focus guiding our definitions. Given the nature of the problem, we rely on information-theoretic measures to define information flow. We motivate this definition through properties, and provide a series of candidate definitions and counterexamples before arriving at our final definition. When defining information flow in such a computational system, we restrict ourselves to ``\emph{observational}'' measures, which can be computed from a sample of all random variables described in the model. We deliberately eschew \emph{interventional} and \emph{counterfactual} measures, as the former require the capability to intervene on the system and change the distributions of the random variables involved, while the latter are a purely theoretical notion that can only be applied in situations where one can ask what \emph{might have occurred} if a specific variable had been different on a particular trial (while keeping the realizations of all other latent sources of randomness fixed).

The approach of building a rigorous theoretical framework that we have adopted in this paper is inspired by two works from biologists titled ``Can a biologist fix a radio?''~\cite{Lazebnik2002Can} and ``Could a neuroscientist understand a microprocessor?''~\cite{Jonas2017Could}. Both these works point to the lack of formal methods, i.e., systematic theory, that could help biologists understand the limitations of their tools and test their assumptions. It is our belief that information theory can help provide the formal methods that are sought in biology, and make an impact in fields such as neuroscience and neuroengineering~\cite{Grover2017Information,Grover2015Information,Venkatesh2017Lower}. In particular, information theory can play an important role in advancing how we understand large computational systems through external measurements and interventions. While developing an understanding of information flow in such systems may not be sufficient for providing a complete description of the nature of computation itself, we believe that it forms an integral component. Going forward, we believe that providing a formal theoretical framework for information flow is but a small part of several larger theoretical questions that are yet to be properly posed: questions such as how one might formalize ``reverse engineering'' the brain, or formalize the notion of ``understanding computation''.

\subsection{Related Work}

Prior work on statistically inferring flows of information in the brain appears under the umbrella of ``functional'' or ``effective connectivity''~\cite{Friston2011Functional,Friston2013Analysing,Bastos2016Tutorial}. These efforts have largely relied on measures of statistical\footnote{We borrow the use of the term ``statistical'' from Pearl~\cite[Sec.~1.5]{Pearl2009Causality}, who contrasts and differentiates ``statistical'' concepts from (strictly) ``causal'' ones.} causal influence such as Granger Causality~\cite{Granger1969Investigating,Bressler2011Wiener}, Massey's Directed Information~\cite{Massey1990Causality,Quinn2011Estimating,Quinn2015Directed,Jiao2013Universal}, Transfer Entropy~\cite{Schreiber2000Measuring} and Partial Directed Coherence~\cite{Baccala2001Partial}. Despite widespread use, these measures have frequently been a subject of debate and disagreement within the neuroscientific community~\cite{David2008Identifying,Roebroeck2011Identification,David2011fMRI,Stokes2017Study,Barnett2018Solved,Faes2017Interpretability,Stokes2017Reply}. In part, these disagreements stem from the widely-acknowledged fact that under non-ideal measurement conditions (e.g.~in the presence of hidden variables~\cite[p.~54]{Pearl2009Causality}, asymmetric noise~\cite{Andersson2005Testing,Nalatore2007Mitigating}, or limited sampling~\cite{Gong2015Discovering}), estimation of these quantities may be erroneous. While these non-idealities may eventually be overcome through improvements in technology, we believe that more fundamental issues still remain. For instance, one basic question that has remained unanswered is: when can statistical causal influence be interpreted as information flow about a message? In previous work, we demonstrated that even under \emph{ideal} measurement conditions, the direction of greater Granger causal influence can be opposite to the direction in which the message is being communicated in certain kinds of feedback communication networks~\cite{Venkatesh2015Direction}. This example points to a more general issue with the use of statistical causal influence measures: there is no direct way to interpret what the influence is \emph{``about''}. While it is understood in certain settings that ``information flow'' refers to information contained in a particular set of \emph{``stimuli''} (as mentioned in the previous section), the aforementioned measures do not incorporate the effect of the stimulus.

The existence of such fundamental issues can be traced back to the fact that there is no underlying model that links information flow (of some message of interest) with the signals that are actually \emph{measured}, leading to a lack of separation between the problems of \emph{defining} information flow and of \emph{estimating} it. The lack of such a computational model also makes it hard to test assumptions and to draw the right interpretations from experimental analyses. We believe that, following Shannon's approach of providing a theoretical foundation for information transmission~\cite{Shannon1948Mathematical}, a solid theoretical treatment of information flow is needed. Such a treatment would begin with a model of the underlying system, give a definition for information flow and describe its properties, and finally end with a suitable estimator. Adopting Shannon's model of defining entropy by stating a set of properties that such a measure must satisfy, we attempt to define information flow by putting forward an intuitive property that we believe is desirable for such a quantity. It is our hope that, by providing a theoretical foundation that separates definition and estimation, along with a concrete model and explicitly-stated assumptions, we can avoid many of the pitfalls encountered by previous approaches to understanding information flow in the brain.

It is useful at this point to mention the key differences between our measure of information flow, and measures based on Granger Causality and its generalizations:
\begin{enumerate}
	\item Our measure depends on a message $M$, that will often be related to the stimulus or the response in a neuroscientific task, whereas tools based on Granger causality do not.
	\item Since Granger causality-based tools use time series modeling to compute an estimate of information flow, they are unable to provide a dynamic, evolving picture of information flow between different areas over time.
	\item Since we start with a \emph{computational framework}, our model provides a direct way to connect information flow with the underlying computation. On the other hand, Granger causality-based tools start with a probabilistic graphical model of the observed nodes, and do not tie the analysis to computation in any way.
\end{enumerate}
While our proposed definition of information flow will also suffer from performance degradation under non-ideal measurement conditions, we believe that it overcomes the fundamental difficulty faced by Granger Causality-based tools: when measurements are ideal, our definition provides a clear and consistent way to interpret information flow about a message, as we illustrate through several examples in Section~\ref{sec:canonical-examples}.

Another line of work that appears within the functional and effective connectivity literature is Dynamic Causal Modeling (DCM)~\cite{Friston2003Dynamic,Friston2013Analysing}. This methodology is, in spirit, much more closely aligned with what we propose here. However, our framework differs from DCM in a few important ways: \begin{enumerate*}[label=(\roman*)]
	\item our underlying framework and model is based on Structural Causal Models rather than dynamical systems, and
	\item we seek to formalize the notion of information flow, not just of effective connectivity.
\end{enumerate*}
However, the style of thinking, which involves starting from theoretical models and incorporating the stimulus and experimental design, is common to both DCM and our approach.

\subsection{Outline of the Paper}

In this paper, we start by giving a mathematical description of a generic computational system, about which inferences are being drawn (Section~\ref{sec:comp-sys}). We then formally define what it means for information about some message to flow on a single edge or on a set of edges in the computational system (Section~\ref{sec:defining-info-flow}). This is done by proposing an intuitive property that we would like such flows to satisfy, along with some candidate definitions, and then examining which candidates satisfy the property. The intuitive property we desire is: \emph{information flow about a message may not completely disappear from the system at a certain time, only to spontaneously reappear at a later point} (formalized in Property~\ref{ppty:broken-telephone}). It emerges that simple and intuitive definitions actually fail to satisfy this basic property, and so a more sophisticated definition is needed. We then show how our definition for information flow about the message satisfies several desirable properties, including guarantees for the existence of so-called ``information paths'' between appropriately defined input and output nodes (Section~\ref{sec:info-flow-properties}). After this, we suggest how one might detect which edges of the computational system have information flow, and provide an ``information path algorithm'', which identifies the aforementioned information paths (Section~\ref{sec:inferring-info-flow}). We also introduce and discuss the concepts of derived information, redundant transmissions and hidden nodes, which allow one to obtain a more fine-grained understanding of information structure in the computational system. To show that our definition of information flow agrees with intuition, we give several canonical examples of computational systems and depict the information flow in each case (Section~\ref{sec:canonical-examples}). Finally, we conclude with discussions on connections with neuroscience, issues related to the difficulty of estimating information flow (along with possible remedies), comparison with the existing directed causal influence literature, connections with fields such as probabilistic graphical models and causality, and a discussion on information volume (Section~\ref{sec:concl-disc}).


\section{The Computational System} 
\label{sec:comp-sys}

Our goal is to develop a rigorous framework for understanding how the information about a message flows in a computational system. To do this, we first need to define the terms ``computational system'', ``message'', ``information about a message'' and ``flow''. In this section, we start with the first two terms, defining the model of the computational system that is used throughout this paper, and explicitly defining the message.

Our model is based on prior art in the information theory literature~\cite{Thompson1980Complexity,Ahlswede2000Network}, and consists of nodes communicating to each other at discrete points in time on a directed graph. At every time instant, each node receives transmissions on its incoming edges and computes a function of these transmissions to send out on its outgoing edges. This function can be random and time-dependent, and can be different for every outgoing edge. We will be interested in the flow of a particular random variable called the ``message'', which will be defined shortly. Since the directed graph forming the computational system may have cycles, the message may flow along a cyclic path. To deal with this possibility while capturing the fact that nodes must be causal\footnote{Causal in the ``Signals and Systems'' sense of the word, where a node cannot make use of future transmissions~\cite{Oppenheim1999Discrete}.}, we define a ``time-unrolled'' graph (in a manner similar to Ahlswede~et~al.\footnote{Although the work of Ahlswede et al.\ (2000) is titled ``Network \emph{Information Flow}'', it actually addresses a different problem: one of the achievable rate region of a broadcast network and the optimal coding strategy that achieves this rate. In contrast to their work, which concentrates on characterizing and achieving the optimal rate, our focus is on understanding how information about a known message flows in an existing computational system.}~\cite{Ahlswede2000Network}), which describes how nodes communicate to each other over time. We define a random variable model for the nodes' transmissions, and demonstrate how each node computes these variables. We also formally define the input nodes of the computational system, through their relationship with the message.


\begin{definition}[Complete directed graph] \label{def:complete-dir-graph}
	A \emph{complete directed graph} $\Gstar = (\Vstar, \Estar)$ is described by a set of nodes and the set of all edges between those nodes (including self-edges). We denote the set of nodes by their indices, $\Vstar = \{1, 2, \ldots, N\}$, where $N$ is a positive integer denoting the number of nodes in the graph. The set of edges in the graph is the set of all ordered pairs of nodes, $\Estar = \Vstar \times \Vstar$.
\end{definition}

Note that \begin{enumerate*}[label=(\roman*)]
	\item edges are directed, so the edge $(A, B) \in \Estar$ describes an edge \emph{from} node $A$ \emph{to} node $B$; and
	\item nodes have self-edges. For every $A \in \Vstar$, there is an edge $(A, A)$ in $\Estar$.
\end{enumerate*}

Moving forward, nodes shall be thought of as performing computations and possessing local memories. We shall interpret the transmission of a node to itself as the variable it stores within its memory\footnote{Instances of directed graphs that are \emph{not complete} and of nodes possessing \emph{no memory} are merely special cases of our model, where the respective edges' transmissions can simply be set to zero.}.


\begin{figure}[tb]
	\centering
	\begin{subfigure}{0.3\textwidth}
		\hfil
		\includegraphics{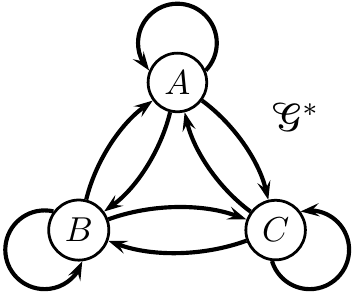}
		\hfil
	\end{subfigure}%
	\quad$\Rightarrow$\quad
	\begin{subfigure}{0.4\textwidth}
		\includegraphics{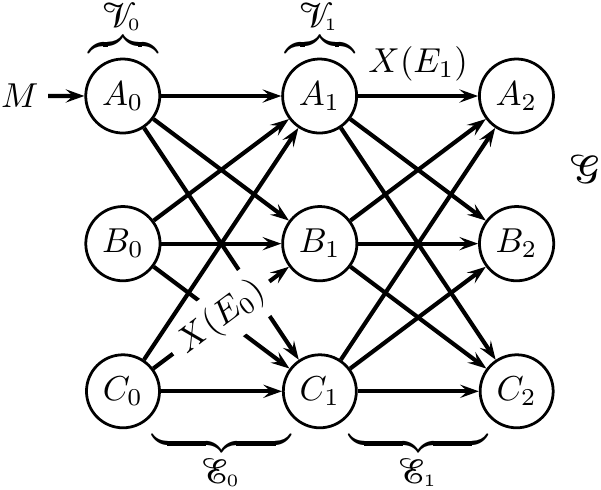}
	\end{subfigure}
	\caption{A diagram showing an example of a how a complete directed graph is unrolled to create a time-unrolled graph. On the left, we show a complete directed graph $\Gstar$ that has three nodes, $\Vstar = \{A, B, C\}$. These nodes are fully connected to each other via edges $\Estar$, including self-edges. \\
	On the right, we show how $\Gstar$ has been unrolled using time indices $\T = \{0, 1, 2\}$ to obtain a time-unrolled graph $\G$. The set of all nodes at time $t=0$ is $\V_0$ and the set of all (outgoing) edges at time $t=0$ is denoted $\E_0$. As an example, we have shown an arbitrary edge $E_0 \in \E_0$ (here, $E_0 = (C_0, B_1)$) and the transmission on that edge, $X(E_0)$. As another example, we show a ``self-edge'' in the time-unrolled graph, $E_1 \in \E_1$, which in this case is $E_1 = (A_1, A_2)$. Also depicted is the transmission $X(E_1)$ on this self-edge, which is interpreted as the contents of the memory of node $A$ from $t=1$ to $t=2$. The message $M$ arrives at the input node $A_0$, but could in general be available at more than one node at $t=0$. \\
	In subsequent illustrations, we do not depict all edges at every time step, even though they are present. This is done only for the sake of clarity.}
	\label{fig:unrolling}
\end{figure}

\begin{definition}[Time-unrolled graph]
	In order to allow nodes to have different transmissions at every time instant, we must provide for the progression of time. Let $\T = \{0, 1, \ldots, T\}$ be a set of time indices, where $T$ is a positive integer representing the maximum time index. Then, a \emph{time-unrolled graph} $\G = (\V, \E)$ is constructed by indexing a complete directed graph $\Gstar$ using the time indices $\T$ as follows:
	\begin{enumerate}
		\item The nodes $\V$ consist of all nodes $\Vstar$ in $\Gstar$, subscripted by time indices in $\T$,
			\begin{equation*}
				\V = \{A_t : A \in \Vstar, t \in \T\};
			\end{equation*}
		\item The edges $\E$ connect nodes of \emph{successive} times in $\V$, so they can be written in terms of the edges in $\Estar$ as
			\begin{equation*}
				\E = \{(A_t, B_{t+1}) : (A, B) \in \Estar, t \in \T\}.
			\end{equation*}
	\end{enumerate}
\end{definition}

For brevity, we denote the set of all nodes at time $t$ by $\V_t$, and the set of all (outgoing) edges at time $t$ by $\Et$. So, for example, we will have $A_1 \in \V_1$ and $(A_1, B_2) \in \E_1$. All of the notation in this section can be visualized in Figure~\ref{fig:unrolling} and is summarized in Table~\ref{tab:notation}.

Once again, note that
\begin{enumerate*}[label=(\roman*)]
	\item edges at time $t$ connect nodes at time $t$ to nodes at time $t+1$; and
	\item since the original graph $\Gstar$ had self-edges, there will always be an edge $(A_t, A_{t+1})$ in $\Et$ for every node $A_t \in \V_t$.
\end{enumerate*}

Also note, we have only presented the complete directed graph in Definition~\ref{def:complete-dir-graph} in order to explicitly define the process of time-unrolling. We do not expect the time-unrolled graph to be ``rolled back'' into a complete directed graph at the end of an information flow analysis. Since we seek a time-evolving picture of information flow between different computational nodes, we will directly view and interpret information flow on the time-unrolled graph. This is illustrated later, through several examples, in Section~\ref{sec:canonical-examples}.


\begin{definition}[Computational System] \label{def:comp-sys}
	A \emph{computational system} $\C = (\G, X, W, f)$ is a time unrolled graph $\G$ that has \emph{transmissions on its edges} which are constrained by \emph{computations at its nodes}. The \emph{input} to the computational system includes a \emph{message}\footnote{The message is the random variable \emph{whose} ``information flow'' we will seek to identify.}, $M$. We now elaborate upon these terms:

	\begin{enumerate}[style=nextline,label=\textit{\thedefinition\alph*)},ref=\thedefinition\alph*]
		\item \emph{Transmissions on Edges}

			We begin by defining a function which maps every edge of $\G$ to a random variable. Let $\mathcal X$ be the set of all random variables in some probability space\footnote{We assume that all probability distributions are such that the mutual information and conditional mutual information between any sets of random variables is well-defined~\cite[Sec.~2.6]{Polyanskiy2017Lecture}.}. Then, let $X : \E \to \mathcal X$ be a function that describes what random variable is being transmitted on a given edge, i.e., $X(E)$ is the random variable corresponding to the transmission on the edge $E$.

			For convenience, we define $X$ applied to a \emph{set of edges} as the set of random variables produced by applying $X$ to each of those edges individually, i.e., for any set $\E' \subseteq \E$,
			\begin{equation} \label{eq:x-of-set}
				X(\E') = \{X(E) : E \in \E'\}.
			\end{equation}
			We extend the use of this notation to other functions of nodes and edges that we define, going forward.

		\item \emph{Computation at a Node} \label{def:comp-at-node}

			Let $A_t \in \V_t$ be a node in the time-unrolled graph $\G$, at some time $t \geq 1$ (recall that $t \in \{0, 1, \ldots, T\}$). Let $\P(A_t)$ be the set of edges entering $A_t$, and $\Q(A_t)$ be the set of edges leaving $A_t$. Further, let us suppose that $A_t$ is able to intrinsically generate the random variable\footnote{$X(E_t)$ and $W(A_t)$ may also be random \emph{vectors} instead of random variables, i.e., an edge may \emph{transmit a vector}. This does not affect the theoretical development presented in this paper; all of our proofs remain unchanged.} $W(A_t)$ at time $t$, where $W(A_t) \independent W(\V\!\setminus\!\{A_t\}) \;\forall\; A_t \in \V$, $W(\V_t) \independent \{M, X(\E_{t-1})\}$ and the symbol ``$\independent$'' stands for independence between random variables. Then, the \emph{computation} performed by the node $A_t$ (for $t \geq 1$) is a deterministic function\footnote{This kind of model is not new, and can be found in the causality literature for instance, under the name ``Structural Equation Models''~\cite[Sec.~1.4.1]{Pearl2009Causality}.} $f_{A_t}$ that satisfies
			\begin{equation} \label{eq:node-comp}
				f_{A_t}\bigl(X(\P(A_t)), W(A_t)\bigr) = X(\Q(A_t)).
			\end{equation}
			Here, $X(\E_{t-1})$, $W(\V\!\setminus\!\{A_t\})$, $W(\V_t)$, $X(\P(A_t))$ and $X(\Q(A_t))$ all make use of the notation described in~\eqref{eq:x-of-set}.

			Note that the definition above does not apply when $t=0$; this is a special case which is discussed below. Also, for convenience, where $\A$ is an arbitrary set of nodes, we will use $f_{\A}$ to denote the ``joint function'' mapping the incoming transmissions of all nodes in $\A$ (along with their intrinsic random variables $W(\A)$) to their respective outgoing transmissions.

		\item \emph{The Message and the Input Nodes} \label{def:input-nodes}

			Each of the nodes in $\V_0$ may receive one or more random variables from the world external to the computational system at time $t=0$. The \emph{message}, $M$, is simply a specific random variable that is of interest to the experimentalist observing the computational system, and for which we shall define information flow. For now, we assume that we are interested in a single message.\footnote{That is, we assume that the message is a single random variable or vector. It is possible to simultaneously examine the information flows of several (possibly dependent) messages, or of sub-messages within a single message. These cases are examined in Section~\ref{sec:multiple-messages}.} We also assume that the message enters the computational system only at time $t=0$, and at no later time instant.

			We formally define the \emph{input nodes} of the system as those nodes of $\G$, at time $t=0$, whose transmissions statistically depend on the message $M$:
			\begin{equation}
				\Vip \coloneqq \{A_0 \in \V_0 : I\bigl(M ; X(\Q(A_0))\bigr) > 0\},
			\end{equation}
			where $\Q(A_0)$ represents the set of edges leaving the node $A_0$.

			To remain consistent with Definition~\ref{def:comp-at-node}, we define the computation performed by an input node $A_0 \in \Vip$ as a function $f_{A_0}$ that satisfies
			\begin{equation}
				f_{A_0}\bigl(M, W(A_0)\bigr) = X(\Q(A_0)),
			\end{equation}
			and the computation performed by a non-input node at time $t=0$, $A_0 \in \V_0\!\setminus\!\Vip$, as a function $f_{A_0}$ that satisfies
			\begin{equation}
				f_{A_0}\bigl(W(A_0)\bigr) = X(\Q(A_0)).
			\end{equation}
			As before, $W(A_0) \independent W(\V_0\!\setminus\!\{A_0\}) \;\forall\; A_0 \in \V_0$ and $W(\V_0) \independent M$.

	\end{enumerate}
\end{definition}

\linespread{1.0}
\begin{table}[!t]
	\caption{Summary of Notation}
	\label{tab:notation}
	\centering
	\tabulinesep=0.75mm
	\ifthenelse{\boolean{arxiv}}{\footnotesize}{}  
	\begin{tabu}{c X}
		\toprule
		Variable(s) & Meaning \\
		\midrule
		$\Gstar = (\Vstar, \Estar)$ & The original complete directed graph, prior to time-unrolling \\
		$\G = (\V, \E)$ & The time-unrolled graph making up the computational system \\
		$\T$ & The set of all time points, $\{0, 1, \ldots, T\}$ \\
		$\V$ & The set of all nodes in the computational system \\
		$\V_t$ & The subset of nodes at time $t$ \\
		$V_t, A_t, B_t, C_t, D_t$ & A node in the graph at time $t$ \\
		$V, A, B, C, D, E$ & A node in the original complete directed graph $\Gstar$, or a node in the computational system at an unspecified time point \\
		$\A, \B$ & Some subset of nodes in $\V$ \\
		$\E$ & The set of all edges in the computational system \\
		$\E_t$ & The set$^\dagger$ of all edges at time $t$ \\
		$\E_t'$ & Some subset$^\ddagger$ of edges in $\E_t$ \\
		$E_t, P_t, Q_t, R_t, S_t$ & An edge in the computational system at time $t$ \\
		$E, P, Q, R, S$ & An edge in the original complete directed graph $\Gstar$, or an edge in the computational system at an unspecified time point \\
		$X(E_t)$ & The random variable representing the transmission on the edge $E_t$ \\
		$X(\{E^{(1)}, E^{(2)}\})$ & Short-hand notation for $\{X(E^{(1)}), X(E^{(2)})\}$ (refer Equation~\eqref{eq:x-of-set})\\
		$\P(V_t)$ & The set of all incoming edges of $V_t$ $(= \V_{t-1} \times \{V_t\} \subseteq \E_{t-1})$ \\
		$\Q(V_t)$ & The set of all outgoing edges of $V_t$ $(= \{V_t\} \times \V_{t+1} \subseteq \E_t)$ \\
		$W(V_t)$ & The intrinsically generated random variable at the node $V_t$ \\
		$M$ & The ``message'', a random variable that enters the system at time $t=0$, and whose information flow we seek to understand (refer Definition~\ref{def:input-nodes}) \\
		$\Vip$ & The input nodes: the subset of nodes at time 0 whose outgoing transmissions depend on the message $M$ (refer Definition~\ref{def:input-nodes}) \\
		$f_{V_t}$ & The function computed by the node $V_t$ (refer Definition~\ref{def:comp-at-node}) \\
		\bottomrule
		\multicolumn{2}{l}{$^\dagger$Script forms typically denote sets} \\
		\multicolumn{2}{l}{$^\ddagger$Primed script forms typically denote subsets}
	\end{tabu}
\end{table}
\ifthenelse{\boolean{arxiv}}{\linespread{1.1}}{\linespread{1.3}}

\paragraph*{Remarks}
\begin{enumerate}
	\item Informally speaking, Definition~\ref{def:comp-sys} is designed to allow each node to generate a randomized function of its incoming transmissions for each of its outgoing transmissions.
	\item The randomization at each node is explicitly captured by its intrinsic random variable $W(\cdot)$, and is assumed to be independent across all nodes of the system.
	\item Furthermore, each node is allowed to send a different transmission on each of its outgoing edges.
	\item Note that the condition imposed by Equation~\eqref{eq:node-comp} introduces dependence between the random variables in the set $X(\E)$.
	\item For the most part, we will not be concerned with the precise form of the computation being performed by every node. We will only make use of information-theoretic measures applied to the message and to the random variables in the computational system.
\end{enumerate}

Throughout the paper, we use the variables $U$, $V$, $A$, $B$, $C$ and $D$ to refer to nodes and $E$, $P$, $Q$, $R$ and $S$ to refer to edges. We use their script forms, e.g.\ $\R$, when referring to sets of nodes and edges, and primed script forms, e.g.\ $\R'$,  when referring to subsets thereof. Once again, the notation we use is summarized in Table~\ref{tab:notation}, and depicted in Figure~\ref{fig:unrolling} for convenience.

Having defined what we mean by the terms ``computational system'' and ``message'', in the following sections we proceed to find a definition for ``information flow'' and identify properties that this definition satisfies in any computational system.


\section{Defining Information Flow} 
\label{sec:defining-info-flow}

Before one can speak of \emph{detecting} information flow in a network, it is first important to \emph{define} what it is that we seek to detect.\footnote{In essence, ``causal influence'' measures such as Granger Causality and Directed Information, while intuitively quantifying transferred information, fail to lay down what \emph{aspect of computation} they actually capture. This is, in part, a result of conflating the stages of defining a quantity we want to understand, and prescribing an estimator for it.} In this section, we focus on arriving at a definition for information flow.

Our goal is to formalize how information about a message flows in a computational system. Ultimately, we expect to find the \emph{path} that the message takes while being processed by the system. Towards this, we start by trying to formally define what it means for information about the message to flow on a given \emph{edge}. This section concludes with a proposal for such a definition: one based on strict positivity of a conditional mutual information. But to provide the intuition behind this choice of definition, we start with several simpler candidate definitions, and show how they fail to satisfy an intuitive property using counterexamples.

After proposing a definition for information flow, in Section~\ref{sec:info-flow-properties}, we discuss the \emph{properties} satisfied by our definition. Then, in Section~\ref{sec:inferring-info-flow}, we specify how the transmissions of the computational system are observed, and describe how information flow might be \emph{inferred} in a real computational system.

\subsection{An intuitive property}

To concretely define what it means for information about a message to flow on an edge, we need some way to assess competing candidate definitions and choose one among them. Towards this goal, we state a straightforward and intuitive property, which we would want any definition of information flow to satisfy.

Suppose that, at a given point in time, there is \emph{no} flow of information about the message across \emph{any} edge of a computational system. Note that this includes self-edges, so no node ``carries'' information about the message within its memory either. Then, we expect that information about the message has ceased to persist in the system, so the information flow about the message \emph{must} be zero on all edges of the computational system, at all future points in time.

\begin{property}[The Broken Telephone\footnote{\texttt{\url{https://en.wikipedia.org/wiki/Telephone_game}}}] \label{ppty:broken-telephone}
	Let $\C$ be a computational system, and let $\F_M : \E \to \{0, 1\}$ be an indicator of the presence of information flow about $M$ on an edge. That is, $\F_M(E) = 1$, if information about $M$ flows on the edge $E \in \E$ and $\F_M(E) = 0$, otherwise. The Broken Telephone Property states that if, at some time $t \in \T$, we have
	\begin{align}
		\F_M(E_t) = 0 &\qquad\forall\; E_t \in \E_t,
		\shortintertext{then}
		\F_M(E_{t'}) = 0 &\qquad\forall\; E_{t'} \in \E_{t'} \;\;\forall\; t' \in \T, t' > t.
	\end{align}
\end{property}

\subsection{Intuiting Information Flow through Counterexamples}

We now propose four candidate definitions, beginning with the simplest. We then construct counterexamples to show how the first three candidate definitions do not satisfy Property~\ref{ppty:broken-telephone}.


\begin{candidate} \label{cd:dependence}
	A simplistic and intuitive definition for information flow might simply stem from dependence. We say that information about the message $M$ flows on an edge $E_t$ if
	\begin{equation*}
		I\bigl(M ; X(E_t)\bigr) > 0.
	\end{equation*}
\end{candidate}


\begin{figure}[tb]
	\centering
	\includegraphics{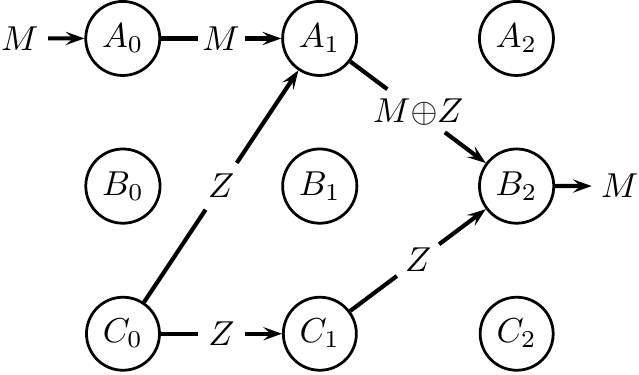}
	\caption{The computational system for Counterexample~\ref{ce:dependence}. We only depict edges relevant to the counterexample here. All other edges in the underlying complete directed graph are still present, but are not shown; their transmissions are assumed to be zero. Observe that no edge at time $t=1$ has information flow as per Candidate Definition~\ref{cd:dependence}, yet the message reappears at time $t=2$.}
	\label{fig:ce-dependence}
\end{figure}

\begin{counterexample} \label{ce:dependence}
	Consider the computational system depicted in Figure~\ref{fig:ce-dependence} (note that, in order to avoid unnecessary clutter, only edges with non-zero transmissions are shown in the figure). $A_0$ is the input node, which has the message $M \sim \text{Ber}(1/2)$ at time $t=0$. The system's goal is to communicate\footnote{This communication can be thought of as computing the identity function, and making the output available at the node $B$.} $M$ to the node $B$. It chooses the following strategy: at $t=0$, $A_0$ ``transmits'' $M$ to $A_1$ (i.e., node $A$ stores $M$ in its memory). $C_0$ independently generates a different random number, ${W(C_0) = Z \sim \text{Ber}(1/2)}$, $Z \independent M$, and sends this message to $A_1$, while also storing it in memory it until $t = 1$. $A_1$ then computes $M \xor Z$ and passes the result to $B_2$, while $C_1$ sends $Z$ to $B_2$. Here, the symbol ``$\xor$'' stands for \textsc{xor}, the exclusive-\textsc{or} operator on two bits. $B_2$ is thus able to recover $M$ by once again \textsc{xor}-ing its inputs, $(M \xor Z)$ and $Z$.

	Note that the output of $B_2$ depends on $M$, even though none of its inputs individually depends on $M$. That is, $I\bigl(M ; X((A_1, B_2))\bigr) = I(M ; M \xor Z) = 0$, and $I\bigl(M ; X((C_1, B_2))\bigr) = I(M ; Z) = 0$, so by Candidate Definition~\ref{cd:dependence}, information about the message flows on \emph{no} edge at time $t=1$. However, information about the message \emph{does} flow out of node $B_2$ at time $t=2$. This violates Property~\ref{ppty:broken-telephone}. Thus, mere \emph{dependence} on the message cannot be a valid definition for flow of information on a single edge.~\hfill\IEEEQED
\end{counterexample}

Communication strategies such as the one in Counterexample~\ref{ce:dependence} frequently arise in cryptography~\cite{Shannon1949Communication}, to prevent an eavesdropper from reading confidential information, and in network coding~\cite{Ahlswede2000Network}, for achieving the communication capacity of a network. Furthermore, a complex computational network may have smaller sub-networks with such topologies. For instance, we observe such a sub-network in the canonical example for network coding: the butterfly network~\cite[Fig.~7b]{Ahlswede2000Network} (this particular example is discussed in detail in Section~\ref{sec:eg-nc-butterfly}). Optimal communication in such a network \emph{requires} the use of such topologies, so Counterexample~\ref{ce:dependence} is far from obscure. In fact, central to the idea of Counterexample~\ref{ce:dependence} is a concept known as ``synergy'', which is well-studied in the literature on Partial Information Decomposition~\cite{Williams2010Nonnegative,Harder2013Bivariate,Bertschinger2014Quantifying} (see~\cite{Lizier2018Information} for a recent review). This is discussed at length in Section~\ref{sec:syn-info-connection}. Even in neuroscience, the concept of synergy is recognized and well-understood~\cite{Schneidman2003Synergy,Latham2005Synergy,Timme2018Tutorial}, and some experimental evidence has appeared in the literature~\cite{Gat1999Synergy}.

Counterexample~\ref{ce:dependence} demonstrates that the information necessary to recover the message (or a function of it) is not necessarily transmitted through individual edges, but jointly across edges. So, we might instead seek to define the ``smallest set of edges'' along which information about the message flows, for every point in time. But if we ultimately wish to isolate \emph{paths} along which information about the message flows, we require an understanding of which edges \emph{specifically} the information flows upon. We therefore continue to think of information as flowing on individual edges.\footnote{It should be noted that the two views---information flowing on individual edges, versus sets of edges---are compatible with each other if we use Definition~\ref{def:info-flow-set} (which will appear shortly) to describe information flow on a set of edges. This equivalence is elaborated upon in Section~\ref{sec:info-flow-set}. Later, in Section~\ref{sec:separation-property}, we attempt to refine our understanding of the aforementioned ``smallest set of edges'' along which information about the message flows.}


We can now update our na\"ive definition to counter the previous counterexample. We start by noting that in Counterexample~\ref{ce:dependence}, although the transmission on edge $(A_1, B_2)$ is independent of $M$, it is not \emph{conditionally} independent of $M$ when given the transmission on $(C_1, B_2)$.

\begin{candidate} \label{cd:cond-one}
	We say that information about the message $M$ flows on an edge $E_t \in \Et$ if one of the following holds:
	\begin{enumerate}
		\item $I\bigl(M ; X(E_t)\bigr) > 0$, or
		\item $\exists\; E_t' \in \Et$ s.t.\ $I\bigl(M ; X(E_t) \given X(E_t')\bigr) > 0$.
	\end{enumerate}
\end{candidate}

\begin{figure}[tb]
	\centering
	\includegraphics{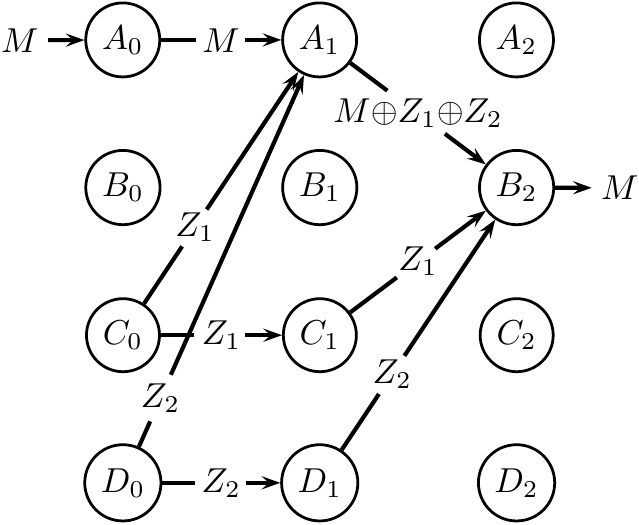}
	\caption{The computational system for Counterexample~\ref{ce:cond-one}. Once again, observe that no edge at time $t=1$ has information flow as per Candidate Definition~\ref{cd:cond-one}, yet the message reappears at time $t=2$. Note that only edges relevant to the counterexample are depicted in the figure. All other edges of the underlying complete directed graph are still present, and their transmissions are assumed to be zero.}
	\label{fig:ce-cond-one}
\end{figure}

\begin{counterexample} \label{ce:cond-one}
	Consider a modified version of Counterexample~\ref{ce:dependence}, shown in Figure~\ref{fig:ce-cond-one}. Now, since there are \emph{two} noise terms, no single extra edge may be conditioned upon to have non-zero information flow at time $t=1$. So, Candidate Definition~\ref{cd:cond-one} also fails to satisfy Property~\ref{ppty:broken-telephone}.~\hfill\IEEEQED
\end{counterexample}


It might seem that a possible rectification is to condition on \emph{all} other edges at time $t$, but we can show that this also fails the test.

\begin{candidate} \label{cd:cond-all}
	We say that information about the message $M$ flows on an edge $E_t \in \Et$ if one of the following holds:
	\begin{enumerate}
		\item $I\bigl(M ; X(E_t)\bigr) > 0$, or
		\item $I\bigl(M ; X(E_t) \given X(\Et\!\setminus\!\{E_t\})\bigr) > 0$.
	\end{enumerate}
\end{candidate}

\begin{figure}[tb]
	\centering
	\includegraphics{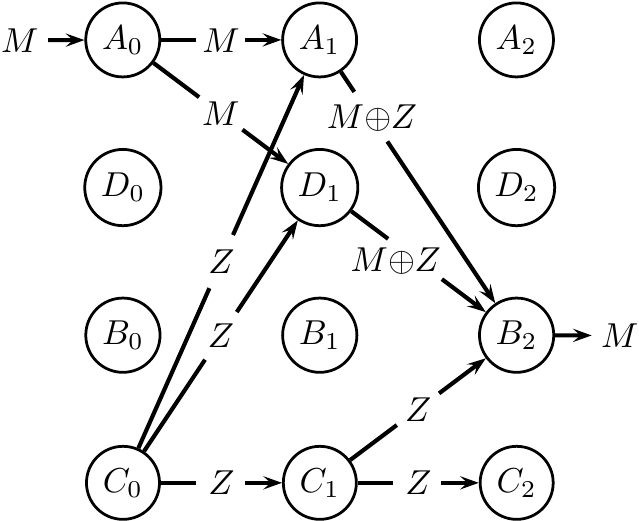}
	\caption{The computational system for Counterexample~\ref{ce:cond-all}. Just as in the previous counterexamples, no edge at time $t=1$ has information flow as per Candidate Definition~\ref{cd:cond-all}, yet the message is reconstructed at time $t=2$. Note that only edges relevant to the counterexample are depicted in the figure. All other edges of the underlying complete directed graph are still present, and their transmissions are assumed to be zero.}
	\label{fig:ce-cond-all}
\end{figure}

\begin{counterexample} \label{ce:cond-all}
	Consider the computational system shown in Figure~\ref{fig:ce-cond-all}. Once again, we have an input node $A_0$ which possesses the message at time $t=0$, and wishes to send this message to node $B$. It does so by mixing $M$ with an independent random variable $Z$ generated at $C_0$, so that the scenario described in Counterexample~\ref{ce:dependence} still holds. But additionally, $A$ communicates to $B$ along a redundant path, through $D_1$. Now, if $E$ is any incoming edge of $B_2$, it is still true that $I\bigl(M ; X(E)\bigr) = 0$. So none of the inputs of $B_2$ individually depends on $M$, thus eliminating the first condition in Candidate Definition~\ref{cd:cond-all}. Furthermore, checking each incoming edge of $B_2$ reveals that the second condition also fails to hold. If we take $E_1 = (A_1, B_2)$, we get
	\begin{equation}
		I\bigl(M ; X(E_1) \given X(\E_1\!\setminus\!\{E_1\})\bigr) = I(M ; M \xor Z \given M \xor Z, Z) = 0.
	\end{equation}
	The same holds true when $E_1 = (D_1, B_2)$ since the transmissions on both edges are identical by construction. Likewise, if we take $E_1 = (C_1, B_2)$, we have
	\begin{equation}
		I\bigl(M ; X(E_1) \given X(\E_1\!\setminus\!\{E_1\})\bigr) = I(M ; Z \given M \xor Z, Z) = 0,
	\end{equation}
	with the same holding true when $E_1 = (C_1, C_2)$. Therefore, no edge at time $t=1$ has any information flow about the message $M$, as per Candidate Definition~\ref{cd:cond-all}. Nevertheless, $B_2$ is able to recover the message at time $t=2$, proving that Property~\ref{ppty:broken-telephone} fails to hold for Candidate Definition~\ref{cd:cond-all}.~\hfill\IEEEQED
\end{counterexample}


\subsection{Information Flow on a Single Edge}


The counterexamples presented in the previous section motivate a new definition for when information about the message can be said to flow on a given edge. Neither \emph{dependence} of $M$ on the transmission of an edge, nor conditional dependence given \emph{one} or \emph{all} other edges, satisfy Property~\ref{ppty:broken-telephone}.

However, in all these counterexamples, given an edge $E_t$ upon which we expect to have non-zero information flow, we observe: there is at least one subset of edges $\Et' \subseteq \Et\setminus\{E_t\}$, such that when given $X(\Et')$, $X(E_t)$ is conditionally dependent\footnote{Equivalently, we could say that there exists at least one subset of edges $\Et' \subseteq \Et$, without explicitly excluding $E_t$, since $I\bigl(M ; X(E_t) \given X(E_t), X(\Et')\bigr) = 0$.} on $M$. In Counterexample~\ref{ce:dependence}, the edge $(A_1, B_2)$, carrying $M \xor Z$, is conditionally dependent on $M$, given $X\bigl((C_1, B_2)\bigr) = Z$. In Counterexample~\ref{ce:cond-one}, $X\bigl((A_1, B_2)\bigr) = M \xor Z_1 \xor Z_2$ is conditionally dependent on $M$, given $\{X\bigl((C_1, B_2)\bigr), X\bigl((D_1, B_2)\bigr)\} = \{Z_1, Z_2\}$. And finally, in Counterexample~\ref{ce:cond-all}, $X\bigl((A_1, B_2)\bigr) = M \xor Z$ is conditionally dependent on $M$, given $X\bigl((C_1, B_2)\bigr) = Z$; note that we do \emph{not} condition on $X\bigl((D_1, B_2)\bigr) = M \xor Z$. Thus, conditioning on a subset of the other edges' transmissions creates dependence between $M$ and the transmission on an edge of interest.

We will shortly prove that Property~\ref{ppty:broken-telephone} holds when information flow is defined as below, so we directly state it as a definition, skipping its candidacy status.

\begin{definition}[$M$-information Flow on a Single Edge] \label{def:info-flow}
	We say that information about the message $M$ flows on an edge $E_t \in \Et$ if
	\begin{equation}
		\exists\; \Et' \subseteq \Et\setminus\{E_t\} \quad\text{s.t.}\quad I\bigl(M ; X(E_t) \given X(\Et')\bigr) > 0.
	\end{equation}
	Henceforth, we refer to ``information flow about the message $M$'' as \emph{$M$-information flow}, and use the phrase ``the edge $E_t$ \emph{has} $M$-information flow'' or ``the edge $E_t$ \emph{carries} $M$-information flow'' to mean that information about $M$ flows on $E_t$ per this definition.
\end{definition}

Note that if $I\bigl(M ; X(E_t) \given X(\E_t')\bigr) > 0$, then $I\bigl(M ; X(\{E_t\} \cup \E_t')\bigr) > 0$. In other words, \emph{there exists} a set of edges that includes $E_t$, whose transmissions depend on $M$. This is why it is important to condition on all possible subsets of $\E_t$. It is not immediately clear, however, whether \emph{every} edge in $\{E_t\} \cup \E_t'$ has $M$-information flow. We return to this point in Section~\ref{sec:separation-property}.

Also, this definition implies that certain edges, such as $(C_1, B_2)$ in Counterexample~\ref{ce:dependence}, may have $M$-information flow, which may seem counter-intuitive. This is discussed further and justified in Section~\ref{sec:orphans}.


\subsection{Information Flow on a Set of Edges}
\label{sec:info-flow-set}

The definition of $M$-information flow for a single edge naturally generalizes to one for a set of edges, at a given time.

\begin{definition}[$M$-information Flow on a Set of Edges] \label{def:info-flow-set}
	We say that information about the message $M$ flows on a set of edges $\E_t' \subseteq \E_t$ if
	\begin{equation}
		\exists\; \R_t' \subseteq \E_t \quad \text{s.t.} \quad I\bigl(M ; X(\E_t') \given X(\R_t')\bigr) > 0.
	\end{equation}
\end{definition}

The definition of $M$-information flow on a set of edges is nearly identical to its single-edge counterpart. Indeed, they are closely related, as the following proposition shows.

\begin{proposition} \label{prop:set-info-equiv}
	A set $\E_t' \subseteq \E_t$ has $M$-information flow if and only if there exists an edge $E_t' \in \E_t'$ that has $M$-information flow.
\end{proposition}

A proof of this proposition can be found in Appendix~\ref{app:set-info-equiv}.

It should be noted that although the counterexamples in this section all employed computational systems which recovered the message $M$ at a new node at a later time, a computational system will in general compute some function of the message. For instance, see the example in Section~\ref{sec:eg-fft}.

\subsection{The Connection with Synergistic Information}
\label{sec:syn-info-connection}

This section connects our definition of $M$-information flow with recent developments on a subject known as ``Partial Information Decomposition'' (PID). Our definition is closely related to the concept of ``Synergistic Information'' that appears in this field. This section exists only for the purpose of providing a deeper intuition for our definition of $M$-information flow, and does not affect the rest of the paper in any significant way. We have attempted to explain this intuition in a way that is accessible to readers unfamiliar the PID literature. However, readers may feel free to skip this section, if desired.

At its core, Counterexample~\ref{ce:dependence} relies on a concept known as ``synergy'', which is described explicitly in the literature on Partial Information Decomposition (PID)~\cite{Williams2010Nonnegative,Harder2013Bivariate,Bertschinger2014Quantifying} (see~\cite{Lizier2018Information} for a recent review, and Appendix~\ref{app:synergistic-info-flow} for a brief introduction). Essentially, this body of literature seeks to decompose the mutual information that two or more variables share about a message, $I\bigl(M ; (Y_1, Y_2, \ldots)\bigr)$, into several individually meaningful, non-negative components. In particular, when discussing the bivariate case---i.e., the case of two variables, $I\bigl(M ; (Y_1, Y_2)\bigr)$---it is understood what the terms in this decomposition should be: \begin{enumerate*}[label=(\roman*)]
	\item information about the message that each variable carries \emph{uniquely}, and which cannot be inferred from the other;
	\item information about the message that the variables share \emph{redundantly}, and which can be extracted from either;
	\item and information about the message that the variables convey \emph{synergistically}, which is revealed only when \emph{both} variables are taken together, and cannot be inferred from either variable \emph{individually}.
\end{enumerate*} Counterexample~\ref{ce:dependence} is the canonical example for synergy, and is known simply as the ``\textsc{xor}'' example in the PID literature. While $M \xor Z$ and $Z$ are \emph{individually} independent of $M$, when taken \emph{together}, $I\bigl(M ; (M \xor Z, Z)\bigr) = H(M)$. This suggests that $M \xor Z$ and $M$ have no unique or shared information about $M$, but convey information synergistically.

While the field has not yet arrived at a consensus on the most appropriate definitions for unique, redundant and synergistic information~\cite{Lizier2018Information}, it is well-understood what properties these quantities must satisfy, at least in the bivariate case (see Appendix~\ref{app:synergistic-info-flow}, specifically, Equations \eqref{eq:pid-decomposition}, \eqref{eq:pid-mi-expansion} and \eqref{eq:pid-gt-zero}). Therefore, even without formal definitions, we can rely on the intuition provided by these properties to understand the implications of PID for $M$-information flow. If a particular edge's transmission contains unique or redundant information about the message (with respect to some other subset of edges at that point in time), then that information will manifest itself in the form of strictly positive mutual information. However, in the absence of positive mutual information between the message and the transmission on a given edge, we need to consider whether said transmission synergistically interacts with another subset of transmissions at that point in time, as this could potentially create dependence with the message through the kind of ``recombination'' described in Counterexample~\ref{ce:dependence}. We then need to decide whether such synergistic interactions ought to be considered to constitute information flow. As we show below, our definition of $M$-information flow \emph{does} consider instances of purely synergistic information to constitute information flow.

Indeed, it is possible to formulate a definition for information flow based on synergy, which is completely equivalent to Definition~\ref{def:info-flow}. The definition below makes use of the PID preliminaries given in Appendix~\ref{app:synergistic-info-flow}.
\begin{definition}[$M$-synergistic information flow] \label{def:syn-info-flow}
	We say that an edge $E_t$ has \emph{$M$-synergistic information flow} if at least one of the following holds:
	\begin{enumerate}
		\item $I\bigl(M ; X(E_t)\bigr) > 0$, or \label{cond:syn-info-cond-1}
		\item $\exists\;\E_t' \subseteq \E_t\setminus\{E_t\} \quad \text{s.t.} \quad CI\bigl(M : X(E_t) ; X(\E_t')\bigr) > 0,$ \label{cond:syn-info-cond-2}
	\end{enumerate}
	where $CI(M : X ; Y)$ represents the synergistic information between $X$ and $Y$ about $M$.
\end{definition}
\begin{proposition}[Equivalence of Information Flow Definitions] \label{prop:syn-info-equiv}
	An edge $E_t$ has $M$-information flow if and only if it has $M$-synergistic information flow. Furthermore, suppose $E_t$ is an edge which satisfies $I\bigl(M ; X(E_t)\bigr) = 0$. Then,
	\begin{equation}
		I\bigl(M ; X(E_t) \given X(\E_t')\bigr) > 0
	\end{equation}
	for some set $\E_t' \subseteq \E_t\setminus\{E_t\}$, if and only if
	\begin{equation}
		CI\bigl(M : X(E_t) ; X(\E_t')\bigr) > 0.
	\end{equation}
	That is, the set $\E_t'$ upon whose transmissions we need to condition is the \emph{same} as the one responsible for providing synergy in the alternate definition.
\end{proposition}
A proof of this proposition is given in Appendix~\ref{app:synergistic-info-flow}.

We should also mention here that it may be possible to leverage the more recent definitions of synergy to supply an intuitive measure of the \emph{volume} of information flow; we discuss this in Section~\ref{sec:info-flow-volume}.


\section{Properties of Information Flow} 
\label{sec:info-flow-properties}

Having defined what it means for information about a message to flow on an edge, we demonstrate that Definition~\ref{def:info-flow} satisfies several intuitively desirable properties, including Property~\ref{ppty:broken-telephone}.

\subsection{The Broken Telephone Property}

\begin{theorem} \label{thm:ppty-def}
	$M$-information flow, as given by Definition~\ref{def:info-flow}, satisfies Property~\ref{ppty:broken-telephone}.
\end{theorem}

Before we prove this theorem, we prove a simpler lemma which directly falls out of Definition~\ref{def:info-flow} and the properties of mutual information.


\begin{lemma} \label{lem:edge-info} \label{lem:dp}
	There is \emph{no} edge in $\E_t$ that carries $M$-information flow if, and only if, $X(\E_t)$ is independent of $M$. In other words,
	\begin{gather}
		I\bigl(M ; X(E_t) \given X(\E_t')\bigr) = 0 \qquad\forall\; E_t \in \E_t,\; \E_t' \subseteq \E_t\!\setminus\!\{E_t\} \label{eq:all-edge-flows-zero}
		\shortintertext{if and only if}
		I\bigl(M ; X(\E_t)\bigr) = 0.
	\end{gather}
	Equivalently, we can state the opposite: $X(\E_t)$ \emph{depends} on $M$ if and only if \emph{at least} one edge in $\E_t$ carries $M$-information flow.
\end{lemma}
\begin{IEEEproof}
	($\Rightarrow$) Suppose that the condition in~\eqref{eq:all-edge-flows-zero} holds. Let $\E_t = \bigl\{E^{(1)}_t, E^{(2)}_t, \ldots, E^{(N^2)}_t\bigr\}$ be any ordering of the edges in $\E_t$. Then,
	\begin{align}
		I\bigl(M ; X(\E_t)\bigr) &\overset{(a)}{=} I\bigl(M ; X(E_{t}^{(1)})\bigr) + I\bigl(M ; X(E_{t}^{(2)}) \given X(E_{t}^{(1)})\bigr) \\
		& \quad + I\bigl(M ; X(E_{t}^{(3)}) \given X(E_{t}^{(1)}), X(E_{t}^{(2)})\bigr) + \cdots \nonumber \\
		&= \sum_{i=1}^{N^2} I\biggl(M ; X(E_{t}^{(i)}) \,\Big\vert\, \bigcup_{j=1}^{i-1} \bigl\{ X(E_{t}^{(j)}) \bigr\}\biggr) \\
		&\overset{(b)}{=} \sum_{i=1}^{N^2} I\biggl(M ; X(E_{t}^{(i)}) \,\Big\vert\, X\Bigl(\bigcup_{j=1}^{i-1} \{E_{t}^{(j)}\}\Bigr)\biggr) \overset{(c)}{=} 0,
	\end{align}
	where (a) follows from the chain-rule of mutual information \cite[Ch.~2]{CoverThomas}, (b) is simply the application of Equation~\eqref{eq:x-of-set}, and (c) follows from the fact that each term in the summation is zero, by~\eqref{eq:all-edge-flows-zero}. This proves the forward implication.

	($\Leftarrow$) Next, suppose $I\bigl(M ; X(\E_t)\bigr) = 0$. Let $E_t$ be any edge in $\E_t$ and let $\E_t'$ be any subset of $\E_t\setminus\{E_t\}$. Also, let $\E_t'' = \E_t\setminus\bigl(\E_t'\cup\{E_t\}\bigr)$. Then,
	\begin{align}
		0 &= I\bigl(M ; X(\E_t)\bigr) \\
		&= I\bigl(M ; X(\E_t')\bigr) + I\bigl(M ; X(E_t) \given X(\E_t')\bigr) + I\bigl(M ; X(\E_t'') \given X(\E_t'), X(E_t)\bigr)
	\end{align}
	by the chain rule. Since (conditional) mutual information is always non-negative \cite[Ch.~2]{CoverThomas}, all three terms on the right hand side must be zero. So in particular,
	\begin{equation}
		I\bigl(M ; X(E_t) \given X(\E_t')\bigr) = 0.
	\end{equation}
	Since $E_t$ and $\E_t'$ are arbitrary, this proves the converse.
\end{IEEEproof}


\begin{IEEEproof}[Proof of Theorem~\ref{thm:ppty-def}]
	We need to prove that $M$-information flow, as given by Definition~\ref{def:info-flow}, satisfies Property~\ref{ppty:broken-telephone}. Explicitly stated, we need to show that if every edge at some time $t$ has zero $M$-information flow, then every edge at all future times $t' > t$ must also have zero $M$-information flow. So suppose that, at time $t$, for every $E_t \in \E_t$ we have
	\begin{equation} \label{eq:all-edge-info-zero}
		I\bigl(M ; X(E_t) \given X(\E_t')\bigr) = 0  \qquad\forall\; \E_t' \subseteq \E_t\setminus\{E_t\}.
	\end{equation}
	By Lemma~\ref{lem:edge-info}, this implies that
	\begin{equation} \label{eq:Et-indept-M}
		I\bigl(M ; X(\E_t)\bigr) = 0.
	\end{equation}
	Now, consider the first future time instant, $t' = t + 1$. For every node $A_{t+1} \in \V_{t+1}$, the definition of computation at a node (Definition~\ref{def:comp-at-node}) states that
	\begin{equation}
		 X(\Q(A_{t+1})) = f_{A_{t+1}}\bigl(X(\P(A_{t+1})), W(A_{t+1})\bigr),
	\end{equation}
	where the reader may recall, $\P(A_{t+1})$ and $\Q(A_{t+1})$ are the edges entering and leaving $A_{t+1}$ respectively. We can collect the individual functions $f_{A_{t+1}}$ across all nodes in $\V_{t+1}$ into a single joint function $f_{\V_{t+1}}$, as described in Definition~\ref{def:comp-at-node}, to obtain
	\begin{equation}
		X(\E_{t+1}) = f_{\V_{t+1}}\bigl(X(\E_t), W(\V_{t+1})\bigr).
	\end{equation}
	Therefore,
	\begin{align}
		0 \leq I\bigl(M ; X(\E_{t+1})\bigr) &= I\bigl(M ; f_{\V_{t+1}}\bigl(X(\E_t), W(\V_{t+1})\bigr)\bigr) \\
		&\overset{(a)}{\leq} I\bigl(M ; X(\E_t), W(\V_{t+1})\bigr) \\
		&= I\bigl(M ; X(\E_t)\bigr) + I\bigl(M ; W(\V_{t+1}) \given X(\E_t)\bigr) \\
		&\overset{(b)}{=} I\bigl(M ; X(\E_t)\bigr) \overset{(c)}{=} 0,
	\end{align}
	where (a) follows from the Data Processing Inequality~\cite[Ch.~2]{CoverThomas}, (b) follows from the fact that $W(\V_{t+1}) \independent \{M, X(\E_t)\}$, and (c) follows from~\eqref{eq:Et-indept-M}. Once again, by non-negativity of mutual information we must have that ${I\bigl(M ; X(\E_{t+1})\bigr) = 0}$. Applying Lemma~\ref{lem:edge-info} once again, we find that for $t' = t + 1$,
	\begin{equation} \label{eq:tprime-all-zero}
		I\bigl(M ; X(E_{t'}) \given X(\E_{t'}')\bigr) = 0 \qquad \forall\; E_{t'} \in \E_{t'}, \E_{t'}' \subseteq \E_{t'}\!\setminus\!\{E_{t'}\}
	\end{equation}
	We have shown that \eqref{eq:all-edge-info-zero}~implies~\eqref{eq:tprime-all-zero}, so induction on $t'$ yields that~\eqref{eq:tprime-all-zero} holds for all future times $t' > t$, completing the proof.
\end{IEEEproof}

\subsection{The Existence of Orphans}
\label{sec:orphans}

Definition~\ref{def:info-flow} also has a very non-intuitive property: an edge leading out of a node may have $M$-information flow, even though \emph{no} edge leading \emph{into} that node has $M$-information flow.
\begin{definition}[$M$-information Orphan] \label{def:orphan}
	In a computational system $\C$, a node $V_t$ is said to be an \emph{$M$-information orphan} if $\Q(V_t)$ has $M$-information flow (as per Definition~\ref{def:info-flow-set}), but $\P(V_t)$ has no $M$-information flow.
\end{definition}
\begin{property} \label{ppty:orphans}
	$M$-information orphans may exist in a computational system.
\end{property}
\begin{IEEEproof}
	Consider the computational system in Figure~\ref{fig:ce-dependence} from Counterexample~\ref{ce:dependence}. The node $C_1$ is an $M$-information orphan, since the edge $(C_1, B_2)$ carries $M$-information flow, whereas none of its incoming edges carries $M$-information flow.
\end{IEEEproof}

The existence of $M$-information orphans, along with the presence of $M$-information flow on $(C_1, B_2)$ in Counterexample~\ref{ce:dependence}, may not be expected, since $Z$ was never computed from $M$. Indeed, $M$-information flow appears to emerge from ``nowhere'' at the node $C_1$, leaving nodes such as $C_1$ orphaned in a view of the graph that contains only edges having $M$-information flow, and hence the name. But closer inspection reveals that in this example, the transmissions arriving at $B_2$ from $A_1$ and $C_1$, i.e.~$M \xor Z$ and $Z$, are \emph{statistically identical}: they are both individually independent of $M$, but when \textsc{xor}'ed, are fully dependent on $M$. In other words, any \emph{purely observational} measure\footnote{i.e., a functional of the joint distribution of $X(\E_t)$} defined on the transmissions at time $t$ that assigns $M$-information flow to $M \xor Z$, must also assign $M$-information flow to $Z$.

Note that, just as $M$-information flow can originate at an $M$-information orphan, $M$-information flow may also terminate at a node---either by simple omission, or as a result of some computation (see Section~\ref{sec:canonical-examples} for such instances). Likewise, multiple outgoing edges of a given node may transmit redundant copies of the same information. Ultimately, we see that there is no ``law of conservation'' for $M$-information flow. In this sense, ``information flow'' is not a typical kind of ``flow'' that is defined on graphs (see, for example,~\cite[Sec.~26.1]{Cormen2009Algorithms}), and well-known results such as the Max-flow Min-cut Theorem~\cite[Thm.~26.6]{Cormen2009Algorithms} do not apply as-is to $M$-information flow.

It is worthwhile to note at this point that the existence of $M$-information orphans such as $C_1$ in Counterexample~\ref{ce:dependence} is not inconsistent with the Data Processing Inequality~\cite[Ch.~2]{CoverThomas}. In fact, a clear example of the Data Processing Inequality in play is seen at the network-level, wherein $M$---$X(\E_t)$---$X(\E_{t+1})$ form a Markov Chain for any time $0 \leq t < T$, and so the information content about $M$ present collectively in all transmissions at time $t+1$ \emph{must} be no more than that present at time $t$. We call this Global Markovity, and state it formally for completeness.
\begin{corollary}[Global Markovity] \label{cor:global-markov}
	At any given time $t$, the following Markov Chain holds: $M$---$X(\E_t)$---$X(\E_{t+1})$.
\end{corollary}
In fact, this Markov condition must hold for every \emph{subset} of nodes, not just for the entire set of nodes, so it is subsumed by the following proposition.
\begin{proposition}[Local Markovity] \label{prop:local-markov}
	For any given subset of nodes $\V_t' \subseteq \V_t$, the following Markov Chain holds: $M$---$X(\P(\V_t'))$---$X(\Q(\V_t'))$.
\end{proposition}
\begin{IEEEproof}
	Since $X(\Q(\V_t')) = f_{\V_t'}\bigl(X(\P(\V_t')), W(\V_t')\bigr)$ by Definition~\ref{def:comp-at-node}, the tuple $\bigl(X(\P(\V_t')), X(\Q(\V_t'))\bigr)$ is also a function of $X(\P(\V_t'))$ and $X(W(\V_t'))$. Hence, the following Markov chain holds:
	\begin{equation*}
		M\text{---}\bigl(X(\P(\V_t')), W(\V_t')\bigr)\text{---}\bigl(X(\P(\V_t')), X(\Q(\V_t'))\bigr).
	\end{equation*}
	By the Data Processing Inequality, this implies that
	\begin{align}
		I\bigl(M ; X(\Q(\V_t')), X(\P(\V_t'))\bigr) &\leq I\bigl(M ; X(\P(\V_t')), W(\V_t')\bigr) \\
		&\overset{(a)}{=} I\bigl(M ; X(\P(\V_t'))\bigr) + I\bigl(M ; W(\V_t') \given X(\P(\V_t'))\bigr) \\
		&\overset{(b)}{=} I\bigl(M ; X(\P(\V_t'))\bigr) + I\bigl(W(\V_t') ; M, X(\P(\V_t'))\bigr) \\
		&\qquad - I\bigl(W(\V_t') ; X(\P(\V_t'))\bigr) \nonumber \\
		&\overset{(c)}{=} I\bigl(M ; X(\P(\V_t'))\bigr) + 0 - 0,
	\end{align}
	where in (a) and (b), we have used the chain rule of mutual information in two different ways, and in (c) we have used the fact that $W(\V_t') \independent \{M, X(\P(\V_t'))\}$. Therefore,
	\begin{equation}
		I\bigl(M ; X(\Q(\V_t')) \given X(\P(\V_t'))\bigr) = 0,
	\end{equation}
	which implies the Markov chain in Proposition~\ref{prop:local-markov}.
\end{IEEEproof}
Since the above also holds for $\V_t' = \V_t$, wherein $\Q(\V_t) = \E_t$, Proposition~\ref{prop:local-markov} implies Corollary~\ref{cor:global-markov}.

Given that these Markov conditions arise directly from the way we have defined the computational system, specifically Definition~\ref{def:comp-at-node}, they may not be very surprising (indeed, they may be considered \emph{properties} of the computational system model itself). However, it is worth noting that Proposition~\ref{prop:local-markov} holds \emph{even at an $M$-information orphan}. Thus, $M$-information orphans do not ``create'' information about $M$, as we would rightly expect, given the Data Processing Inequality.

\subsection{The Existence of Information Paths}

We now show that if the outgoing transmissions of any given node depend on the message, then we can find a path leading to that node from one or more input nodes, along which $M$-information flows. Before we demonstrate this property, we formally define what we mean by the terms ``path'' and ``cut''.

\newcommand{\Vin}{\V^{\text{src}}}
\newcommand{\Vout}{\V^{\text{sink}}}
\newcommand{\Vop}{V_{\text{op}}}
\newcommand{\Ein}{\E^{\text{src}}}
\newcommand{\Eout}{\E^{\text{sink}}}
\newcommand{\Ecut}{\E^{\text{cut}}}
\newcommand{\Pout}{\P^{\text{sink}}}
\newcommand{\Pcut}{\P^{\text{cut}}}

\begin{definition}[Path] \label{def:path}
	In any computational system $\C$, suppose $\A$ and $\B$ are two disjoint sets of nodes in $\V$. Then, a \emph{path} from $\A$ to $\B$ is any ordered set of nodes $\{V^{(0)}, V^{(1)}, \ldots, V^{(L)}\}$ that satisfies
	\begin{enumerate*}[label=(\roman*)]
		\item $V^{(0)} \in \A$;
		\item $V^{(L)} \in \B$; and
		\item $(V^{(i-1)}, V^{(i)}) \in \E$ for every $1 \leq i \leq L$,
	\end{enumerate*}
	where $L$ is a positive integer indicating the length of the path. We refer to the set $\{(V^{(i-1)}, V^{(i)})\}_{i=1}^L$ as the \emph{edges of the path}.
\end{definition}

\begin{definition}[$M$-Information Path] \label{def:info-path}
	Continuing from Definition~\ref{def:path}, we define an \emph{$M$-information path} from $\A$ to $\B$ as any path from $\A$ to $\B$, each of whose edges carries $M$-information flow. That is, if $(V^{(i-1)}, V^{(i)}) = E_{t_i} \in \E_{t_i}$ for some $t_i \in \T$, then for every $1 \leq i \leq L$,
	\begin{equation}
		\exists\;\E_{t_i}' \subseteq \E_{t_i} \quad \text{s.t.} \quad I\bigl(M ; X(E_{t_i}) \given X(\E_{t_i}')\bigr) > 0.
	\end{equation}
\end{definition}

\begin{definition}[Cut] \label{def:cut}
	In any computational system $\C$, suppose $\A$ and $\B$ are two disjoint sets of nodes in $\V$. Then, a \emph{cut} separating $\A$ and $\B$ is any pair of sets $(\Vin, \Vout)$, such that
	\begin{enumerate*}[label=(\roman*)]
		\item $\Vin \cup \Vout = \V$;
		\item $\Vin \cap \Vout = \emptyset$;
		\item $\A \subseteq \Vin$; and
		\item $\B \subseteq \Vout$.
	\end{enumerate*}
	We refer to the set of edges \emph{going from $\Vin$ to $\Vout$}, i.e.~$\E \cap (\Vin \times \Vout)$, as the \emph{edges in the cut set}\footnote{Note that it is not necessary for us to assume that, individually, $\Vin$ and $\Vout$ are \emph{connected} sets of nodes. For instance, there may be an isolated subset of $\Vout$, surrounded only by nodes in $\Vin$. Our theorems and proofs remain unaffected, even in such a scenario.}.
\end{definition}

\begin{definition}[Zero--$M$-information Cut] \label{def:zero-info-cut}
	Continuing from Definition~\ref{def:cut}, we say that a cut $(\Vin, \Vout)$ is a \emph{zero--$M$-information cut} if every edge in its cut set has zero $M$-information flow. That is, for every $E_t \in \E \cap (\Vin \times \Vout)$,
	\begin{equation} \label{eq:zero-info-cut-eqn}
		I\bigl(M ; X(E_t) \given X(\E_t')\bigr) = 0 \quad \forall\; \E_t' \subseteq \E_t\setminus\{E_t\}.
	\end{equation}
\end{definition}

\paragraph*{Remark} In Definition~\ref{def:zero-info-cut}, we require that Equation~\eqref{eq:zero-info-cut-eqn} hold for every edge $E_t$ in $\E \cap (\Vin \times \Vout)$. However, the edges in this set may belong to several different time points, since the cut is not restricted to any particular time (e.g., see Figure~\ref{fig:proof-outline}). The time $t$ used in Equation~\eqref{eq:zero-info-cut-eqn}, therefore, is determined by the time of the edge $E_t$, and varies for each $E_t$ that we check in $\E \cap (\Vin \times \Vout)$.


\begin{property}[Existence of an Information Path] \label{ppty:info-path}
	In any computational system $\C$, suppose that at some time $t_\text{op} \in \T$, there is an ``output node'' $\Vop \in \V$ whose outgoing edges $\Q(\Vop)$ satisfy $I\bigl(M ; X(\Q(\Vop))\bigr) > 0$. Then, there must exist an $M$-information path from the input nodes $\Vip$ to $\Vop$.
\end{property}
\begin{theorem} \label{thm:info-path}
	Definition~\ref{def:info-flow} satisfies Property~\ref{ppty:info-path}.
\end{theorem}

While the theorem seems obvious on the surface, the proof is in fact non-trivial because of the nature of our definition of $M$-information flow. Due to Property~\ref{ppty:orphans}, $M$-information flowing \emph{out of} a node does \emph{not} imply that $M$-information must flow \emph{into} that node. Therefore, a straightforward application of the Data Processing Inequality at every node fails to prove the theorem, and we must resort to a more rigorous cut-set-based approach.

\renewcommand{\IEEEQED}{\IEEEQEDopen}
\begin{IEEEproof}[Proof outline]
	We shall prove the contrapositive of the theorem, i.e., we will show that if there exists no $M$-information path from $\Vip$ to $\Vop$, then the outgoing transmissions of $\Vop$ are independent of $M$. We first connect the absence of any $M$-information path with the presence of a zero--$M$-information cut. This is achieved in Lemma~\ref{lem:path-cut}, which we present before the proof of Theorem~\ref{thm:info-path}.

	The proof itself proceeds by induction over time. We divide the proof into two steps: initialization and continuation. Starting with the first nodes that come after the cut (temporally) in the initialization step, we systematically show that all nodes to the right of the cut have outgoing transmissions that are independent of the message $M$ through induction. In this proof outline, we show these steps intuitively using Figure~\ref{fig:proof-outline}, where the dashed black line denotes the cut.

	\textit{Initialization.} Here, node $C_1$ is the first node to the right of the cut, and all of its incoming edges must come from across the cut (depicted by lines in red). Because the cut is a zero--$M$-information cut, none of its incoming transmissions have $M$-information flow. Furthermore, the intrinsically generated random variable $W(C_1)$ is independent of $M$. Using these two facts along with the Data Processing Inequality, we can show that the transmissions on $C_1$'s outgoing edges, $X(\Q(C_1))$, are also independent of $M$.

	\textit{Continuation.} At the second time instant to the right of the cut, nodes $B_2$ and $C_2$ receive their incoming transmissions from either $C_1$ (shown in orange) or from across the cut (shown in blue). Once again, the transmissions coming from across the cut can have no information flow, and we have shown that the transmissions coming from $C_1$ are independent of $M$. Also, $W(B_2)$ and $W(C_2)$ are independent of $M$ and all incoming transmissions. This suffices to show that the outgoing transmissions of $B_2$ and $C_2$, $X\bigl(\Q(B_2) \cup \Q(C_2)\bigr)$, are independent of $M$. Applying this argument repeatedly over time shows that the transmissions of all nodes to the right of the cut are independent of $M$.

	Therefore, if there is a node $\Vop$ whose outputs depend on $M$, we can be assured that there exists no zero--$M$-information cut separating $\Vip$ from $\Vop$. Therefore, by Lemma~\ref{lem:path-cut}, there exists an $M$-information path from $\Vip$ to $\Vop$.
\end{IEEEproof}
\renewcommand{\IEEEQED}{\IEEEQEDclosed}

A few nuances are omitted in this outline, such as how the definition of $\Vip$ plays a role precisely. These subtleties are better elucidated in the full proof.

Before proceeding to the formal proof of Theorem~\ref{thm:info-path}, we first state and prove the lemma we alluded to earlier, which shows how the absence of an $M$-information path implies the presence of a zero--$M$-information cut, and vice versa.

\begin{lemma} \label{lem:path-cut}
	Let $\A$ and $\B$ be two disjoint sets of nodes in the computational system $\C$. There exists no $M$-information path from $\A$ to $\B$ if and only if there is a zero--$M$-information cut separating $\A$ and $\B$.
\end{lemma}
\begin{IEEEproof}
	($\Rightarrow$) Suppose there exists no $M$-information path from $\A$ to $\B$. Consider the set of all nodes to which there exists at least one $M$-information path from $\A$. Let $\Vin$ be the collection of all such nodes, along with the nodes in $\A$, i.e.,
	\begin{equation}
		\Vin \coloneqq \A \cup \{V_t \in \V: \exists \text{ an $M$-information path from $\A$ to $V_t$}\}.
	\end{equation}
	Let $\Vout = \V \setminus \Vin$, so that $\Vout$ consists of nodes to which there is no $M$-information path from $\A$. Then, we must have $\B \subseteq \Vout$, since it is known that there are no $M$-information paths from $\A$ to $\B$. Therefore, $(\Vin, \Vout)$ is a cut that separates $\A$ and $\B$, such that no edge in the cut set has $M$-information flow. In other words, by Definition~\ref{def:zero-info-cut}, this is a zero--$M$-information cut separating $\A$ and $\B$.

	($\Leftarrow$) Next, suppose that there \emph{is} an $M$-information path $\{V^{(i)}\}_{i=0}^L$ from $\A$ to $\B$. Then, we claim that there can exist no zero--$M$-information cut separating $\A$ and $\B$. Let $(\Vin, \Vout)$ be any cut separating $\A$ and $\B$. By Definition~\ref{def:path}, we must have $V^{(0)} \in \Vin$ and $V^{(L)} \in \Vout$. So, there must be at least one edge going from $\Vin$ to $\Vout$ which lies on the path. This implies that at least one edge in the cut set carries $M$-information flow. Since the conditions of Definition~\ref{def:zero-info-cut} are not satisfied, this cut is \emph{not} a zero--$M$-information cut. Since this is true for every cut separating $\A$ and $\B$, the claim holds.
\end{IEEEproof}

\begin{IEEEproof}[Proof of Theorem~\ref{thm:info-path}]
	As mentioned in the proof outline, we prove the contrapositive of the theorem. Suppose there exists no $M$-information path from the input nodes $\Vip$ to $\Vop$. Then, by Lemma~\ref{lem:path-cut}, there exists a zero--$M$-information cut\footnote{Note that, in general, this cut may be arbitrarily complex, spanning several nodes and multiple time instants.} separating $\Vip$ and $\Vop$. We use this to prove that the transmissions of $\Vop$ are independent of $M$.

\begin{figure}
	\centering
	\includegraphics{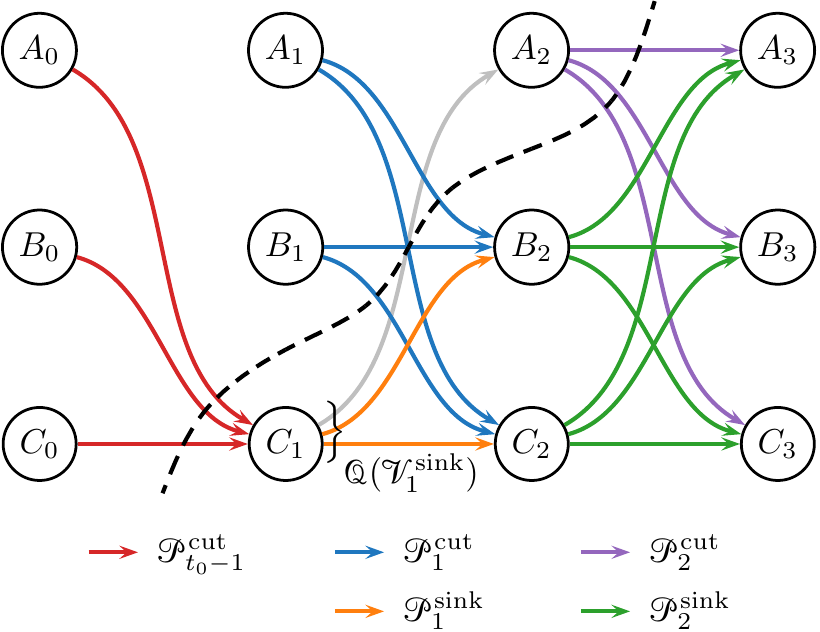}
	\caption{A generic computational system used in the proof outline and to explain certain steps in the proof of Theorem~\ref{thm:info-path}. For the purposes of the proof outline, it suffices to note that the black dashed line denotes the cut. All variable names can be ignored at this point of time. \\ For the purposes of the formal proof, note that in this figure, $\Ecut$ is essentially the union of the red, blue and purple edges, while $\Eout$ is the union of the orange and green edges. From this, it is evident that $\P(\Vout_t) = \Pcut_{t-1} \cup \Pout_{t-1}$ for any time $t$, i.e., the incoming edges of $\Vout$ at time $t$ must either come from nodes in $\Vout$ or from nodes across the cut. Secondly, it should be clear that $\Pout_{t-1} = \Q(\Vout_{t-1}) \cap \Eout$, i.e., the \emph{incoming} edges of $\Vout_t$ that originate from nodes in $\Vout$ are simply the \emph{outgoing} edges of $\Vout_{t-1}$ which terminate at nodes in $\Vout$. This is seen best at time $t=1$ in the graph above, where the orange and grey lines together represent $\Q(\Vout_1)$, the orange and green edges together make up $\Eout$, and $\Pout_1$ is given by the orange edges, which is the intersection of the two sets.}
	\label{fig:proof-outline}
\end{figure}

	\textit{Setup.} Let the cut separating $\Vip$ and $\Vop$ be given by $(\Vin, \Vout)$, so that $\Vip \subseteq \Vin$ and $\Vop \in \Vout$. Then, the cut divides $\E$ into the following sets: $\Ein = \E \cap (\Vin \times \Vin)$, the edges between the nodes in $\Vin$; $\Eout = \E \cap (\Vout \times \Vout)$, the edges between nodes in $\Vout$; and $\Ecut = \E \cap (\Vin \times \Vout)$, the edges going from $\Vin$ to $\Vout$ (the edges going from $\Vout$ to $\Vin$ will not be relevant to our discussion). From the previous paragraph, Lemma~\ref{lem:path-cut} implies that $(\Vin, \Vout)$ is a zero--$M$-information cut, so by Definition~\ref{def:zero-info-cut}, we have that for all $E_t \in \Ecut$,
	\begin{equation}
		I\bigl(M ; X(E_t) \given X(\E_t')\bigr) = 0 \quad \forall\; \E_t' \subseteq \E_t\setminus\{E_t\}.
	\end{equation}
	Note that the edges in $\Ecut$ may belong to different time instants. In particular, the time instant $t$ in the equation above corresponds to the time of the edge $E_t$, whose flow is in question.\footnote{In fact, this is one of the central factors that prevents us from recursively applying the Data Processing Inequality at every node, leading from $\Vip$ to $\Vop$.}

	Order the nodes in $\Vout$ by time, and let $\Vout_t$ be the subset of nodes in $\Vout$ at time $t$. Let $\P(\Vout_t)$ and $\Q(\Vout_t)$ respectively be the sets of edges \emph{collectively} entering and leaving all nodes in $\Vout_t$. We shall prove that the outgoing transmissions of every node in $\Vout$, including those of $\Vop$, must be independent of the message, i.e.,
	\begin{equation}
		I\bigl(M ; X(\Q(V))\bigr) = 0 \quad \forall\; V \in \Vout.
	\end{equation}

	\textit{Initialization.} Let $t_0$ be the first time instant $t$ for which $\Vout_t$ is non-empty. Then, we encounter two cases: either $t_0 = 0$, in which case the nodes in $\Vout_{t_0}$ have \emph{no} incoming edges, or $t_0 > 0$, and the nodes in $\Vout_{t_0}$ \emph{have} incoming edges. We shall first prove that in \emph{both} cases, the outgoing transmissions of $\Vout_{t_0}$ are independent of the message, i.e.\ $I\bigl(M ; X(\Q(\Vout_{t_0}))\bigr) = 0$.

	(Case I) When $t_0 = 0$, $\Vout_0 \cap \Vip = \emptyset$. This is because the cut separates $\Vip$ from $\Vop$, with $\Vip \subseteq \Vin$, so no nodes in $\Vout_0$ can be input nodes. So, by the definition of (non-)input nodes (Definition~\ref{def:input-nodes}), we must have
	\begin{align}
		I\bigl(M ; X(\Q(\Vout_0))\bigr) &= I\bigl(M ; f_{\Vout_0}(W(\Vout_0))\bigr) \\
		&\overset{(a)}{\leq} I\bigl(M ; W(\Vout_0)\bigr) \\
		&\overset{(b)}{=} 0, \label{eq:t0-eq-0-works}
	\end{align}
	where step (a) uses the data processing inequality and step (b) makes use of the fact that $W(\V_0) \independent M$.

	(Case II) When $t_0 > 0$, the definition of $t_0$ implies that all nodes at time $t_0 - 1$ are in $\Vin$, so all incoming edges of $\Vout_{t_0}$ must lie in the cut set, i.e., $\P(\Vout_{t_0}) \subseteq \Ecut$. Since the cut is a zero--$M$-information cut, we have that for all $E_{t_0-1} \in \P(\Vout_{t_0})$,
	\begin{align}
		I\bigl(M ; X(E_{t_0-1}) \given X(\E_{t_0-1}')\bigr) &= 0 \quad \forall\; \E_{t_0-1}' \subseteq \E_{t_0-1}.
	\intertext{By the definition of $M$-information flow for a set of edges (Definition~\ref{def:info-flow-set}) and Proposition~\ref{prop:set-info-equiv}, we have}
		I\bigl(M ; X(\P(\Vout_{t_0})) \given X(\E_{t_0-1}')\bigr) &= 0 \quad\forall\; \E_{t_0-1}' \subseteq \E_{t_0-1}. \label{eq:p-t0}
	\end{align}
	Once again, considering $\Q(\Vout_{t_0})$, we have
	\begin{align}
		I\bigl(M ; X(\Q(\Vout_{t_0}))\bigr) &= I\bigl(M ; f_{\Vout_{t_0}}\bigl(X(\P(\Vout_{t_0})), W(\Vout_{t_0})\bigr)\bigr) \\
		&\overset{(a)}{\leq} I\bigl(M ; X(\P(\Vout_{t_0})), W(\Vout_{t_0})\bigr) \\
		&\overset{(b)}{=} I\bigl(M ; X(\P(\Vout_{t_0}))\bigr) + I\bigl(M ; W(\Vout_{t_0}) \given X(\P(\Vout_{t_0}))\bigr) \\
		&\overset{(c)}{=} 0, \label{eq:t0-gt-0-works}
	\end{align}
	where (a) and (b) follow from the Data Processing Inequality and the chain rule of mutual information respectively. In step (c), the first expression in the sum goes to zero by taking $\E_{t_0-1} = \emptyset$ in~\eqref{eq:p-t0} and the second expression is zero since $W(\Vout_{t_0}) \independent \{M, X(\E_{t_0-1})\}$, and $\P(\Vout_{t_0}) \subseteq \E_{t_0-1}$ (refer Definition~\ref{def:comp-at-node}). So, from equations~\eqref{eq:t0-eq-0-works} and~\eqref{eq:t0-gt-0-works}, we have that for all values of $t_0$,
	\begin{equation}
		I\bigl(M ; X(\Q(\Vout_{t_0}))\bigr) = 0.
	\end{equation}

	\textit{Continuation.} Now, suppose that for some $t > t_0$, we have $I\bigl(M ; X(\Q(\Vout_{t-1}))\bigr) = 0$. We shall prove that this implies $I\bigl(M ; X(\Q(\Vout_t))\bigr) = 0$. First, observe that
	\begin{equation}
		\P(\Vout_t) = (\P(\Vout_t) \cap \Ecut) \cup (\P(\Vout_t) \cap \Eout)
	\end{equation}
	For convenience, let $\Pcut_{t-1} \coloneqq \P(\Vout_t) \cap \Ecut$ and $\Pout_{t-1} \coloneqq \P(\Vout_t) \cap \Eout$. We have used the subscript $t-1$ here to remind the reader that $\P(\Vout_t)$, which are the \emph{incoming} edges of $\Vout_t$, are a subset of $\E_{t-1}$. Then, we have
	\begin{equation}
		\P(\Vout_t) = \Pcut_{t-1} \cup \Pout_{t-1}.
	\end{equation}
	Since the cut is a zero--$M$-information cut, we have that for every $E_{t-1} \in \Pcut_{t-1}$,
	\begin{equation}
		I\bigl(M ; X(E_{t-1}) \given X(\E_{t-1}')\bigr) = 0 \quad \forall\; \E_{t-1}' \subseteq \E_{t-1}.
	\end{equation}
	Therefore, by Definition~\ref{def:info-flow-set} and Proposition~\ref{prop:set-info-equiv},
	\begin{equation} \label{eq:pcut-zero}
		I\bigl(M ; X(\Pcut_{t-1}) \given X(\E_{t-1}')\bigr) = 0 \quad \forall\; \E_{t-1}' \subseteq \E_{t-1}.
	\end{equation}
	Secondly, $\Pout_{t-1} = \Q(\Vout_{t-1}) \cap \Eout$. This is depicted in Figure~\ref{fig:proof-outline}, and explained in the caption. So,
	\begin{align}
		I\bigl(M ; X(\Pout_{t-1})\bigr) &= I\bigl(M ; X(\Q(\Vout_{t-1}) \cap \Eout)\bigr) \\
		&\overset{(a)}{\leq} I\bigl(M ; X(\Q(\Vout_{t-1}))\bigr) \overset{(b)}{=} 0 \label{eq:pout-zero}
	\end{align}
	where (a) follows from the fact that considering more random variables can only increase mutual information, and (b) follows from the induction assumption. Finally, consider how $X(\Q(\Vout_t))$ depends on $M$:
	\begin{align}
		I\bigl(M ; X(\Q(\Vout_t))\bigr) &= I\bigl(M ; f_{\Vout_t}\bigl(X(\Pout_{t-1} \cup \Pcut_{t-1}), W(\Vout_t)\bigr)\bigr) \\
		&\overset{(a)}{\leq} I\bigl(M ; X(\Pout_{t-1}), X(\Pcut_{t-1}), W(\Vout_t)\bigr) \\
		&\overset{(b)}{=} I\bigl(M ; X(\Pout_{t-1})\bigr) + I\bigl(M ; X(\Pcut_{t-1}) \given X(\Pout_{t-1})\bigr)\\
		&\quad + I\bigl(M ; W(\Vout_t) \given X(\Pout_{t-1}), X(\Pcut_{t-1})\bigr) \nonumber \\
		&\overset{(c)}{=} 0,
	\end{align}
	where once again, (a) and (b) follow from the data processing inequality and the chain rule respectively. In step (c), the first and second terms go to zero by equations~\eqref{eq:pout-zero} and~\eqref{eq:pcut-zero} respectively, while the third term is zero since $W(\Vout_t) \independent \{M, X(\E_{t-1})\}$ and $\Pout_{t-1} \cup \Pcut_{t-1} \subseteq \E_{t-1}$.

	The proof follows from induction on $t$, so
	\begin{align}
		I\bigl(M ; X(\Q(\Vout_t))\bigr) &= 0 \quad \forall\; t \geq t_0,
	\shortintertext{which in turn implies that}
		I\bigl(M ; X(\Q(V))\bigr) &= 0 \quad \forall\; V \in \Vout.
	\end{align}
	If there exists an output node whose transmissions depend on $M$, then there can exist no cut consisting of edges with zero $M$-information flow, and hence by Lemma~\ref{lem:path-cut}, there must be a path consisting of edges that carry $M$-information flow between the input nodes and the output node in question.
\end{IEEEproof}


\subsection{The Separability Property}
\label{sec:separation-property}

Finally, we state a property that may be of interest to obtain a deeper understanding of the nature of $M$-information flow, as given by Definitions~\ref{def:info-flow} and~\ref{def:info-flow-set}.

\begin{proposition}[Separability] \label{prop:separation}
	Let $\C$ be a computational system. Then, at any given point in time $t$, there exist two sets $\R_t, \S_t \subseteq \E_t$, such that all of the following conditions hold:
	\begin{enumerate}
		\item $\R_t \cup \S_t = \E_t$
		\item $\R_t \cap \S_t = \emptyset$
		\item Either $\R_t = \emptyset$, or for every $R_t \in \R_t$ there exists a subset $\R_t' \subseteq \R_t \setminus \{R_t\}$ such that
			\begin{equation} \label{eq:Rt-cond}
				I\bigl(M ; X(R_t) \given X(\R_t')\bigr) > 0.
			\end{equation}
		\item Either $\S_t = \emptyset$, or for every $\E_t' \subseteq \E_t$,
			\begin{equation} \label{eq:St-cond}
				I\bigl(M ; X(\S_t) \given X(\E_t')\bigr) = 0.
			\end{equation}
	\end{enumerate}
\end{proposition}

A proof of this proposition can be found in Appendix~\ref{app:separability-proof}.

Proposition~\ref{prop:separation} shows that at any given point in time $t$, it is possible to partition $\E_t$ into two sets: $\R_t$, consisting only of edges that have $M$-information flow, and $\S_t$, comprising edges that have no $M$-information flow. Furthermore, when considering the $M$-information flow of edges in $\R_t$, it suffices to condition on the transmissions of edges within $\R_t$ to ascertain the presence of $M$-information flow. Conditioning upon the transmissions of edges in $\S_t$ will not change the mutual information between the message and the transmissions of edges in $\R_t$.


\section{Inferring Information Flow} 
\label{sec:inferring-info-flow}

Having discussed the definition and the properties of $M$-information flow, we now consider how these flows of information might be inferred in a real computational system. We first discuss an observation model that describes which random variables are observed and how they are sampled. Under this model, we show how existing techniques from the literature can be used to identify which edges carry $M$-information flow. As in previous sections, we restrict our attention to detecting \emph{whether or not} a given edge has $M$-information flow, relegating quantification of these flows to future work. Quantification is briefly discussed in the form of an example in Section~\ref{sec:eg-sk}, and again in Section~\ref{sec:info-flow-volume}.

We then describe an algorithm that recovers all $M$-information paths between the input nodes and a given output node, by leveraging the knowledge of which edges have $M$-information flow. We also explain how one might attain a fine-grained characterization of the structure of information flow, by introducing the concept of ``derived information''. This is useful for understanding which transmissions are ``derived'' from others, allowing one to find transmissions that are redundant and discover the presence of hidden nodes. Finally, we explain how flows of information about multiple messages can be inferred in our framework.

\subsection{The Observation Model}
\label{sec:obs-model}

Before we can describe how information flow and information paths can be identified, we must provide a statistical description of the random variables that are observed. Let $\C$ be a computational system under observation. We then make the following assumptions:
\begin{enumerate}
	\item Transmissions on all edges, including self-edges, are observed. The random variables that are intrinsically generated at each node are \emph{not} observed, unless they are also transmitted on an edge (which could be a self-edge).
	\item Several trials\footnote{The word ``trial'' is borrowed from the neuroscience literature, wherein a neuroscientist will often conduct multiple trials in a single experiment. In each trial, a human participant or an animal under study is presented with one of a set of carefully chosen stimuli (corresponding to a realization of the message $M$ in our setting), and neural activity is recorded using some modality. Scientific inferences are then drawn by making use of the activity from all trials.} are observed, each of which corresponds to an independent realization of all random variables in the model\footnote{In reality, trials are not independent in neuroscientific experiments. Indeed, neurons are known to ``adapt'' their responses from trial to trial, often showing suppressed activity when presented the same stimulus multiple times. This, in part, is considered to be evidence of \emph{learning} in neural circuitry. However, for simplicity, we restrict our attention here to computational systems that do not learn or show trial-to-trial adaptation.}. Every trial uses a realization of $M$ which is independently drawn from a distribution determined by the experimentalist\footnote{A more detailed discussion of this distribution can be found in Section~\ref{sec:multiple-messages}.}. For every node $V \in \V$, the intrinsically generated random variable $W(V)$ is also assumed to be independently and identically distributed across trials.
	\item \label{itm:noiseless-assmptn} Observations are made noiselessly, in that the realization of each transmission in every trial is observed as-is, without being further corrupted by random noise of any kind. The implications of noisy measurements will be the subject of future work.
\end{enumerate}
Under these conditions, we discuss statistical tests for information flow that are consistent in the asymptotic limit of infinite trials. It should be noted that these assumptions may be valid to varying degrees in different contexts. This is discussed further in Section~\ref{sec:assumptions-in-model}.

\subsection{Detecting Information Flow}
\label{sec:estimating-info-flow}

Given a sample of all random variables described in the observation model, our next task is to identify which edges have $M$-information flow. In other words, we need to describe how the conditions given by Definition~\ref{def:info-flow} can be rigorously tested, and how we might assert with some confidence that a certain set of edges has information flow at each point in time.

According to Definition~\ref{def:info-flow}, in order to check whether a particular edge $E_t$ carries $M$-information flow at time $t$, we need to test whether at least one of several conditional mutual information quantities is strictly positive. The standard statistical approach for solving this problem is to frame it as a set of ``hypothesis tests'', which in this case is a set of ``conditional independence tests''. In general, a hypothesis test formalizes the problem of making an informed decision about the value of some functional of a joint distribution, when observing a sample of data from it. A good conditional independence testing procedure will seek to maximize ``statistical power'', i.e.\ the probability of \emph{correctly} identifying the presence of conditional dependence, while keeping the probability of an incorrect identification fixed below some ``level'' $\alpha$ that is picked beforehand. One intuitive way to do this might be to construct an estimator for the appropriate conditional mutual information, and ``reject'' the ``null'' hypothesis of conditional independence if the conditional mutual information was sufficiently larger than some threshold, $\epsilon > 0$. This threshold would have to be chosen so that, on average, the probability of falsely rejecting the null hypothesis is at most $\alpha$. However, there are usually better ways of performing this test, i.e., it is often possible to attain higher power at the same level \emph{without} actually estimating the conditional mutual information.
	
While it would be impossible to provide a comprehensive list of papers that have researched the problem of conditional independence testing, it has received (and continues to receive) much attention in the statistics, causality, and information theory communities~\cite{Bergsma2004Testing,Zhang2011Kernel,Huang2010Testing,Su2007Consistent,Huang2016Flexible,Sen2018Mimic}. In its most general form, conditional independence testing is considered to be a hard problem for continuous random variables~\cite{Shah2018Hardness}. However, if we ignore issues associated with the practical difficulty of estimation (discussed later in Section~\ref{sec:difficulty-of-estimation}), these works provide consistent tests under reasonable assumptions on the joint distribution of the variables involved~\cite{Huang2010Testing,Su2007Consistent,Huang2016Flexible}.

Although we mentioned that there are better ways to test for conditional dependence than to estimate the conditional mutual information, there may be instances when one might want to estimate the conditional mutual information anyway. For instance, in an example that will appear shortly in Section~\ref{sec:eg-sk}, we rely on an \emph{estimate} of the conditional mutual information to \emph{quantify} the amount of $M$-information flowing on a given edge. While our paper has only defined $M$-information flow in terms of \emph{whether or not} it is present at an edge $E_t$, it is also extremely useful to know \emph{how much} $M$-information flow there is. We defer further discussion of this topic until Sections~\ref{sec:eg-sk} and \ref{sec:info-flow-volume}. For now, we note that several papers have considered how to \emph{estimate} mutual information and conditional mutual information, both of which might be essential for an understanding of \emph{quantification} of $M$-information flow~\cite{Paninski2003Estimation,Gao2017Estimating,Kraskov2004Estimating,Liu2012Exponential}.

For completeness, we now present a description of how we expect information flow will be detected in practice. We assume that we have samples of observations from every edge of the computational system, at every point in time. If not, appropriate assumptions may need to be made, as discussed later in Section~\ref{sec:assumptions-in-model}. At every instant of time $t$, consider the set of all edges $\E_t$ present in the network. For every edge $E_t \in \E_t$, use the following process to determine whether it has $M$-information flow:
\begin{enumerate}
	\item First test whether the mutual information between its transmission and the message is greater than zero, i.e., $I\bigl(M ; X(E_t)\bigr) > 0$. If so, declare that $E_t$ has $M$-information flow.
	\item If not, test for conditional dependence between its transmission and the message, given each of the other edges $E_t'$, i.e., $I\bigl(M ; X(E_t) \given X(E_t')\bigr) > 0$, $\forall\; E_t' \in \E_t \setminus \{\E_t\}$. If any of these tests rejects the null hypothesis, declare that $E_t$ has $M$-information flow.
	\item If not, test for conditional dependence between $X(E_t)$ and $M$, given subsets of other edges, while sequentially taking edges taken pairwise, then in threes, etc. If any of these tests rejects the null, declare that $E_t$ has $M$-information flow.
	\item If none of the above tests rejects the null hypothesis, declare that $E_t$ carries no $M$-information flow.
\end{enumerate}
Note that we have not discussed the level, $\alpha$, at which we should reject the null in each of the above tests. In general, since we are performing multiple hypothesis tests simultaneously, some manner of ``\emph{correction}'' is required to ensure that we do not find, what is effectively, a spurious correlation. This is discussed at length in Section~\ref{sec:difficulty-of-estimation}.

\subsection{Discovering Information Paths}
\label{sec:info-path-algo}

\begin{algorithm}[t] 
	\caption{Information Path Algorithm: Finds all paths from $\Vip$ to $\Vop$}
	\label{alg:info-path}
	\begin{algorithmic}[1]
		\State Initialize an empty graph $\H$ \Comment{$\H$ will store valid paths from $\Vip$ to $\Vop$}
		\Statex \Comment{$\H$ currently contains no nodes or edges}
		\State \Call{FindInfoPaths}{$\C$, $\Vop$, $\H$} \Comment{Call a function (defined below) to populate $\H$}
		\If{$\Vop$ is marked ``invalid''}
			\State \textbf{raise} Error \Comment{No path from $\Vip$ to $\Vop$ was found} \label{ln:error-2}
		\EndIf
		\Statex
		\Function{FindInfoPaths}{$\C$, $V_t$, $\H$}
			\If{$\P(V_t)$ is empty} \Comment{$V_t$ has no inputs $\Rightarrow t=0$}
				\If{$V_t \in \Vip$}
					\State Mark $V_t$ ``valid''
					\State Add $V_t$ to $\H$
				\Else \Comment{We somehow reached a non-input node at $t=0$}
					\State \textbf{raise} Error \label{ln:error-1}
				\EndIf
			\Else \Comment{$V_t$ has inputs}
				\ForAll{$(U_{t-1}, V_t) \in \P(V_t)$}
					\If{$(U_{t-1}, V_t)$ has $M$-information flow}
						\If{$U_{t-1}$ is unmarked}
							\State \Call{FindInfoPaths}{$\C$, $U_{t-1}$, $\H$} \Comment{This will mark $U_{t-1}$}
						\EndIf
						\If{$U_{t-1}$ is marked ``valid''}
							\State Mark $V_t$ ``valid''
							\State Add $V_t$ and $(U_{t-1}, V_t)$ to $\H$
						\EndIf
					\EndIf
				\EndFor
				\If{$V_t$ is still unmarked} \Comment{No input of $V_t$ was ``valid''}
					\State Mark $V_t$ ``invalid''
				\EndIf
			\EndIf
		\EndFunction
	\end{algorithmic}
\end{algorithm}

Next, we discuss an algorithm that discovers all $M$-information paths leading from the input nodes to a given output node, $\Vop$, in any computational system. As discussed in Section~\ref{sec:info-flow-properties}, whenever the transmissions $\Q(\Vop)$ of the output node depend on the message, Theorem~\ref{thm:info-path} guarantees that at least one $M$-information path exists.

Algorithm~\ref{alg:info-path}, which we propose for recovering all $M$-information paths, is an adaptation of the well-known Depth-First Search\footnote{It is also possible to discover all $M$-information paths using an adaptation of Breadth-First Search~\cite[Sec.~22.2]{Cormen2009Algorithms}, but doing so would require some mechanism to prune $M$-information paths that do not lead to the input nodes $\Vip$. So we prefer to use Depth-First Search for simplicity of exposition.} method~\cite[Sec.~22.3]{Cormen2009Algorithms}. It takes as its input a computational system $\C$ in which all edges having $M$-information flow have been identified, the output node $\Vop$, and an empty graph $\H$ that is completely devoid of nodes and edges. The algorithm returns the set of all $M$-information paths in the form of a directed subgraph $\H$ of the time-unrolled graph $\G$. Starting from $\Vip$, following \emph{any} path in $\H$ will lead one to $\Vop$, provided at least one $M$-information path exists.

The algorithm works by recursively visiting nodes, starting from the output node $\Vop$. It traverses only edges that carry $M$-information flow, and uses a marking scheme to avoid revisiting nodes. The same marking scheme is also used to designate nodes to which there are $M$-information paths from $\Vip$. As the algorithm passes through each node, it marks the node ``valid'' whenever an $M$-information path exists between $\Vip$ and that node. If no such path exists, then the node is marked ``invalid''. The objective of the algorithm, therefore, reduces to one of finding a path of ``valid'' nodes from $\Vip$ to $\Vop$. The algorithm's recursive function can be expressed as follows: \emph{A node $V_t \in \V$ is ``valid'' if and only if there exists a node $U_{t-1} \in \V$ such that $U_{t-1}$ is valid, and the edge $(U_{t-1}, V_t)$ has $M$-information flow.} This is a recursive expression since checking the validity of a node at time $t$ involves finding valid nodes at time $t-1$. The only nodes that are considered valid by default are the input nodes $\Vip$.

The algorithm sequentially checks the validity of nodes $V_t \in \V$, starting from the output node $\Vop$. The function \textsc{FindInfoPaths}, when called on any given node $V_t$, checks the validity of $V_t$. This involves checking each of the incoming edges of $V_t$ for $M$-information flow. If $U_{t-1}$ is a node from which $M$-information flows to $V_t$, then the algorithm immediately checks the validity of $U_{t-1}$ by calling the function \textsc{FindInfoPaths} again. Eventually, if in this recursive process, we arrive at an input node in $\Vip$, then that node is marked ``valid'', and added to the output subgraph $\H$. Once every node $U_{t-1}$ from which $M$-information flows to $V_t$ has been marked ``valid'' or ``invalid'', the validity of $V_t$ can be ascertained. For every ``valid'' node $U_{t-1}$ from which $M$-information flows to $V_t$, the edge $(U_{t-1}, V_t)$ and the node $V_t$ are added to the output subgraph $\H$, and $V_t$ is marked ``valid''. If there are no such nodes leading to $V_t$, then $V_t$ is marked ``invalid'' and does not fall on an $M$-information path.

This recursive logic yields the set of all $M$-information paths leading from the input nodes to $\Vop$. The two lines at which errors are returned correspond to scenarios that should not occur if the conditions of Theorem~\ref{thm:info-path} hold. In line~\ref{ln:error-1}, we visit a non-input node at time $t = 0$. But such a node should never have been reached in the recursion, since we only followed edges that have $M$-information flow. Its presence, therefore, would contradict the computational system model. In line~\ref{ln:error-2}, $\Vop$ is marked ``invalid'', implying that there is no path leading to it from the input nodes. Once again, this can only occur if the computational system model is violated, or if the conditions of Theorem~\ref{thm:info-path} do not hold.

\subsubsection*{On Computational Complexity}

The complexity of this algorithm is exactly that of Depth-first Search, $\mathcal O(\abs{\V} + \abs{\E})$~\cite[Sec.~22.3]{Cormen2009Algorithms}. To be precise, we consider the computational system to extend until the time of the output node, i.e., we take $T = t_\text{op}$. So the complexity of the algorithm is $\mathcal O(\abs{\Vstar}t_\text{op} + \abs{\Estar}t_\text{op})$. This is easily verified: if we assume that all edges in the system have $M$-information flow, then all edges and nodes must be traversed by the search. At each node, we must execute lines 7 through 14, and 26 through 28, which take a constant amount of time. Since we have $\abs{\Vstar}$ nodes over $t_\text{op}$ time points, this adds up to $\mathcal O(\abs{\Vstar}t_\text{op})$ steps. We also need to execute the loop in lines 15 through 24, which counts the number of incoming edges at every node. For all nodes combined, this adds up to $\mathcal O(\abs{\Estar}t_\text{op})$ steps.

If the graph is fully connected as described in Section~\ref{sec:comp-sys}, then $\abs{\Vstar} = N$ and $\abs{\Estar} = N^2$, so the effective complexity is just $\mathcal O(N^2 t_\text{op})$. However, if we \emph{know} that the underlying graph is sparse (e.g., because of anatomical priors in neuroscience), then we may have $\abs{\Estar} = \mathcal O(N \log N)$, or even $\abs{\Estar} = \mathcal O(N)$, bringing down the complexity of the search. It should be noted that in either case, the complexity of identifying \emph{which} edges have $M$-information flow is potentially exponential in $N$, as discussed later in Section~\ref{sec:difficulty-of-estimation}. This is much larger than the complexity of tracing out information paths, so \emph{finding edges} with $M$-information flow is, in fact, the ``hard part'' of the problem.

\subsection{Derived Information and Redundancy}
\label{sec:derived-info-redundancy}

The framework we develop for information flow allows one to obtain a more fine-grained understanding of information structure in a computational system, especially when compared with classical tools such as correlation and phase synchrony~\cite{Lachaux1999Measuring,Varela2001Brainweb}. This allows the experimentalist to better investigate the nature of the computation being performed. A concept that we believe will be extremely useful in this regard is one we call ``derived information'', which is defined below.

\begin{definition}[Derived $M$-Information] \label{def:derived-info}
	In a computational system $\C$, a transmission $X(Q_t)$ is said to be \emph{derived $M$-information} of a different transmission $X(P_{t'})$ if $M$---$X(P_{t'})$---$X(Q_t)$ forms a Markov chain. That is, the following condition must hold:
	\begin{gather}
		I\bigl(M ; X(Q_t) \given X(P_{t'})\bigr) = 0,
		\shortintertext{implying that}
		H\bigl(M \given X(P_{t'})\bigr) = H\bigl(M \given X(P_{t'}), X(Q_t)\bigr).
	\end{gather}
	So, $X(Q_t)$ adds no new information about $M$, when given $X(P_{t'})$. The same definition extends to transmissions on sets of edges. Note that, as far as the definition is concerned, $t$ and $t'$ may be any two arbitrary points in time. However, we will typically consider cases when $t \geq t'$.
\end{definition}

One potential use-case scenario for derived information arises in the context of redundant flows. Consider the computational system presented in Figure~\ref{fig:ce-cond-all}, originally described under Counterexample~\ref{ce:cond-all}. We see two edges sending the same transmission to the node $B_2$. This is an example of what we call ``redundant transmissions''. In general, since we only consider information about $M$ to be relevant, the exact transmissions communicated over two edges at a given point in time may be different. But if they convey the \emph{same information about $M$} to a given node, then we view them as essentially redundant. Definition~\ref{def:info-flow}, when applied to this system, will detect both these edges as having $M$-information flow, since given $X((C_1, B_2))$, their transmissions depend on $M$. In the notation of the Separability property mentioned earlier (Proposition~\ref{prop:separation}), both edges $(A_1, B_2)$ as well as $(D_1, B_2)$ will belong in the set $\R_1$.

Derived information provides a general methodology to understand when transmissions on certain edges may be redundant. Naturally, if the transmissions on two edges $Q_t$ and $P_{t'}$ are redundant, then they must be derived $M$-information of one another. This amounts to checking two more conditional independence relationships, for which consistent tests exist in the limit of infinite trials, as discussed in Section~\ref{sec:estimating-info-flow}.

In the following section, we shall see another application of derived information; when applied to specific sets, it can in some cases be used to detect the presence of hidden (unobserved) nodes. Later, in Section~\ref{sec:eg-sk}, we discuss an example where the notion of derived information helps us make a new kind of inference about the fine structure of information flow, one that would not be possible using tools such as Granger Causality and Directed Information.

\subsection{Hidden Nodes}
\label{sec:hidden-nodes}

In Section~\ref{sec:info-path-algo}, we showed how the Information Path Algorithm may fail to discover $M$-information paths if one of the assumptions of the computational system model or the observation model breaks in some way. Here, we discuss one specific situation in which the observation model may break, i.e., when not all nodes are observed. We call these unobserved nodes ``hidden nodes'', and assume that we do not see transmissions on incoming or outgoing edges of these nodes.

\begin{definition}[Hidden nodes]
	Consider a computational system $\C = (\G, X, W, f)$ defined on the time-unrolled graph $\G = (\V, \E)$ as before. Suppose that only a subset of nodes in this graph are observed. Specifically, if $\Vstar$ was the original set of nodes in $\Gstar$, prior to time-unrolling, then we observe only the nodes $\widetilde\V^* = \Vstar \setminus \H^*$, where $\H^* = \{H^{(0)}, H^{(1)}, \ldots, H^{(K-1)}\}$ is a set of unobserved nodes called \emph{hidden nodes}.

	To describe the observed component of the computational system, we define $\widetilde\E^* = \widetilde\V^* \times \widetilde\V^*$, $\widetilde\V = \{V_t : V \in \widetilde\V^*, t \in \T\}$ and $\widetilde\E = \{(A_t, B_{t+1}) : (A, B) \in \widetilde\E^*, t \in \T\}$. Also let $\H = \{H_t : H \in \H^*, t \in \T\}$. Finally, we set up the observed component of the computational system as before: $\widetilde\C = (\widetilde\G, X, W, f)$. Thus, we only observe the transmissions on edges in $\widetilde\E$. As usual, we denote the set of all hidden nodes at time $t$ by $\H_t$, and the set of all observed nodes at time $t$ by $\widetilde\E_t$.
\end{definition}

\begin{figure}
	\centering
	\includegraphics{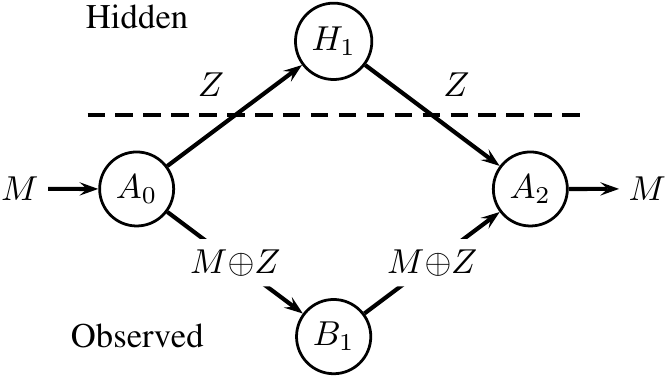}%
	\qquad%
	\includegraphics{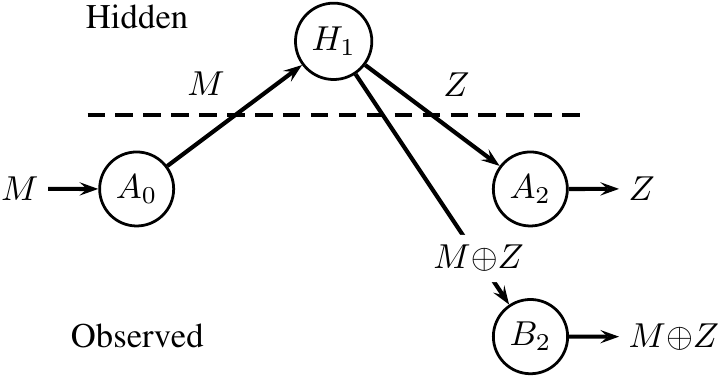}
	\caption{Two simple examples showing how hidden nodes may prevent one from being able to discover $M$-information paths in a computational system. In both cases shown here, $H_1$ is a hidden node, and we do not observe its incoming or outgoing transmissions. On the left is an example where a transmission that we might need to condition upon to discover $M$-information flow passes through the hidden node, and therefore cannot be seen. On the right, the hidden node itself generates the source of randomness $Z$.}
	\label{fig:hidden-basic}
\end{figure}

The presence of hidden nodes of this nature implies that much of the theory we have developed will not apply. Lemma~\ref{lem:edge-info} no longer truly holds, in that information about $M$ may persist in the system by passing through the hidden node, even if \emph{no} observed edge has $M$-information flow. So, naturally, Property~\ref{ppty:broken-telephone} also fails to hold. Hence, we are not guaranteed to be able to identify all edges with $M$-information flow, and discover all $M$-information paths as before. For example, refer to the cases shown in Figure~\ref{fig:hidden-basic}, where we no can longer find $M$-information paths because of the presence of a hidden node.

Fortunately, at least in some cases, the concept of derived information (Definition~\ref{def:derived-info}) provides a simple way to tell whether or not a hidden node exists. Specifically, if at some time $t$, a hidden node transmits information about $M$ which is unavailable within the system at that time, and which is utilized by some node at the next time instant, then the set of all observed transmissions $X(\widetilde\E_t)$ will \emph{not} be derived $M$-information of the set of all transmissions at time $t-1$. In other words, the Global Markovity condition (Corollary~\ref{cor:global-markov}) on the observed graph, $M$---$X(\widetilde\E_{t-1})$---$X(\widetilde\E_t)$, will break. Unfortunately, the notion of ``utilization'' is difficult to express mathematically, without resorting to the use of ideas from causality that are based on intervention. The result we prove, therefore, is a simpler sufficiency argument, which guarantees the presence of a hidden node if the aforementioned Markov condition is observed to break. This result is proved in Proposition~\ref{prop:hidden-markov}, but first, we define some adjectives.

\begin{figure}
	\centering
	\includegraphics{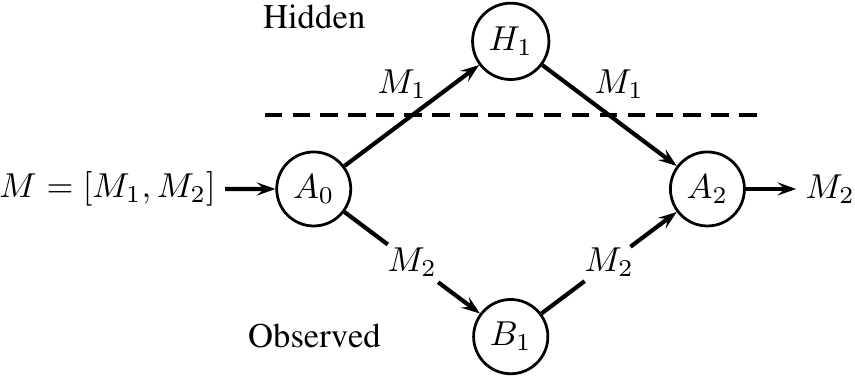}
	\caption{A computational system serving as a counterexample to the converse of Proposition~\ref{prop:hidden-markov}. Here, the hidden node $H_1$ is $M$-relevant because its outgoing transmission, $M_1$, is not present in any of the observed transmissions at time $t=1$. However, since $A_2$ chooses to ignore $M_1$ at its output, the Markov chain $M$---$X(\widetilde\E_1)$---$X(\widetilde\E_2)$ boils down to $M$---$M_2$---$M_2$, which obviously holds. Thus, at least based on our current definitions, there may be $M$-relevant hidden nodes in the system even if Global Markovity continues to hold.}
	\label{fig:hidden-markov-ce}
\end{figure}

\begin{definition}[$M$-relevant hidden node] \label{def:relevant-hidden-node}
	A hidden node $H_t$ is said to be \emph{$M$-relevant} if $\Q(H_t)$ carries $M$-information flow in $\G$. Similarly, a subset of hidden nodes $\H_t' \subseteq \H_t$ is said to be $M$-relevant if $\Q(\H_t')$ carries $M$-information flow in $\G$.
\end{definition}
\begin{definition}[$M$-derived hidden node] \label{def:derived-hidden-node}
	A hidden node $H_t$ is said to be \emph{$M$-derived} if the Markov chain $M$---$X(\widetilde\E_t)$---$X(\Q(H_t))$ holds. Similarly, a subset of hidden nodes $\H_t' \subseteq \H_t$ is said to be $M$-derived if the Markov chain $M$---$X(\widetilde\E_t)$---$X(\Q(\H_t'))$ holds.
\end{definition}
\begin{lemma} \label{lem:hidden-rel-derived}
	If a subset of hidden nodes is \emph{not} $M$-derived, then it \emph{is} $M$-relevant.\footnote{If this lemma appears to be somewhat strong, it is only because of the nomenclature ``$M$-derived''. For our purposes, a hidden node whose transmissions are independent of the message is also $M$-derived, since it satisfies the aforementioned Markov condition.}
\end{lemma}
\begin{proposition} \label{prop:hidden-markov}
	In a computational system $\C$ with hidden nodes, if Global Markovity on the observed graph, $\widetilde\G$, fails to hold from time $t$ to $t+1$, i.e.\ if $I\bigl(M ; X(\widetilde\E_{t + 1}) \given X(\widetilde\E_t)\bigr) > 0$, then the hidden nodes $\H_t$ at time $t$ are \emph{not} $M$-derived.
\end{proposition}
Proofs of Lemma~\ref{lem:hidden-rel-derived} and Proposition~\ref{prop:hidden-markov} are very straightforward, and are provided in Appendix~\ref{app:hidden-node-proofs}. As a direct consequence of these two results, if Global Markovity fails to hold on the observed nodes from time $t$ to $t+1$, then $\H_t$ is $M$-relevant. By Proposition~\ref{prop:set-info-equiv}, this simply means that there exists at least one $M$-relevant hidden node at time $t$.

\begin{figure}
	\centering
	\begin{subfigure}[t]{0.48\textwidth}
		\centering
		\includegraphics{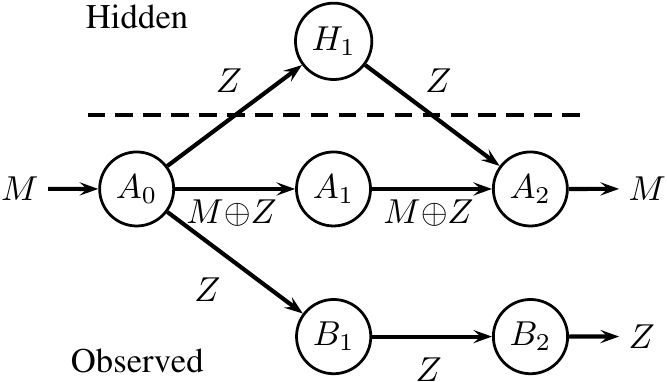}
		\vspace{0.63cm} 
		\caption{Here, although Global Markovity holds, one could argue that testing for Local Markovity at each node (or at various subsets of nodes) could help uncover the presence of a hidden node.}
		\label{fig:hidden-redundant-1}
	\end{subfigure}%
	\hfill%
	\begin{subfigure}[t]{0.48\textwidth}
		\centering
		\includegraphics{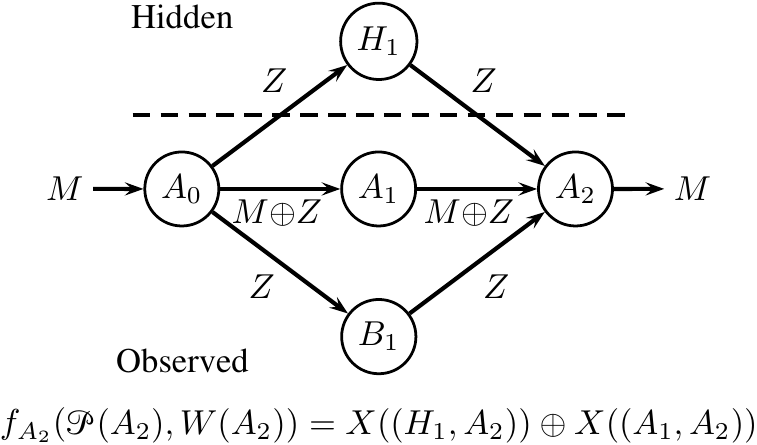}
		\caption{In this case, the hidden node breaks neither Global nor Local Markovity. However, the \emph{function} computed by $A_2$ makes \emph{use} of only the hidden node's transmission. As a result, the hidden node has a causal effect on the output of the system, since destroying the outgoing edge of the hidden node would change the output. Such a hidden node is likely undetectable using only observational methods.}
		\label{fig:hidden-redundant-2}
	\end{subfigure}
	\caption{Examples of computational systems with an $M$-derived hidden node. In both of these systems, the hidden node's transmission at time $t=1$ has an affect on the output at $A_2$. However, Global Markovity continues to hold from $t=1$ to $t=2$, because the observed transmissions, $M \xor Z$ and $Z$, contain all information necessary to explain the output, $M$.}
	\label{fig:hidden-redundant}
\end{figure}

Although Proposition~\ref{prop:hidden-markov} appears to provide a straightforward mechanism to test whether or not hidden nodes exist, it does not always work. If a hidden node's transmissions have no $M$-information flow, then the node will not be detected. But in this case, it could be argued that such a hidden node does not change whether information paths can be identified, and so can be subsumed by one or more of the intrinsic random variables $W(\cdot)$. Such a hidden node is, therefore, classified by Definition~\ref{def:relevant-hidden-node} as \emph{not} $M$-relevant.

However, to make matters worse, the converse of Proposition~\ref{prop:hidden-markov} does not hold. In particular, there may exist an \emph{$M$-relevant} hidden node at time $t$, whose transmission is \emph{ignored} by the node that received it, so that the Markov chain $M$---$X(\widetilde\E_t)$---$X(\widetilde\E_{t+1})$ continues to hold (see Figure~\ref{fig:hidden-markov-ce}). Such a hidden node may \emph{still} be considered largely innocuous.

The most serious case of a hidden node going undetected is one that contains an \emph{$M$-derived} hidden node, whose transmission \emph{is used} by the receiving node while performing its computation; however the hidden node's transmission is ``masked'' by a redundant transmission from an observed node (see Figure~\ref{fig:hidden-redundant}). In this case, Global Markovity on $\widetilde\G$ will \emph{not} break, yet the hidden node's transmission may be instrumental in producing a certain output distribution. In some instances, such hidden nodes can be detected by checking for Local Markovity (Proposition~\ref{prop:local-markov}; see Figure~\ref{fig:hidden-redundant-1}). However, there are still cases where if we were somehow able to intervene and delete the transmission of the hidden node, then the computational system's output may not remain the same, despite the existence of a redundant transmission from an observed node (see Figure~\ref{fig:hidden-redundant-2}). Indeed, the presence of redundancy in such a scenario does not guarantee that the computational system will actually leverage it.

\subsection{On Multiple Messages and the Distribution of the Message}
\label{sec:multiple-messages}

Just as we can infer information flow and information paths for a single message, we can examine the flows of multiple messages in the same computational system. Consider a case where we wish to understand the information flows of two messages, $M_1$ and $M_2$. An neuroscientific example of this might be information flow about two independent components of a visual stimulus, e.g., shape and color (such as in~\cite{Almeida2013Tool}). If $M_1 \independent M_2$, then we could separately identify edges and paths that have $M_1$-information flow and $M_2$-information flow, by applying the theory and algorithm as-is for each message individually.

\begin{figure}
	\centering
	\includegraphics{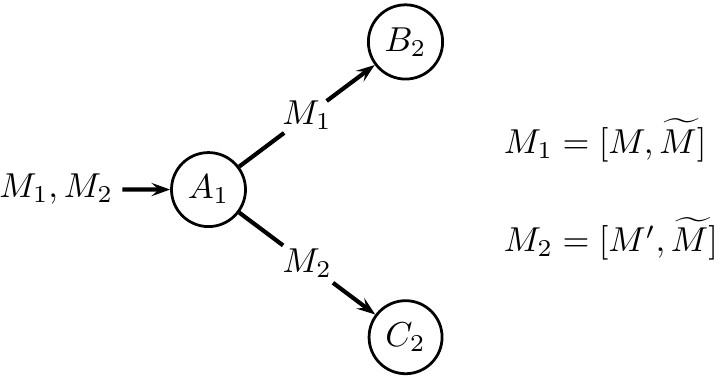}
	\caption{A simple example demonstrating the importance of having independent messages (or sub-messages) when exploring the flows of multiple messages in a computational system. As $M_1$ and $M_2$ both redundantly contain information about $\widetilde M$, both edges shown here have $M_1$- as well as $M_2$-information flow. Thus, we are unable to detect the fact that $M_1$ and $M_2$ take different paths in the system, because of our choice of stimuli.}
	\label{fig:multiple-msgs}
\end{figure}

However, if the two messages are \emph{dependent} on one another, one could end up confounding their information flows, based on how they depend on each other, and how the computational system's transmissions carry their joint information. As a simple example, consider the system shown in Figure~\ref{fig:multiple-msgs}, where $M_1 = [M, \widetilde M]$ and $M_2 = [M', \widetilde M]$, with $M, M', \widetilde M \sim \text{i.i.d.~Ber}(1/2)$. Clearly, $M_1$ and $M_2$ both share some redundant information in $\widetilde M$, and $I(M_1 ; M_2) = 1$~bit. Thus, we will see $M_1$-information flow as well as $M_2$-information flow on both edges, since the transmission of each edge $E$ satisfies $I\bigl(M_i ; X(E)\bigr) > 0$ for $i \in \{1, 2\}$.

Consider what this means for the aforementioned example of shape and color of a visual stimulus. If a neuroscientist expects that the information paths corresponding to shape and color in the brain are different from each other, what is the best way to design stimuli so as to bring out this difference? Suppose they decided to present a total of four different stimuli, $M \in \{0, 1, 2, 3\}$, with two different shapes and two different colors. Let $M_1$ be the first bit of the binary representation of $M$, denoting shape, and $M_2$ be the second bit, denoting color. Now if the neuroscientist chose to present stimuli with a uniform distribution over $M$, i.e., if each shape-color combination was shown for one-quarter of all trials, then $M_1$ and $M_2$ would be independent of each other, and their individual flows could be tracked separately. However, if the neuroscientist chose to present the four possible stimuli with probabilities $\{1/2, 1/4, 1/8, 1/8\}$ respectively, then $M_1$ and $M_2$ are no longer independent of each other, and it may become hard to separate their individual flows as in the example in Figure~\ref{fig:multiple-msgs}.

These examples suggest that, when trying to understand the flows of different messages in a computational system, it helps if they are independent of one another. So from the perspective of experiment design in a neuroscientific context, it is often more sensible to design stimuli so that the two messages of interest are independent of one another. Even when considering a single message that takes one of several values, it becomes important to appropriately choose a distribution over these values to ensure that any sub-messages that are of interest remain independent of one another. This would allow the experimentalist to better understand how ``independent dimensions'' of the stimulus are processed in the brain.

However, there are also situations where the experimental paradigm necessitates a statistical distribution of stimuli that makes two sub-messages of interest dependent on one another. For instance, the Posner experimental paradigm for attention~\cite{Posner1980Orienting} only works when the proportion of ``valid'' trials (a certain \emph{type} of trial specific to this paradigm) is roughly 70\%. Similarly, during data preprocessing, it is common to discard trials that are excessively noisy, based on some predetermined metric: this process could skew the distribution of the message, even if the original distribution was uniform. If it is still of interest to understand the individual flows of sub-messages in this case, then a possible solution might then be to sub-select experimental trials in such a way as to keep the two sub-messages independent of one another.


\section{Canonical Computational Examples} 
\label{sec:canonical-examples}

In this section, we provide a few canonical examples for computational systems from various contexts. In each case, we discuss what the message $M$ is, and identify which edges carry $M$-information flow. We also explain how the path recovered by the information path algorithm might be the intuitive choice in each example.

\subsection{The Butterfly Network from Network Coding}
\label{sec:eg-nc-butterfly}

\begin{figure}
	\centering
	\includegraphics{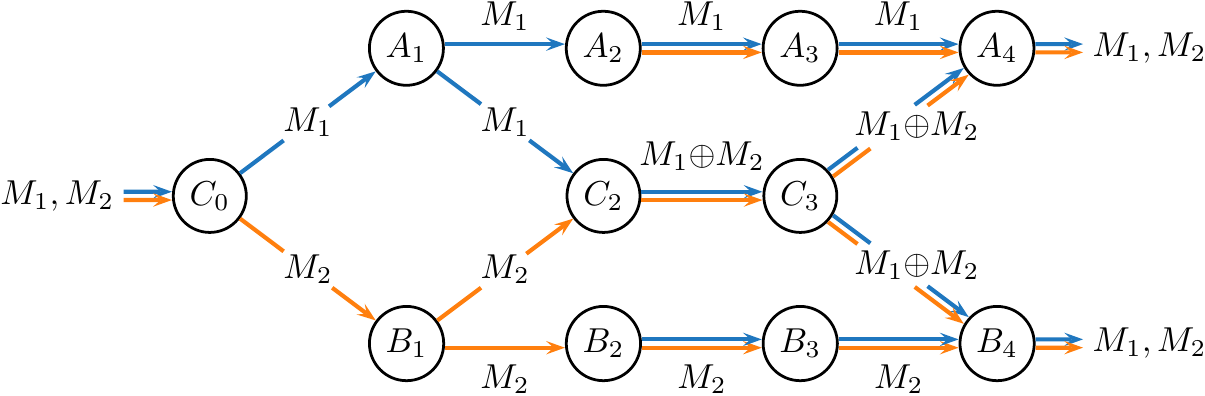}
	\caption{A depiction of the butterfly network discussed in Section~\ref{sec:eg-nc-butterfly}. There are two messages, $M_1$ and $M_2$, each with its own information flow. All edges with $M_1$-information flow are shown in blue and all edges with $M_2$-information flow are shown in orange. After time $t=2$, all edges shown have both $M_1$- and $M_2$-information flow. Once the system computes $M_1 \xor M_2$, edges transmitting $M_1$ have information flow about both $M_1$ and $M_2$, since $M_2$ can now be decoded from $M_1 \xor M_2$ and $M_1$. Furthermore, observe the $M_1$- and $M_2$-information paths in this system. In particular, there are two possible $M_1$-information paths to $A_4$, but only one possible $M_2$-information path, which flows through the middle link. The same applies to the $M_1$-information path to $B_4$. This may suggest the importance of the middle link in enabling this computation.}
	\label{fig:nc-butterfly}
\end{figure}

For our first example, we cover the butterfly network from network coding literature~\cite[Fig.~7b]{Ahlswede2000Network}, reproduced here in Figure~\ref{fig:nc-butterfly}. We consider two different messages, $M_1, M_2 \sim \text{i.i.d.~Ber}(1/2)$, provided as input to the system. Edges along which information about $M_1$ flows are colored in blue, while edges along which information about $M_2$ flows are colored in orange. The reader may identify these using Definition~\ref{def:info-flow} and the transmission on each edge shown in Figure~\ref{fig:nc-butterfly}.

An important feature to observe is that when $C_2$ mixes information by computing the \textsc{xor} of $M_1$ and $M_2$, we see information about $M_1$ spontaneously beginning to flow on $(B_2, B_3)$ and similarly, information about $M_2$ beginning to flow on $(A_2, A_3)$. This is expected, since $M_2$ is relevant for decoding $M_1$ at this stage, and indeed, it is exactly this idea which is used to decode $M_1$ at $B_4$. All of this is true, despite the fact that $M_1 \xor M_2$ is independent of $M_1$ and $M_2$ individually. This is once again, a prime example of synergy in action.

Applying the information path algorithm for the message $M_1$ at $A_4$ will reveal two paths: the ``upper path'' $(C_0, A_1, A_2, A_3, A_4)$, and the ``middle path'' $(C_0, A_1, C_2, C_3, A_4)$. However, applying the information path algorithm for the message $M_2$ at $A_4$ reveals that $M_2$ exclusively used the ``middle path'', $(C_0, B_1, C_2, C_3, A_4)$, to arrive at $A_4$ from the input nodes.

\subsection{The Fast Fourier Transform}
\label{sec:eg-fft}

\begin{figure}
	\centering
	\includegraphics{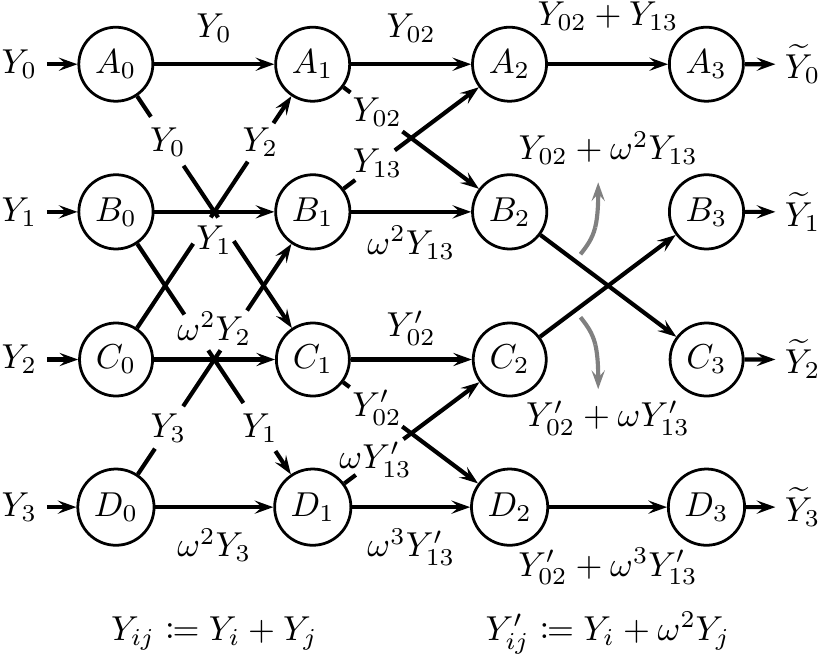}
	\caption{The computational system of the 4-point Fast Fourier Transform. For brevity, we have set $\omega \coloneqq e^{-j\frac{2\pi}{4}}$.}
	\label{fig:fft-base}
\end{figure}

The Fast Fourier Transform (FFT) is a well-known computational network that provides an intuitive setting for examining information flow. In general, the $N$-point FFT is an implementation of the $N$-point Discrete Fourier Transform (DFT), given by
\begin{equation}
	\widetilde Y_k = \sum_{i=0}^{N-1} Y_i e^{-j \frac{2 \pi k}{N} i}, \quad k \in \{0, 1, \ldots, N-1\}.
\end{equation}
The DFT is a basis transformation of a discrete-time signal $Y$, which is usually assumed to be periodic with period $N$. The $N$-point DFT represents such a signal in the complex-exponential Fourier basis, yielding the Fourier coefficients $\widetilde Y$. We consider a simple 4-point DFT, i.e.\ $N = 4$. The FFT implements this transform using the computational system shown in Figure~\ref{fig:fft-base}. We refer the reader to~\cite[Ch.~9]{Oppenheim1999Discrete} for details. For notational convenience, we have set $\omega = e^{-j\frac{2\pi}{N}} = e^{-j\frac{2\pi}{4}}$.

\begin{figure}
	\centering
	\includegraphics{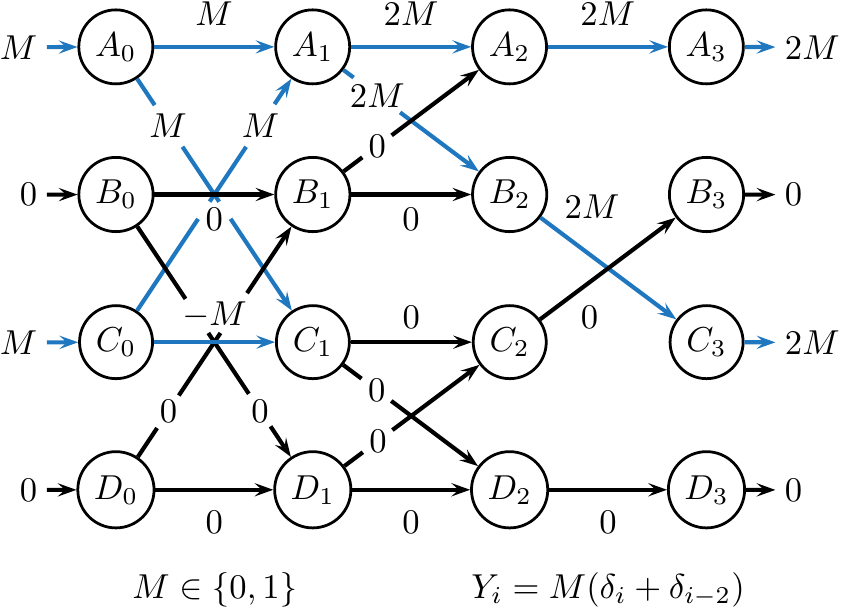}
	\caption{An example of information flow in the 4-point FFT, when the message determines which of two signals is supplied to the system: $Y = [0, 0, 0, 0]$ or $Y = [1, 0, 1, 0]$. Observe that, since $M$ is encoded in the even part of $Y$, only the ``even component'' of the FFT network is active. Furthermore, only the DC component, $\widetilde Y_0$ and the first harmonic, $\widetilde Y_2$ are active, as we would expect based on the two input signals.}
	\label{fig:fft-eg-1}
\end{figure}

We use this example to demonstrate how the definition of the message is important in determining information flow. First, suppose the message is one of two signals: $Y = [0, 0, 0, 0]$, or $Y = [1, 0, 1, 0]$. This can be written as $M \in \{0, 1\}$ and $Y_i = M(\delta_i + \delta_{i-2})$, where $\delta_i = \mathbb I\{i = 0\}$ is the Kronecker Delta function, and we assume $M \sim \text{Ber}(1/2)$. The full computational system, along with the random variables computed on all edges, is shown in Figure~\ref{fig:fft-eg-1}. The edges that have $M$-information flow are highlighted in blue. Since $M$ is encoded into the \emph{even} part of $Y$ (observe that $Y_i = Y_{-i} \;\;\forall\;M$), we notice that only the ``even component'' of the FFT system (corresponding to the 2-point FFT on the even indices of $Y$) is active~\cite[Sec.~9.3]{Oppenheim1999Discrete}. Furthermore, only $\widetilde Y_0$, the DC component, and $\widetilde Y_2$, the first harmonic, show variation with $M$ at the output, as we would expect based on the two input signals.

\begin{figure}
	\centering
	\includegraphics{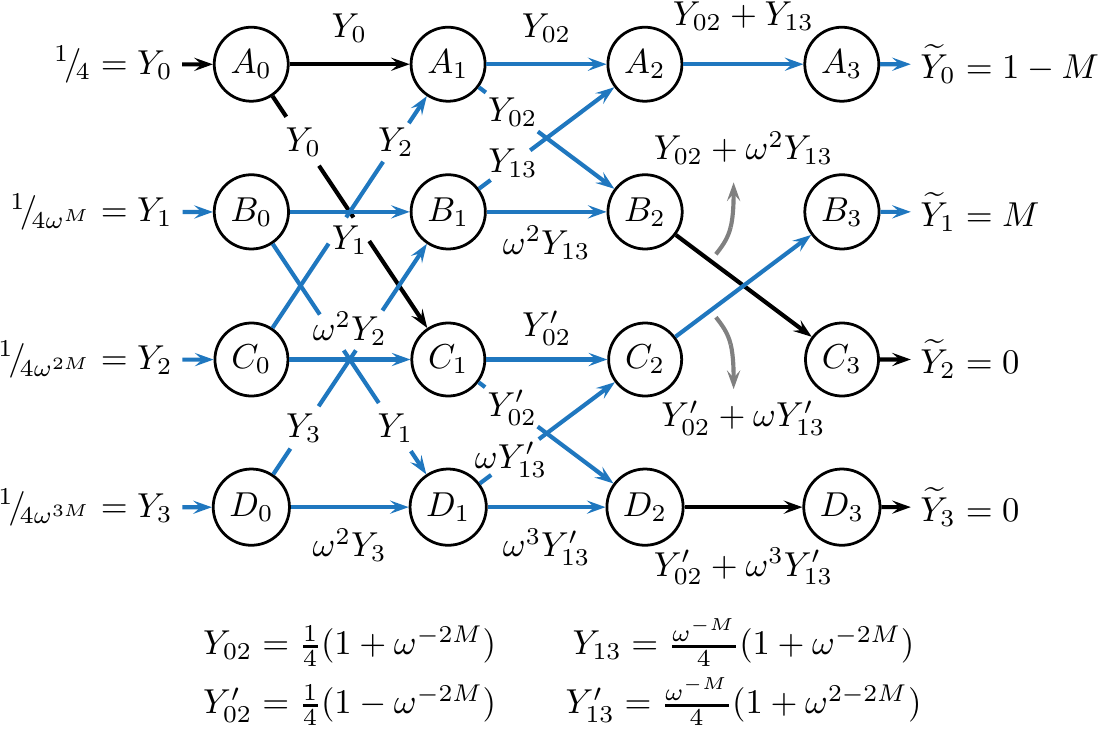}
	\caption{Another example of information flow in the 4-point FFT, when the message determines which of two signals is supplied to the system: $Y = [1, 1, 1, 1]$ or $Y = [1, 1/\omega, 1/\omega^2, 1/\omega^3]$. The $M$-information paths are different from those in Figure~\ref{fig:fft-eg-1}, showing how the choice of the message can have a strong impact on the flows within the same computational system.}
	\label{fig:fft-eg-2}
\end{figure}

As a second example, consider the case shown in Figure~\ref{fig:fft-eg-2}. Here, the message is again one of two signals: $Y = [1, 1, 1, 1]$, or $Y = [1, 1/\omega, 1/\omega^2, 1/\omega^3]$. These signals can be jointly expressed in terms of the binary message random variable $M \sim \text{Ber}(1/2)$ as $Y_i = 1 / \omega^{iM}$. The two signals are flat in their magnitude spectra and differ only in their phase, creating $\delta$-functions in the Fourier domain that are frequency-shifted with respect to one another: $\widetilde Y_k = \delta_{k - M}$. Once again, the edges in the network that carry $M$-information flow are demarcated in blue. A detailed derivation of the values of the transmissions in the computational system can be found in Appendix~\ref{app:fft-eg-2-derivation}.

These two examples make it clear that, based on how the message is defined, the $M$-information paths in the system can be very different. Indeed, if the message were as general as possible, by placing a probability distribution over all possible values of $Y$ in $\mathbb R^4$, we know that \emph{all} edges in the computational system would have $M$-information flow. However, selectively restricting $M$ to just a few signals helps reveal some kind of structure within the FFT network.

Another feature that can be observed in these examples is how the output of the computational system can be a function of the message. Although only very simple functions of the message have been shown at the outputs here, the FFT demonstrates that, in principle, more complex functions of the message may also be generated.

\subsection{The Schalkwijk and Kailath Scheme}
\label{sec:eg-sk}

The Schalkwijk and Kailath scheme~\cite{Schalkwijk1966Coding} is an efficient strategy for communicating a message in the presence of a noisy feedforward channel and a noiseless feedback channel. We have previously used this scheme as a counterexample~\cite{Venkatesh2015Direction}, to show that comparing Granger causal influences in forward and backward directions can lead to erroneous inferences on the direction in which the message is being sent in this feedback system. We first provide a brief overview of the scheme, then recapitulate our previous result, and finally demonstrate what the information flow framework developed in this paper has to offer in the case of this example.

\begin{figure}
	\centering
	\includegraphics{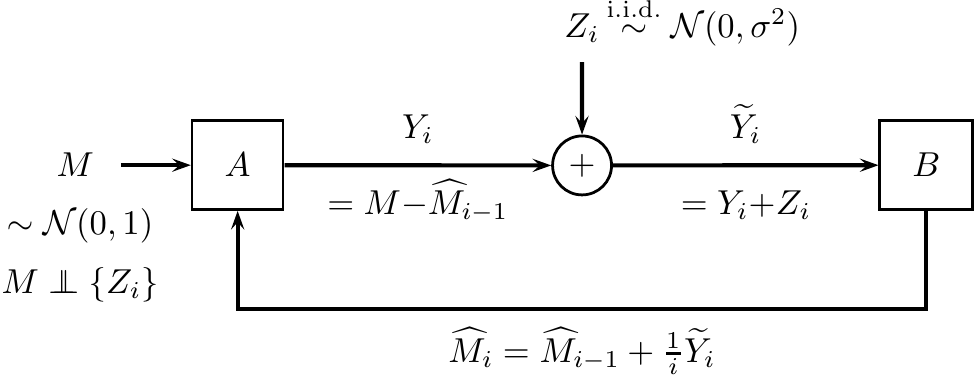}
	\caption{A communication system depicting the Schalkwijk and Kailath scheme. Alice, represented by node $A$, communicates a message $M$ to Bob, represented by node $B$, in the presence of a noisy feedforward channel and a noiseless feedback channel. In the $i^{th}$ iteration, Alice transmits the error in Bob's most recent estimate of the message, $Y_i$, but her transmission is corrupted by the noise $Z_i$. Bob updates and transmits his estimate, $\widehat M_i$, which reach Alice noiselessly.}
	\label{fig:sk-scheme}
\end{figure}

Consider the communication system depicted in Figure~\ref{fig:sk-scheme}, which shows the schematic of a simplified version of the Schalkwijk and Kailath scheme. For convenience, let us denote the transmitter, $A$, and receiver, $B$, by Alice and Bob respectively. Alice is attempting to communicate a message $M$ to Bob over an additive Gaussian channel, but in the presence of noiseless feedback. Alice starts by transmitting the message $Y_1 = M$ to Bob, over the noisy feedforward channel. Bob receives a corrupted version of $M$, given by $\widetilde Y_1 = Y_1 + Z_1$, and computes an estimate $\widehat M_1$. He sends this estimate back to Alice over the noiseless feedback channel. In the iterations that follow, Alice computes the error in Bob's most recent estimate, $Y_i = M - \widehat M_{i-1}$, and sends this to Bob over the noisy feedforward channel. Meanwhile, Bob updates his estimate based on Alice's noisy transmissions $\widetilde Y_i = Y_i + Z_i$, using the following rule:
\begin{align}
	\widehat M_i &= \widehat M_{i-1} + \frac{1}{i} \widetilde Y_i
	\shortintertext{It can be shown that this rule implies}
	\widehat M_i &= M + \frac{1}{i} \sum_{j=1}^i Z_i  \label{eq:sk-mhat-final}
\end{align}
Thus, this strategy ensures that Bob's estimate $\widehat M_i$ converges to $M$ in mean squared sense~\cite{Venkatesh2015Direction}.

Intuitively, one might expect that, since the message $M$ is being transmitted in the forward direction, the Granger Causal influence from Alice to Bob is greater than that from Bob to Alice. However, our earlier result~\cite{Venkatesh2015Direction} showed that, in fact, the opposite is true. In other words, even though the message is being communicated from Alice to Bob, the Granger Causal influence from Bob to Alice is greater; in fact, the Granger Causal index from Bob to Alice is \emph{infinite}. The reason for this is that, while Alice's past transmissions do not perfectly predict Bob's transmissions (due to the presence of noise in the feedforward link), Bob's past transmissions \emph{perfectly} predict Alice's transmissions (since the latter are a simple function of the former). Therefore, the Granger Causal index from Alice to Bob, which measures the relative predictive gain of including Alice's past transmissions in the autoregression for Bob's transmissions, remains finite; while the Granger Causal index from Bob to Alice becomes infinite.

Our earlier paper on this subject~\cite{Venkatesh2015Direction} concluded that the direction of greater Granger Causal influence could be opposite to the ``direction of information flow'' in the Schalkwijk and Kailath scheme. There, ``information flow'' was being used purely in an intuitive sense, to mean the direction in which the message was being communicated in that system. The intent of our previous paper was to explain that it is not always possible to interpret a larger Granger causal influence in a certain direction to mean that a specific message is being communicated in that direction. In contrast, this paper presents a refined theoretical framework that defines information flow about a message $M$ for a specific \emph{edge} in a computational system. Now, we no longer speak of \emph{one specific direction} in which information flows; rather, we describe \emph{which edges} carry information about the message in their transmissions \emph{at each point in time}. This leads to a more nuanced understanding of information flow in the Schalkwijk-Kailath setting.

\begin{figure}
	\centering
	\includegraphics{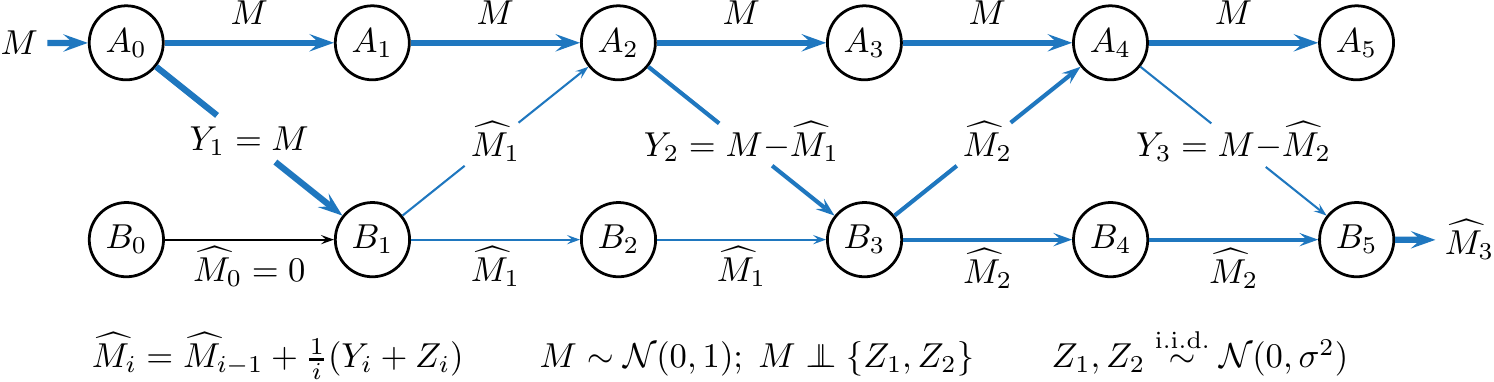}
	\caption{A computational system describing the first few iterations of the Schalkwijk and Kailath scheme. Almost every edge shown here has $M$-information flow. However, the \emph{quantity} of $M$-information flow (shown using line thickness) reveals the asymmetry between Alice and Bob: Alice has the message to begin with, and her transmissions have a larger volume of $M$-information flow. In contrast, Bob's initial transmissions are poor estimates and have small volumes of $M$-information flow, but they get better over a few iterations, and eventually come close to the true message. Furthermore, we also reveal an asymmetry between Alice and Bob using the concept of derived information: each of Bob's transmissions is $M$-derived from Alice's previous transmissions, whereas Alice's transmissions are \emph{not} $M$-derived from Bob's previous transmissions. Both these facts point towards the idea that Alice is slowly sending information about $M$ to Bob.}
	\label{fig:sk-comp-sys}
\end{figure}

Before we can analyze the $M$-information flows in the Schalkwijk-Kailath scheme, we need to fit the scheme within the computational system framework. Figure~\ref{fig:sk-comp-sys} shows the time-unrolled computational system corresponding to two feedforward and feedback iterations of the simplified Schalkwijk-Kailath scheme described before. In order to translate the communication system into our computational system model while remaining consistent with our earlier work~\cite{Venkatesh2015Direction}, we have merged the process of noise addition with the receiver, i.e., Bob. This exposes the edges with Alice's and Bob's \emph{transmissions}, making them observable, as was assumed in our previous paper~\cite{Venkatesh2015Direction}. This is also consistent with what \emph{would have been} observable if $A$ and $B$ were neurons (or neural populations) whose outputs a neuroscientist were to measure.\footnote{From a wireless communication system perspective, as well, it is more reasonable to assume noise to be a part of the receiver's node, since the additive noise in a signal is usually considered to be the result of thermal noise in the receiver's circuitry.} Note that one full iteration of the Schalkwijk-Kailath scheme takes two time steps in this model, so the iteration index $i$ advances once for every two time steps $t$. Also, note that this does \emph{not} make $\widetilde Y$ or $Z$ ``hidden nodes'', since the function computed at $B_t$ can be defined purely in terms of its inputs, $(Y_i, \widehat M_{i-1})$, and its intrinsic random variable, $W(B_t)$ (which absorbs $Z_i$), as follows:
\begin{equation}
	f_{B_{2i - 1}}\bigl(Y_i, \widehat M_{i-1}, W(B_{2i-1})\bigr) = \widehat M_{i-1} + \frac{1}{i}\bigl(Y_i + W(B_{2i-1})\bigr)
\end{equation}
where $W(B_{2i-1}) = Z_i$ takes the role of the noise in the communication system. Also, to understand the time index for node $B$, note that in the first step of iteration $i$, Alice transmits to Bob, i.e., node $A_{2i-2}$ transmits to $B_{2i-1}$ (see Figure~\ref{fig:sk-comp-sys}).

Now, we first show that all edges depicted in blue in Figure~\ref{fig:sk-comp-sys} carry $M$-information flow, based on Definition~\ref{def:info-flow}. Specifically, both Alice's feedforward transmissions \emph{and} Bob's feedback transmissions have $M$-information flow. This should not be surprising for the following intuitive reasons: Alice's transmissions convey information about $M$ which Bob uses to improve his estimate; meanwhile, Bob's transmissions are estimates of $M$, and therefore must depend on $M$. In fact, we can take this intuitive argument further: suppose we were to \emph{quantify} $M$-information flow by using the following natural extension of our definition,
\begin{equation}
	\F_M(E_t) \coloneqq \max_{\E_t' \subseteq \E_t} I\bigl(M ; X(E_t) \given X(\E_t')\bigr).
\end{equation}
Noting that Definition~\ref{def:info-flow} only specified \emph{whether or not} a given edge $E_t$ had information flow, all that we have now done is to take the maximum over the subsets of edges used to discover $M$-information flow in that definition. This quantification is fully consistent with our definition of $M$-information flow, since it goes to zero if and only if the $M$-information flow on an edge goes to zero. Now, using this quantitative notion of information flow, we can ask how the $M$-information flow on a given link---feedforward or feedback---varies with time. In particular, it should be intuitively clear that the $M$-information content in Bob's transmissions, i.e.\ $\widehat M_i$, \emph{increases} over time as his estimate improves. This is depicted as an increase in the thickness of the edges carrying Bob's transmissions with time. Meanwhile, the information content in Alice's transmissions \emph{decreases} with time. To understand why the latter is true, note that, after the first iteration, Alice's transmission represents the noise in Bob's estimate. Therefore, just as in Counterexample~\ref{ce:dependence}, when conditioned on Bob's estimate, Alice's transmissions depend on the message. At the initial iterations, when Bob's estimate is poor, we must have that $I(M ; \widehat M_i)$ is very small. Hence, we see that:
\begin{align}
	I(M ; Y_i \given \widehat M_{i-1}) &= I(M ; M - \widehat M_{i-1} \given \widehat M_{i-1}) \\
									   &\overset{(a)}{=} H(M \given \widehat M_{i-1}) + H(M \given M - \widehat M_{i-1}, \widehat M_{i-1}) \\
									   &= H(M \given \widehat M_{i-1}) \\
									   &= H(M) - I(M ; \widehat M_{i-1}) \overset{(b)}{\approx} H(M),
\end{align}
where in (a), the second term is zero because $M$ is a constant when given $\widehat M_{i-1}$ and $M - \widehat M_{i-1}$; while in (b), $I(M ; \widehat M_{i-1})$ is assumed to be approximately zero when Bob's estimate is poor (as we might expect if the noise is large, for instance). Hence, for the first few iterations, the quantified $M$-information flow of Alice's transmissions is close to $H(M)$, which is as large as the flow can get. However, as Bob's estimate improves, $I(M ; \widehat M_{i-1})$ becomes closer to $H(M)$, and therefore $I(M ; Y_i \given \widehat M_{i-1})$ becomes close to zero. At the same time, $I(M ; Y_i)$ is equal to zero, since $Y_i$ carries only information about the noise in $\widehat M_{i-1}$ (after the first iteration), which is independent of $M$. Thus, the quantified $M$-information flow of Alice's transmissions decreases over time. Correspondingly, this is depicted using edges whose thickness decreases over time in Figure~\ref{fig:sk-comp-sys}.

Quantifying the $M$-information flows of the feedforward and feedback links thus reveals an asymmetry between Alice and Bob that strongly suggests that the message is being transmitted from Alice to Bob. However, we can get a more nuanced understanding of information flow in this system by asking whether Bob's transmissions are \emph{derived} from Alice's, or vice versa. First, consider whether Bob's transmissions are derived $M$-information of Alice's previous transmissions: this can be expressed in terms of the Markov chain $M$---$[M, \; M - \widehat M_1]$---$\widehat M_2$. Observe that this Markov chain holds trivially:
\begin{equation}
    I(M ; \widehat M_2 \,\vert\, M, \, M - \widehat M_1) = 0.
\end{equation}
However, if we consider whether Alice's transmissions are derived $M$-information of Bob's past transmissions, it can be shown that $M$---$[\widehat M_1, \; \widehat M_2]$---$(M - \widehat M_2)$ is not a valid Markov chain (see Appendix~\ref{app:sk-markov-proof} for a detailed derivation). Hence, we see that Bob's transmissions are derived $M$-information of all of Alice's past transmissions, however, Alice's transmissions are \emph{not} derived $M$-information of all of Bob's past transmissions. In conjunction with the fact that the volume of $M$-information flow in Alice's transmissions slowly decreases from $H(M)$ with time, while the volume of $M$-information flow in Bob's transmissions slowly increases to $H(M)$ with time, this suggests that Alice has some information about the message $M$ that Bob slowly receives from Alice.

This example shows how a measure that quantifies information flow, along with derived information, can be used to understand some finer computational structure present within the computational system. In general, however, care needs to be exercised in applying derived $M$-information: one must choose what Markov condition to check in a principled manner. In the specific case of the Schalkwijk-Kailath example, we had the advantage of being in a two-node setting, where the derived information expressions we examined had clear interpretations. It may be that analyzing information flow first, to understand which variables transmit information about $M$ to one another, can help guide the choice of variables to examine when applying derived $M$-information.

\subsection{A Message Defined at the Output of a System}
\label{sec:eg-msg-op}

\begin{figure}
	\centering
	\begin{subfigure}{0.45\textwidth}
		\hfill
		\includegraphics{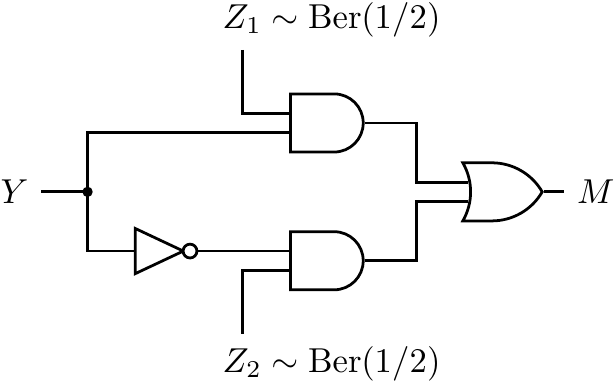}
	\end{subfigure}%
	\hfill%
	\begin{subfigure}{0.45\textwidth}
		\includegraphics{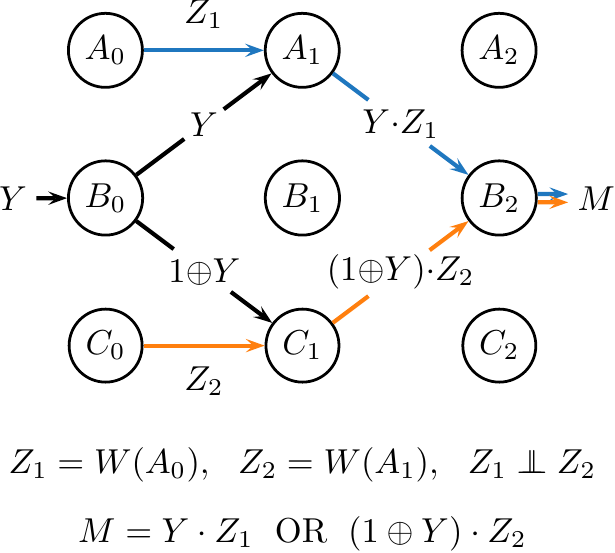}
		\hfill
	\end{subfigure}
	\caption{A boolean circuit demonstrating a message defined at the output of the computational system. Note that ``$\xor$'' refers to bitwise-\textsc{xor}, ``OR'' refers to bitwise-\textsc{or}, and ``$\cdot$'' refers to bitwise-\textsc{and}. We see that information paths may lead from an internal node, that generates an intrinsic random variable, to the output node. Furthermore, this path may change with the ``external parameters'' of the system.}
	\label{fig:eg-msg-op}
\end{figure}

We now describe an example where the message is defined at the \emph{output} of a computational system, instead of at the input. Although Definition~\ref{def:input-nodes} defines the message to be a random variable available at the input nodes, it is also possible to define the message at the output of the computational system. In this scenario, the input nodes are no longer well-defined as per Definition~\ref{def:input-nodes}. Instead, we would define \emph{output nodes} in the same manner.\footnote{Note, however, that the corresponding ``opposite'' of Theorem~\ref{thm:info-path} (wherein the places of ``input'' and ``output'' nodes are switched) does not hold in this case. That is, it is \emph{not true} that if at some previous time instant, an ``\emph{input}'' node's outgoing transmissions depend on the message, then there is an information path connecting that input node to the aforementioned output nodes. The reason this fails is that there could be a ``source'' node at an even earlier time instant, which provides information about $M$ to \emph{both} the input node under consideration, and the output nodes, via two separate, diverging paths. Therefore, there may be no path from the input node to the output nodes.}

Consider the computational system shown in Figure~\ref{fig:eg-msg-op}. The system on the right executes the function depicted by the boolean circuit shown on the left. $Y \in \{0, 1\}$ is an external parameter, which is taken to be a fixed constant. When $Y = 1$, the \textsc{and} gate at the top is activated while the \textsc{and} gate at the bottom is deactivated, so the message depends only on $Z_1$. In this case, only the edges shown in blue have $M$-information flow. On the other hand, when $Y = 0$, the opposite happens, and the message $M$ depends only on $Z_2$. Now, only edges shown in orange have $M$-information flow. If $Y$ was not a deterministic external parameter, but a random variable itself, then all edges shown in the figure would have $M$-information flow, since $M$ would depend on all their values.

So, we see that when the message is defined at the output, the ``origin'' of the message may be from within the computation system itself, in the form of one or more intrinsically generated random variables: here, either $Z_1 = W(A_0)$ or $Z_2 = W(C_0)$. The notion of information flow and information paths can thus help us identify where the message originates within the computational system.

Furthermore, just as information paths can change depending upon how the message is defined (as in Section~\ref{sec:eg-fft}), information paths may also change depending on external parameters: inputs such as $Y$ that are fed into the computational system, which are not part of the message. These inputs essentially shape the nature of the computation being performed, and so, naturally, they can affect information paths.


\section{Discussion} 
\label{sec:concl-disc}

This paper presented a theoretical framework for defining and studying information flow of a specific message in a computational system. The core contribution of our paper was a definition for information flow that is concretely grounded in the \emph{computational task} and intimately tied to a \emph{specific message}. This relied on another important contribution: the development of an underlying computational model, which enables the interpretation of statistical analyses. After providing a clearly-defined model for a computational system, we presented several candidate definitions for information flow along with counterexamples and showed that our definition, which is based on positivity of a conditional mutual information expression, satisfies several intuitive properties, whereas other candidate definitions do not. We then examined these properties in detail and showed, in particular, that our definition naturally leads to the existence of ``information paths''. We also discussed how information flow can be inferred through conditional independence testing, and provided an algorithm for recovering the information paths in a given system. Finally, we studied some canonical examples of computational systems from different contexts, and showed that our definition of information flow is intuitive in each case.

We proceed to discuss several important assumptions and simplifications in our model. We also discuss existing literature related to estimation of causal influence in neuroscience, and how our computational system model leads to a significantly different measure of information flow. Similarly, we discuss how our framework is very different from the field of Probabilistic Graphical Models.

\subsection{Neuroscientific Concerns}
\label{sec:assumptions-in-model}

\subsubsection{Observing edges vs.\ nodes}

The observation model stated in Section~\ref{sec:obs-model} makes a crucial assumption, namely, that transmissions on each \emph{edge} can be observed. In neuroscientific experiments, however, we often record activity from single neurons (as in the case of electrophysiological recordings), or aggregate activity from groups of neurons (as with Local Field Potentials measured in Electrocorticography and Electroencephalography). These neurons, or groups of neurons, are considered to be nodes communicating to one another in a network. It may not be known which nodes are connected to which other nodes, let alone the recipient of each transmission at every time instant. This is a marked departure from our assumption that transmissions on edges can be observed. To some extent, it is possible to incorporate a ``node-centric'' model within our computational system by assuming that all nodes broadcast their transmissions. However, that still leaves unanswered the question of which nodes actually ``hear'' another's transmissions. A possible resolution to that question might arise from an understanding of \emph{receiver response}. That is, we consider a revised model in which an edge exists if a receiving neuron \emph{uses} the information transmitted by some neuron at the previous time instant. This issue is beyond the scope of the current work, and will be addressed in subsequent studies.

We also note that, although tools based on Granger causality implicitly assume that nodes are measured and not edges, they do not resolve the issue of which node is ``talking'' to which other node. For example, if two different nodes $A_1$ and $B_1$ communicate the same information to a third node, $C_2$, any regression based analysis will assign a weight of one-half to each of $A_1$ and $B_1$. However, the true function, $f_{C_2}$, may be using only the information coming from $A_1$, or only the information coming from $B_1$, or using the two in some other unequal proportion. Such cases may only be identifiable through an interventional approach.

\subsubsection{Observing memories}

Another important assumption in the observation model is that \emph{memories} of nodes are observed as transmissions on self-edges. If these transmissions are implemented in the form of some internal state at each node, then they might be difficult to observe in practice.\footnote{If every node represents a \emph{group} of neurons, however it may just be that their internal state is represented in the form of \emph{communication between these neurons}. In that case, perhaps observing their internal state is just a matter of having more spatially refined measurements.} It remains to be fully understood whether one can compensate for not observing memories in some manner, e.g., by assuming that the memory of a node is the full history of its transmissions and receptions. While this means that intrinsically generated random variables that are \emph{not} propagated to other nodes will never be observed, it could be argued that such variables could have no impact on the system (save for acting as ``computational noise''). So perhaps it suffices to observe only transmissions between \emph{different} nodes (and not self-edges). Further work is required to understand what ramifications such an assumption has on identifying information flows and information paths.

Conversely, our work may suggest to neuroscientists that inferences about information flow are more reliably obtained if one can measure transmissions on edges in the graph, rather than transmissions of nodes. This may call for newer imaging modalities, or new uses of existing modalities, such as treating axons as targets for invasive recordings, perhaps at nodes of Ranvier. Further, perhaps if one wishes to observe memories, it is important to measure not only spikes, but also membrane voltages (e.g.\ using voltage-sensitive dyes~\cite{Grinvald2004VSDI} or, less directly, through measurements of changes in neurotransmitter concentrations outside a cell~\cite{Watson2006Invivo}).

\subsubsection{Discretization of time}

Yet another implicit assumption in our computational system model is that transmissions occur at discrete points in time. This assumption is justified for synchronous digital circuits used commonly today, or if the computational system of interest is a trained artificial neural network, for instance. However, this is not a perfect model of the brain, because neural spiking (among other processes), does not occur only at multiples of some fundamental unit of time. This issue might be partially mitigated by assuming that neural computation happens at a certain time scale, and by using a sufficiently high sampling rate so that Nyquist-rate-type arguments apply. This may not be possible in certain modalities (e.g.\ Calcium imaging and functional Magnetic Resonance Imaging) that are inherently slow, however, so it would be interesting to understand what inferences we are no longer capable of making. Alternatively, if the sampling rate is too high, it may be useful to consider windows within which to look for $M$-information flow. The exact implications of using such preprocessing methods will also need to be studied in greater detail, in future.

\subsubsection{Message enters at $t=0$}

Another assumption in our framework is that the message enters the system at, and only at, time $t = 0$. This is essential, given the way we have defined input nodes: nodes at time $t = 0$, whose outputs depend on the message (and which have no other shared source of randomness). However, this assumption does not allow for a dynamically evolving stimulus, which is also common in neuroscientific experiments. Suppose we allow the message to enter the system at a later time instant, say at some node $U_t$, for $t > 0$, i.e., $U_t$ may compute a function not just of its inputs, but also of $M$. Then, if we want the information path theorem to continue to hold, we must also add $U_t$ to the set of input nodes.\footnote{We should also expect that any Local Markovity conditions at time $t$ (see Proposition~\ref{prop:local-markov}) that involve the node $U_t$ will no longer hold.} Thus, if we see dependence at some other node $V_{t'}$, at a later time instant $t' > t$, the information paths leading to $V_{t'}$ may arise from the original input nodes \emph{or} from $U_t$, or both. As we might intuitively expect, the more time points we allow the message to enter \emph{at}, the more such information paths we will likely see, making the results of our analysis harder to interpret.

\subsubsection{Experimental design and the message}

An important aspect of our work is that it explicitly incorporates the message, which in neuroscientific experiments is often some information contained in the stimulus. This aids the neuroscientist in designing experiments, for example, in understanding what stimuli will help them make a certain inference about information flow. In particular, one needs to use at least two different stimuli in order to obtain any determination about information flow. While this is implicitly understood in neuroscience, as evidenced by comparisons with baselines, or by the use of permutation tests to scramble stimulus-trial correlations for a null model, our framework provides a more direct method for identifying and interpreting stimulus-related information flow.

\subsection{The Difficulty of Estimation}
\label{sec:difficulty-of-estimation}

A strategy for detecting edges that have $M$-information flow was presented in Section~\ref{sec:estimating-info-flow}. In practice, however, there are several issues associated with employing such a strategy. These are discussed below.

Firstly, we currently assume that observations are noiseless (see Section~\ref{sec:obs-model}, Assumption~\ref{itm:noiseless-assmptn}). It is unclear, exactly, to what extent noisy observations will impact the inference of information flow. In particular, it is worth understanding whether small amounts of observation noise can be tolerated if all edges with $M$-information flow have a sufficiently large ``volume'' of information (i.e., the corresponding mutual or conditional mutual information is sufficiently large). As was described intuitively in Section~\ref{sec:estimating-info-flow}, if the information volume is large, then even under noisy conditions, we might expect the test statistic to clear the threshold, so the presence of $M$-information flow can still be detected consistently. But small volumes of information that aggregate over time---e.g.\ information about $M$ ``trickling'' over time from one node to another---could still pose issues. Such $M$-information flow could go undetected, as has been shown to occur in different contexts~\cite{Andersson2005Testing}, using different measures of flow. It is possible that Derived Information, in particular, is hard to infer in the presence of noise. This could make the task of detecting the presence of a hidden node difficult (consider the case of a ``trickling'' hidden node), as well as that of identifying redundant links.

Secondly, detecting whether each edge at time $t$ has information flow involves checking \emph{all} subsets of $\E_t$. For $N$ nodes and $N^2$ edges, this implies $2^{N^2}$ subsets of edges that need to be searched. This could be seen as being prohibitively difficult for $N^2 \geq 30$, or for $N$ greater than about 5 or 6 nodes. However, in reality, graphs in neuroscience are often known to be edge-sparse~\cite{Bassett2006Small,Achard2006Resilient,Russo2013Brain}. For example, in the brain, a well-established 11-node network is the reward network~\cite{Russo2013Brain}. Most nodes in this network typically have just one incoming and one outgoing connection. The two most important nodes have five incoming edges each, with two and four outgoing edges respectively. Further, it is known which connections are inhibitory and which are excitatory, which could further help with testing for information flow. A fully connected network would have had 121 edges, but the underlying connectivity of the circuit only allows for a total of 17 edges in this network. So in reality, anatomical priors help reduce the number of edges to well within the range of what is computable. Nevertheless, it remains of interest to find methods by which nodes and/or edges can be excluded from the search, and this could be another topic for further research.

Another statistical issue that crops up when attempting to simultaneously perform several conditional independence tests is the problem of \emph{multiple comparisons}~\cite{Shaffer1995Multiple}. Simply put, when performing a large number of independent hypotheses tests, say $N$, at some fixed false alarm rate $\alpha$, on average, we should expect $\alpha N$ of these tests to erroneously reject the null. In the context of information flow, we might wish to set the null hypothesis to be the absence of $M$-information flow on a given edge. Then, to test for $M$-information flow on this edge, we need to perform a large number of conditional independence tests---call this number $N$---at some false alarm rate $\alpha$. These tests are, in fact, \emph{not independent} of one another; nevertheless, very loosely put, if we choose a false alarm rate $\alpha \approx 1/N$, we may find that the probability of \emph{at least one} false alarm is too high. This would make us erroneously infer that this particular edge has $M$-information flow; moreover, since this argument applies to any edge, if $\alpha$ is not chosen conservatively enough, we may erroneously infer that \emph{all} edges have $M$-information flow. This multiple hypothesis testing problem is better posed as a ``Global Null test'' (e.g., see~\cite{Duan2019Interactive}), wherein the global null is the hypothesis that \emph{all} of the conditional independence tests are individually null (i.e., that there is \emph{no} $M$-information flow on the given edge), and the global alternative is the hypothesis that \emph{at least one} of the conditional independence tests is non-null (i.e., that there \emph{is} $M$-information flow on the given edge). As mentioned before, however, the conditional independence tests dictated by Definition~\ref{def:info-flow} are, in general, \emph{dependent} on one another. Furthermore, it might not be easy to describe the manner of dependence, so when choosing methods that control the family-wise error rate, it is essential to choose those that work under arbitrary dependence. A simple example of such a test is the well-known Bonferroni correction, which uses a level $\alpha' = \alpha / N$ for each test (where $\alpha$ is the desired false alarm rate for the overall global null test); but we may find that such methods have insufficient statistical power. A potential solution to this problem might involve combining multiple global null tests in some meaningful way: for example, one could imagine designing a procedure that controls the False Discovery Rate\footnote{These methods control the expected proportion of false discoveries, i.e., the proportion of null hypotheses that are falsely rejected.}~\cite{Benjamini1995Controlling} on the \emph{identification of edges} with $M$-information flow.\footnote{Care is needed when doing this, however, since tests for $M$-information flow on different edges at the same time instant are \emph{also} dependent on one another.} Another approach might be to find ways of directly testing information \emph{paths}, wherein the hypothesis tested would be that a certain $M$-information \emph{path} exists in the system, rather than requiring every \emph{edge} with $M$-information flow be identified first. All of these ideas are potential avenues for future work.

\subsection{The Limitations of Granger Causality and Related Tools}

Mapping directed functional connectivity and information flow in the brain has been a hot topic for several years, as evidenced by the large body of work in this direction~\cite{Friston2011Functional,Friston2013Analysing,Bastos2016Tutorial}. Approaches for statistically mapping functional connectivity often rely on variations of Granger Causality~\cite{Bressler2011Wiener} and, more recently, Directed Information~\cite{Quinn2011Estimating,Quinn2015Directed,Jiao2013Universal}, which we here collectively refer to as ``Granger Causality-based tools''. These approaches lack a systematic framework that ties the statistical analysis to the underlying computation, however, and the interpretations drawn from their use have often been questioned~\cite{David2008Identifying,David2011fMRI,Stokes2017Study,Venkatesh2015Direction,Andersson2005Testing,Nalatore2007Mitigating,Gong2015Discovering}.

In particular, a crucial difference between our approach and that of Granger Causality-based tools is that the latter do not have an explicit description of the message. Instead, they provide mechanisms to condense a pair of time series into a single statistic. There are no concrete models that can be used to interpret what this statistic means for the flow of \emph{information} about the message. Furthermore, if one is interested in the information flow of multiple messages, Granger Causality-based tools do not provide an immediate solution. This is why a tool that ties information flow directly with a message is of great interest to practitioners.

The absence of an underlying computational framework with well-defined assumptions inherently makes it very hard to draw sound inferences through the application of Granger Causality-based tools. A striking example of this is a recent result of ours~\cite{Venkatesh2015Direction} that shows, using a feedback communication system, that the direction of greater Granger-causal influence can be opposite to the direction in which the message is communicated, even in the absence of hidden nodes and measurement noise. The time-unrolled graph framework presented here has been specifically designed to address this issue, and present a clear understanding of information flow, even in the presence of feedback. The example given in Section~\ref{sec:eg-sk} demonstrates a potential resolution to this issue.

Granger Causality was originally developed for the study of time-series that occur only once, such as in economics~\cite{Granger1969Investigating}. An artifact of this development is that it was not designed to incorporate multiple trials of the same process. Instead, it assumes stationarity to help estimate parameters of the random variables that control the process. In the neuroscientific context, stationarity is often a very poor assumption, since the segment of time-series data corresponding to each trial may be short, and often sees some kind of stimulus presentation. Naturally, presentation of the stimulus changes the underlying parameters of the time-series and destroys stationarity; indeed, this is the quintessential aspect of the experiment. Thus, in order to understand processing in such stimulus-driven tasks, one needs to be able to infer time-dependent information flows from data. While information-theoretic extensions of Granger Causality such as Transfer Entropy and Directed Information do not assume stationarity, they nevertheless fail to provide a dynamically evolving picture of information flow.

Lastly, it is unclear whether the directional influences estimated using Granger Causality-based tools have any correspondence with the rigorous notion of information flow we have derived here, under special assumptions, e.g., Gaussianity. This is a promising future direction as well, since it is important to understand in which situations these methods recover meaningful flows of information, and in which cases we must be careful with interpretation.

\subsection{Probabilistic Graphical Models and Pearl's Causality}

There is one important difference that distinguishes our work from the perspective adopted in the field of probabilistic graphical models (PGMs)~\cite{Koller2009Probabilistic}, and the representations therein. In our framework, nodes represent computational units, whereas in PGMs, nodes represent the random variables themselves, and edges capture the conditional independence relationships between these variables. While it might be possible to construct a PGM that is equivalent to our computational model, this would likely eliminate any intuitive structure captured by the computational graph.

It remains to be understood whether and how Pearl's notions of causality~\cite{Pearl2009Causality} can be seamlessly merged with the understanding of information flow developed here. We expect that some formal application of causality will be needed in going from an edge-centric model (as presented here) to a more node-centric one (discussed in Section~\ref{sec:assumptions-in-model}), in order to identify which transmissions influenced a given node's output.

There are several works in the literature that discuss measures of information flow in probabilistic graphical models~\cite{Ay2008Information,Janzing2013Quantifying}, but they are heavily inspired by causality and largely center around an interventionist approach. In contrast, our definition of information flow is based on a computational system model that translates more readily to neuroscience, and we assume that the experimentalist is restricted to making observations.

\subsection{Future Directions for Theoretical Development}
\label{sec:info-flow-volume}

A natural question that arises from this paper is: how can our definition of information flow on an edge be extended to a more generic information \emph{measure}, which also quantifies the volume of flow? Finding such a measure will involve aggregating the conditional mutual information for each subset of edges into a single value (one example of such a measure was provided in Section~\ref{sec:eg-sk}, though it was not developed from first-principles). It is as yet unclear how this might be achieved, while still gelling well with our intuition of what this information flow volume ought to be. We believe that the right approach is to start by designating a set of properties that we would like information flow volumes to satisfy, and then to propose a measure through the use of representative examples and counterexamples.

A second direction that emerges is related to Partial Information Decomposition (PID)~\cite{Williams2010Nonnegative,Harder2013Bivariate,Bertschinger2014Quantifying}, which was discussed earlier in Section~\ref{sec:syn-info-connection}. $M$-information flow is very closely related to the PID: while Candidate Definition~\ref{cd:dependence} checks for positivity of mutual information between $M$ and $X(E_t)$, and hence implying the presence of unique and/or redundant information, our definition also detects the presence of purely synergistic information. Since our definition is closely tied to computation and is strongly motivated through the goal of finding unbroken information paths, the close relationship between PID and our definition suggests that PID might be the right toolset for obtaining a more fine-grained understanding of information flow, as well as computation. In particular, it would be useful to know how the understanding of computation is enhanced through a PID analysis, which describes the unique, redundant and synergistic components of the message in different nodes' transmissions. Finally, we note that the PID could also help inform the discussion on a definition for information volume. Providing a useful definition of information volume based on current definitions of unique, redundant and synergistic information, and asking whether the problem of information flow can inform the PID literature, will also be the subject of future research.

A third direction has to do with alternate definitions of information flow: the properties we stated in this paper are not sufficient to uniquely specify our definition of information flow. For example, the all-zero function as well as the all-ones function satisfy the Broken Telephone property, although they are not particularly useful definitions of information flow. Thus, it would be useful to understand what other properties we should impose so as to arrive at a unique definition of information flow. As a crude and preliminary example, we demonstrate how this might be done in Appendix~\ref{app:uniqueness}.

\subsection{Concluding Remarks}

We conclude by describing some of our general impressions in working on the theoretical development presented in this paper. As such, these points merely highlight some of our opinions on how theory---and more specifically, information theory---may be applied in neuroscience.

As mentioned in the introduction, we drew inspiration from two papers that discuss how experimentalists understand systems in biology and neuroscience~\cite{Lazebnik2002Can,Jonas2017Could}. Both these works advocate for theory by arguing that we need new analytical tools, and that the accumulation of empirical knowledge alone does not constitute \emph{understanding}. Lazebnik~\cite{Lazebnik2002Can}, in particular, mentions how terminology in biology tends to be vague and non-committal. We feel that an important reason for this is the absence of concrete underlying models, with clearly-stated assumptions. In other words, we think that theory and modeling can go a long way in providing a \emph{language} that will enable well-grounded discussions. This language, in turn, arises through the development of theoretical models and formal definitions.

Another point made by both the aforementioned papers is that we should attempt to understand large computational systems by first examining smaller models, and models in which the ground truth is already known. This approach allows us to create new analytical tools that can be thoroughly vetted, so that the interpretations drawn from their use in experimental practice is unambiguous and undebated. We also believe that when trying to understand large computational systems, it is essential to start with toy models such as Counterexample~\ref{ce:dependence}. This philosophy of starting with toy models, and abstracting out meaningful ideas that hold more generally in large systems, is well-entrenched in the field of information theory, and can become a useful export in fields such as neuroscience.


\section*{Acknowledgments} 

We have many people to thank for extremely useful discussions. A non-exhaustive list follows: Mayank Bakshi, Marlene Behrmann, Todd Coleman, Uday Jagadisan, Haewon Jeong, Rob Kass, Gabe Schamburg, Tsachy Weissman. We also thank the anonymous reviewers whose comments improved our exposition substantially.

Praveen Venkatesh was supported, in part, by a Fellowship in Digital Health from the Center for Machine Learning and Health at Carnegie Mellon University. Pulkit Grover was supported, in part, by an NSF CAREER Award.




\ifthenelse{\boolean{arxiv}}{
	\bibliographystyle{unsrt}  
	\bibliography{references}
}{
	\bibliographystyle{IEEEtran}
	\bibliography{IEEEabrv,references}
}



\appendices 

\section{Proof of Proposition~\ref{prop:set-info-equiv}}
\label{app:set-info-equiv}

\begin{IEEEproof}[Proof of Proposition~\ref{prop:set-info-equiv}]
	($\Rightarrow$) Suppose there exists some $E_t' \in \E_t'$ that has $M$-information flow. That is,
	\begin{equation} \label{eq:et-prime-flow}
		\exists\;\E_t'' \subseteq \E_t\setminus\{E_t'\} \quad \text{s.t.} \quad I\bigl(M ; X(E_t') \given X(\E_t'')\bigr) > 0.
	\end{equation}
	Then,
	\begin{align}
		I\bigl(M ; X(\E_t') \given X(\E_t'')\bigr) &= I\bigl(M ; X(E_t') \given X(\E_t'')\bigr) + I\bigl(M ; X(\E_t'\!\setminus\!\{E_t'\}) \given X(\E_t''), X(E_t')\bigr) \\
		&\overset{(a)}{\geq} I\bigl(M ; X(E_t') \given X(\E_t'')\bigr) \overset{(b)}{>} 0
	\end{align}
	where (a) follows from the non-negativity of conditional mutual information and (b) from~\eqref{eq:et-prime-flow}. Taking $\mathcal R_t' \coloneqq \E_t''$ in Definition~\ref{def:info-flow-set}, we see that set $\E_t'$ has $M$-information flow.

	($\Leftarrow$) Next, suppose that the set $\E_t'$ has $M$-information flow, as per Definition~\ref{def:info-flow-set}. That is, there exists a set $\mathcal R_t' \subseteq \E_t$ such that
	\begin{equation}
		I\bigl(M ; X(\E_t') \given X(\mathcal R_t')\bigr) > 0.
	\end{equation}
	Also, let $\{E_t^{(1)}, E_t^{(2)}, \ldots E_t^{(K)}\}$ be any ordering of the nodes in $\E_t'$ (where $K = \abs{\E_t'}$). Then by the chain rule of mutual information,
	\begin{align}
		0 &< I\bigl(M ; X(\E_t') \given X(\mathcal R_t')\bigr) \\
		&= \sum_{k=1}^K I\biggl(M ; X(E_t^{(k)}) \,\Big\vert\, X(\mathcal R_t'), X\Bigl(\bigcup_{j=1}^{k-1} \{E_t^{(j)}\}\Bigr)\biggr).
	\end{align}
	By the non-negativity of conditional mutual information, at least one of the terms in the summation must be strictly positive. Let the index of this term be $k^*$. Hence, there exists $E_t' \coloneqq E_t^{(k^*)}$ and $\E_t'' \coloneqq \mathcal R_t' \cup \{E_t^{(1)}, \ldots E_t^{(k^*-1)}\}$, such that
	\begin{equation}
		I\bigl(M ; X(E_t') \given X(\E_t'')\bigr) > 0.
	\end{equation}
	In other words, there exists an edge $E_t' \in \E_t'$ that has $M$-information flow.
\end{IEEEproof}

\section{Proof of Proposition~\ref{prop:separation}}
\label{app:separability-proof}

\begin{IEEEproof}[Proof of Proposition~\ref{prop:separation}]
	Consider the set of all $E_t \in \E_t$ that have $M$-information flow. That is, $E_t$ must satisfy
	\begin{equation} \label{eq:edge-info}
		\exists\; \E_t' \subseteq \E_t \quad \text{s.t.} \quad I\bigl(M ; X(E_t) \given X(\E_t')\bigr) > 0.
	\end{equation}
	Define
	\begin{equation} \label{eq:Rt-St-defn}
		\begin{gathered}
			\R_t \coloneqq \{E_t \in \E_t : \text{\eqref{eq:edge-info} holds}\}, \\
			\S_t \coloneqq \E_t \setminus \R_t.
		\end{gathered}
	\end{equation}
	Then, we claim that $\R_t$ and $\S_t$ satisfy equations \eqref{eq:Rt-cond} and \eqref{eq:St-cond}.

	First, note that if $\S_t \neq \emptyset$, then for every $S_t \in \S_t$, we must have that
	\begin{equation} \label{eq:single-St-cond}
		\forall\; \E_t' \subseteq \E_t, \quad I\bigl(M ; X(S_t) \given X(\E_t')\bigr) = 0.
	\end{equation}
	If not, then $S_t \in \R_t$ by~\eqref{eq:Rt-St-defn}, which implies that $S_t \notin \S_t$, which is a contradiction. Hence, we see that no edge in $\S_t$ has $M$-information flow. Therefore, by Proposition~\ref{prop:set-info-equiv}, the set $\S_t$ has no $M$-information flow. This directly implies the condition in~\eqref{eq:St-cond}.

	Next, we claim that if $\R_t \neq \emptyset$, then for every $R_t \in \R_t$, if $\E_t' \subseteq \E_t$ is a set that satisfies
	\begin{equation} \label{eq:Etprime-satisfies}
		I\bigl(M ; X(R_t) \given X(\E_t')\bigr) > 0,
	\end{equation}
	then $\R_t' \coloneqq \E_t' \cap \R_t$ satisfies
	\begin{equation}
		I\bigl(M ; X(R_t) \given X(\R_t')\bigr) > 0.
	\end{equation}
	Let $\S_t' \coloneqq \E_t' \setminus \R_t'$, so that $\S_t' \subseteq \S_t$. Then,
	\begin{equation} \label{eq:Rtprime-Stprime-neq-0}
		I\bigl(M ; X(R_t) \given X(\R_t'), X(\S_t')\bigr) > 0
	\end{equation}
	by~\eqref{eq:Etprime-satisfies}. So,
	\begin{align}
		I\bigl(M ; X(R_t) \given X(\R_t')\bigr) &\overset{(a)}{=} I\bigl(M ; X(R_t), X(\S_t') \given X(\R_t')\bigr) - I\bigl(M ; X(\S_t') \given X(\R_t'), X(R_t)\bigr) \\
		&\overset{(b)}{=} I\bigl(M ; X(R_t), X(\S_t') \given X(\R_t')\bigr) \\
		&\overset{(c)}{=} I\bigl(M ; X(R_t) \given X(\R_t'), X(\S_t')\bigr) + I\bigl(M ; X(\S_t') \given X(\R_t')\bigr) \\
		&\overset{(d)}{=} I\bigl(M ; X(R_t) \given X(\R_t'), X(\S_t')\bigr) \\
		&\overset{(e)}{>} 0, \nonumber
	\end{align}
	where (a) and (c) follow from the chain rule, (b) and (d) follow from~\eqref{eq:St-cond}, and (e) follows from~\eqref{eq:Rtprime-Stprime-neq-0}. Thus, condition~\eqref{eq:Rt-cond} also holds.
\end{IEEEproof}


\section{Synergistic Information Flow}
\label{app:synergistic-info-flow}

\subsection{Partial Information Decomposition preliminaries}
\label{app:pid-preliminaries}

The literature on Partial Information Decomposition seeks to find a decomposition for the mutual information between a message, $M$, and a set of random variables, $\{X_1, X_2, \ldots \}$ into several individually meaningful, non-negative terms~\cite{Lizier2018Information}. For our purposes, it suffices to consider the \emph{bivariate} case, i.e., the decomposition of $I(M ; X, Y)$ into non-negative components. In the bivariate case, it is well-understood \emph{how many} components there ought to be, and what these quantities \emph{intuitively represent}, but as yet, there is no consensus on a single set of definitions~\cite{Lizier2018Information}.

There is, however, consensus on a basic set of properties that we expect these components to satisfy. For our purposes, we will only make use of the basic properties stated here, so that \emph{any definition} of the aforementioned components which satisfies these properties suffices for our theory.

In the bivariate case, the mutual information between $M$ and $(X, Y)$ is decomposed into four components: information about $M$ which is \begin{enumerate*}[label=(\roman*)]
	\item unique to $X$ and not present in $Y$,
	\item unique to $Y$ and not present in $X$,
	\item redundantly present in both $X$ and $Y$, and
	\item synergistically present in $X$ and $Y$.
\end{enumerate*}
In the notation of~\cite{Bertschinger2014Quantifying}, the decomposition is written as:
\begin{equation} \label{eq:pid-decomposition}
	I\bigl(M ; (X, Y)\bigr) = UI(M : X \setminus Y) + UI(M : Y \setminus X) + SI(M : X ; Y) + CI(M : X ; Y),
\end{equation}
where the components are ordered exactly as stated above. Note that $SI$ refers to ``shared'', and hence redundant, information, while $CI$ refers to ``complementary'', and hence synergistic, information. We shall continue to use the terms ``redundant'' and ``synergistic'', however, since they are more meaningful in this context. Also, in what follows, we shall assume that $SI$ and $CI$ are symmetric in $X$ and $Y$. This is usually an additional condition that is imposed when defining these quantities, but here, we take it as given.

Given what we want the four components to represent, we would also expect the following to hold:
\begin{equation} \label{eq:pid-mi-expansion}
	\begin{aligned}
		I(M ; X) &= UI(M : X \setminus Y) + SI(M : X ; Y), \\
		I(M ; Y) &= UI(M : Y \setminus X) + SI(M : X ; Y).
	\end{aligned}
\end{equation}
As a natural consequence, this means that the conditional mutual information will satisfy:
\begin{equation} \label{eq:pid-cmi-expansion}
	\begin{aligned}
		I(M ; X \,\vert\, Y) &= I\bigl(M ; (X, Y)\bigr) - I(M ; Y) \\
		&= UI(M : X \setminus Y) + CI(M : X ; Y), \\
		I(M ; Y \,\vert\, X) &= I\bigl(M ; (Y, X)\bigr) - I(M ; X) \\
		&= UI(M : Y \setminus X) + CI(M : X ; Y).
	\end{aligned}
\end{equation}
Finally, we want each of these components to always be non-negative:
\begin{equation} \label{eq:pid-gt-zero}
	\begin{aligned}
		UI(M : X \setminus Y) &\geq 0 \quad & \quad SI(M : X ; Y) &\geq 0 \\
		UI(M : Y \setminus X) &\geq 0 \quad & \quad CI(M : X ; Y) &\geq 0.
	\end{aligned}
\end{equation}
It is not obvious that a consistent definition of these four quantities which also satisfies the equations stated above even \emph{exists}, but in fact, additional properties are required to obtain a unique definition. For instance, see~\cite{Bertschinger2014Quantifying} for one such development.

As stated before, our theory only relies on the properties stated in this section. As a result, our theorem on the equivalence of information flow definitions holds irrespective of what definition is used, exactly, for synergistic information. It only matters that the definition used satisfies the basic properties presented here.

\subsection{Equivalence of information flow definitions}

\begin{IEEEproof}[Proof of Proposition~\ref{prop:syn-info-equiv}]
	($\Rightarrow$) Suppose the edge $E_t$ has strictly positive $M$-information flow. Then,
	\begin{equation} \label{eq:defn-equiv-case-1}
		\exists\;\E_t'\subseteq\E_t \quad \text{s.t.} \quad I\bigl(M ; X(E_t) \given X(\E_t')\bigr) > 0.
	\end{equation}
	If $I\bigl(M ; X(E_t)\bigr) > 0$ with $\E_t' = \emptyset$ in~\eqref{eq:defn-equiv-case-1}, then condition~\ref{cond:syn-info-cond-1} in Definition~\ref{def:syn-info-flow} holds, so nothing remains to be shown. If not, then $I\bigl(M ; X(E_t)\bigr) = 0$, so~\eqref{eq:defn-equiv-case-1} implies that there must exist some $\E_t' \neq \emptyset$ such that
	\begin{gather}
		I\bigl(M ; X(E_t) \given X(\E_t')\bigr) > 0,
		\shortintertext{which, by~\eqref{eq:pid-cmi-expansion}, is equivalent to}
		UI\bigl(M : X(E_t) \setminus X(\E_t')\bigr) + CI\bigl(M : X(E_t) ; X(\E_t')\bigr) > 0.
	\end{gather}
	However, since $I\bigl(M ; X(E_t')\bigr) = 0$, we must have $UI\bigl(M : X(E_t) \setminus X(\E_t')\bigr) = 0$ by~\eqref{eq:pid-mi-expansion} and~\eqref{eq:pid-gt-zero}. Hence,
	\begin{equation}
		\exists\;\E_t' \subseteq \E_t\setminus\{E_t\} \quad \text{s.t.} \quad CI\bigl(M : X(E_t) ; X(\E_t')\bigr) > 0.
	\end{equation}
	So the implication in the forward direction holds.

	($\Leftarrow$) For the converse, suppose that $E_t$ has no $M$-information flow. That is,
	\begin{gather}
		I\bigl(M ; X(E_t) \given X(\E_t')\bigr) = 0 \quad \forall\;\E_t' \subseteq \E_t\setminus\{\E_t\}.
		\shortintertext{By~\eqref{eq:pid-cmi-expansion}, this implies that}
		UI\bigl(M : X(E_t) \setminus X(\E_t')\bigr) + CI\bigl(M : X(E_t) ; X(\E_t')\bigr) = 0 \quad \forall\;\E_t' \subseteq \E_t\setminus\{\E_t\}.
	\end{gather}
	Since $UI$ and $CI$ are both non-negative by~\eqref{eq:pid-gt-zero}, we must have that
	\begin{equation}
		CI\bigl(M : X(E_t) ; X(\E_t')\bigr) = 0 \quad \forall\;\E_t' \subseteq \E_t\setminus\{\E_t\}.
	\end{equation}
	This proves the converse.
\end{IEEEproof}

\section{Miscellaneous Proofs from Section~\ref{sec:inferring-info-flow}}
\label{app:hidden-node-proofs}

\subsection{Proof of Lemma~\ref{lem:hidden-rel-derived}}

\begin{IEEEproof}[Proof of Lemma~\ref{lem:hidden-rel-derived}]
	Consider a subset of hidden nodes $\H_t' \subseteq \H_t$ that is not $M$-relevant. Then, by Definition~\ref{def:relevant-hidden-node}, $\Q(\H_t')$ carries no $M$-information flow in $\G$. This means that
	\begin{equation}
		\forall \; \E_t' \subseteq \E_t, \quad I\bigl(M ; X(\Q(\H_t')) \given X(\E_t')\bigr) = 0.
	\end{equation}
	Specifically, taking $\E_t' = \widetilde\E_t$, we have
	\begin{equation}
		I\bigl(M ; X(\Q(\H_t')) \given X(\widetilde\E_t)\bigr) = 0.
	\end{equation}
	Therefore, by Definition~\ref{def:derived-hidden-node}, $\H_t'$ is $M$-derived. Thus, if $\H_t'$ is \emph{not} $M$-relevant, it \emph{is} $M$-derived. Taking the contrapositive, if $\H_t$ is \emph{not} $M$-derived, then it \emph{is} $M$-relevant.
\end{IEEEproof}

\subsection{Proof of proposition~\ref{prop:hidden-markov}}

\begin{IEEEproof}[Proof of Proposition~\ref{prop:hidden-markov}]
	We are given that
	\begin{equation} \label{eq:hidden-markov-proof-1}
		I\bigl(M ; X(\widetilde\E_{t+1}) \given X(\widetilde\E_t)\bigr) > 0,
	\end{equation}
	and must prove that the hidden nodes at time $t$, $\H_t$, are \emph{not} $M$-derived.

	First note that, since $\Q(\widetilde\V_{t+1}) = \widetilde\E_{t+1} \cup (\widetilde\V_{t+1} \times \H_{t+2})$, we must have
	\begin{align}
		I\bigl(M ; X(\Q(\widetilde\V_{t+1})) \given X(\widetilde\E_t)\bigr) &= I\bigl(M ; X(\widetilde\E_{t+1}), X(\widetilde\V_{t+1} \times \H_{t+2}) \given X(\widetilde\E_t)\bigr) \label{eq:hidden-markov-proof-3} \\
																			&= I\bigl(M ; X(\widetilde\E_{t+1}) \given X(\widetilde\E_t)\bigr) + I\bigl(M ; X(\widetilde\V_{t+1} \times \H_{t+2}) \given X(\widetilde\E_{t+1}), X(\widetilde\E_t)\bigr) \\
																			&\geq I\bigl(M ; X(\widetilde\E_{t+1}) \given X(\widetilde\E_t)\bigr) \\
																			&> 0, \label{eq:hidden-markov-proof-2}
	\end{align}
	where the last line follows from the fact that conditional mutual information is non-negative, and from~\eqref{eq:hidden-markov-proof-1}.

	Next, observe that Local Markovity conditions (Proposition~\ref{prop:local-markov}) \emph{must} hold on the \emph{entire} graph $\G$, which consists of both observed and hidden nodes. If we apply the Local Markovity condition to $\widetilde\V_{t+1}$, we have $M$---$X(\P(\widetilde\V_{t+1}))$---$X(\Q(\widetilde\V_{t+1}))$, or in other words
	\begin{equation}
		I\bigl(M ; X(\Q(\widetilde\V_{t+1})) \given X(\P(\widetilde\V_{t+1}))\bigr) = 0.
	\end{equation}
	Note that $\P(\widetilde\V_{t+1}) = \widetilde\E_t \cup \widetilde\Q(\H_t)$, where $\widetilde\Q(\H_t) \coloneqq \H_t \times \widetilde\V_{t+1}$ is the subset comprising outgoing edges of $\H_t$ that go to $\widetilde\V_{t+1}$. Therefore,
	\begin{equation}
		I\bigl(M ; X(\Q(\widetilde\V_{t+1})) \given X(\widetilde\E_t), X(\widetilde\Q(\H_t))\bigr) = 0.
	\end{equation}
	Expanding this conditional mutual information, we get
	\begin{equation}
		I\bigl(M ; X(\Q(\widetilde\V_{t+1})), X(\widetilde\Q(\H_t)) \given X(\widetilde\E_t)\bigr) - I\bigl(M ; X(\widetilde\Q(\H_t)) \given X(\widetilde\E_t)\bigr) = 0.
	\end{equation}
	So we have
	\begin{align}
		I\bigl(M ; X(\widetilde\Q(\H_t)) \given X(\widetilde\E_t)\bigr) &= I\bigl(M ; X(\Q(\widetilde\V_{t+1})), X(\widetilde\Q(\H_t)) \given X(\widetilde\E_t)\bigr) \\
																		&= I\bigl(M ; X(\Q(\widetilde\V_{t+1})) \given X(\widetilde\E_t)\bigr) + I\bigl(M ; X(\widetilde\Q(\H_t)) \given X(\Q(\widetilde\V_{t+1})), X(\widetilde\E_t)\bigr) > 0,
	\end{align}
	where the final inequality follows from~\eqref{eq:hidden-markov-proof-2} and the fact that conditional mutual information is non-negative. Finally, since $\widetilde\Q(\H_t) \subset \Q(\H_t)$, we have that $I\bigl(M ; X(\Q(\H_t)) \given X(\widetilde\E_t)\bigr) > 0$, just as we in equations \eqref{eq:hidden-markov-proof-3}--\eqref{eq:hidden-markov-proof-2}. Hence, the Markov chain $M$---$X(\tilde\E_t)$---$X(\Q(\H_t))$ does not hold, so by Definition~\ref{def:derived-hidden-node}, $\H_t$ are not $M$-derived.
\end{IEEEproof}


\section{On the Uniqueness of Our Definition of Information Flow}
\label{app:uniqueness}

From the perspective of designing an axiomatic framework, it is desirable to find a minimal set of properties that gives rise to a unique definition of information flow. Although Property~\ref{ppty:broken-telephone} helped us motivate a definition for information flow, it did not uniquely specify a definition. Indeed, the all-zero function as well as the all-ones function also satisfy the property, although they are not particularly useful definitions of information flow.

In this section, we provide a set of properties that uniquely leads to our definition of information flow. However, we must acknowledge that we arrived at these properties with the benefit of hindsight, after having proved many other properties of our definition. As such, they are mathematically very similar to our definition, and one might feel uncomfortable with the idea of imposing such a set of properties at the very outset. Our goal here is only to begin a discussion in this direction: a search for a more abstract set of properties that leads to a unique definition of information flow would be a worthy endeavour in future.
\begin{property} \label{ppty:uniqueness}
	Let $\C$ be a computational system, and let $\F_M : \E \to \{0, 1\}$ be an indicator of the presence of information flow about $M$ on an edge. That is, $\F_M(E) = 1$, if information about $M$ flows on the edge $E \in \E$, and $\F_M(E) = 0$ otherwise. We now state three conditions $\F_M$ must satisfy, which naturally leads to our definition of information flow (Definition~\ref{def:info-flow}):
	\begin{enumerate}[label=\textit{\theproperty\alph*)},ref=\theproperty\alph*]
		\item $I\bigl(M ; X(E_t)\bigr) > 0 \quad \Rightarrow \quad \F_M(E_t) = 1$, \label{ppty:dependence}
		\item $\exists\; \E_t' \subseteq \E_t\!\setminus\!\{E_t\} \;\text{ s.t. }\; I\bigl(M ; X(\E_t') \given X(E_t)\bigr) > I\bigl(M ; X(\E_t')\bigr) \quad \Rightarrow \quad \F_M(E_t) = 1$, \label{ppty:cond-inc-info}
		\item $I\bigl(M ; X(E_t) \given X(\E_t')\bigr) = 0 \;\forall\; \E_t' \subseteq \E_t \quad \Rightarrow \quad \F_M(E_t) = 0$. \label{ppty:total-indep}
	\end{enumerate}
\end{property}

Property~\ref{ppty:dependence} is a very natural and intuitive requirement for information flow. Property~\ref{ppty:cond-inc-info} states that an edge should be considered to carry information about $M$, if upon conditioning, its transmission \emph{increases} the information that some set $X(\E_t')$ conveys about $M$. Property~\ref{ppty:total-indep} is reminiscent of the separability property from Proposition~\ref{prop:separation}, and states that if an edge has no dependence with $M$, no matter what other transmission is conditioned upon, then it can carry no information flow about $M$.

Effectively, Property~\ref{ppty:dependence} states that if an edge has unique or redundant information about $M$, then it must carry information flow, while Property~\ref{ppty:cond-inc-info} states that if an edge has synergistic information about $M$ along with some other set of transmissions, then it must carry information flow. Finally, Property~\ref{ppty:total-indep} states that if all three of these components are absent, then that edge carries no information flow. This also explains how, if any one of these three properties is absent, our definition is no longer unique.

As we acknowledged previously, some of these properties could be seen as too restrictive or contrived, and a more abstract set of properties is certainly desirable. Nevertheless, these properties do uniquely identify our definition of information flow.
\begin{proposition}[Uniqueness] \label{prop:uniqueness}
	If $\F_M$ is an indicator of information flow that satisfies the conditions in Property~\ref{ppty:uniqueness}, then $\F_M(E_t) = 1$ if and only if $E_t$ has $M$-information flow, per Definition~\ref{def:info-flow}.
\end{proposition}
\begin{IEEEproof}
	($\Rightarrow$) Suppose the edge $E_t$ has no $M$-information flow per Definition~\ref{def:info-flow}. This directly implies the condition in Property~\ref{ppty:total-indep}. Hence, $\F_M(E_t) = 0$. This proves that if $\F_M(E_t) = 1$, the edge $E_t$ must have $M$-information flow.

	($\Leftarrow$) Suppose the edge $E_t$ \emph{has} $M$-information flow per Definition~\ref{def:info-flow}. Then,
	\begin{equation} \label{eq:uniqueness-proof-1}
		\exists\; \E_t' \subseteq \E_t\!\setminus\!\{E_t\} \quad\text{s.t}\quad I\bigl(M ; X(E_t) \given X(\E_t')\bigr) > 0.
	\end{equation}
	If $\E_t' = \emptyset$, $I\bigl(M ; X(E_t)\bigr) > 0$, so by Property~\ref{ppty:dependence}, $\F_M(E_t) = 1$. If $I\bigl(M ; X(E_t)\bigr) = 0$, then~\eqref{eq:uniqueness-proof-1} guarantees the existence of some $\E_t' \neq \emptyset$ such that
	\begin{alignat}{2}
		&& I\bigl(M ; X(E_t) \given X(\E_t')\bigr) &> 0 \\
		&\Rightarrow\qquad & I\bigl(M ; X(\E_t')\bigr) + I\bigl(M ; X(E_t) \given X(\E_t')\bigr) &\overset{(a)}{>} I\bigl(M ; X(\E_t')\bigr) \\
		&\Rightarrow\qquad & I\bigl(M ; X(E_t), X(\E_t')\bigr) &\overset{(b)}{>} I\bigl(M ; X(\E_t')\bigr) \\
		&\Rightarrow\qquad & I\bigl(M ; X(E_t)\bigr) + I\bigl(M ; X(\E_t') \given X(E_t)\bigr) &\overset{(c)}{>} I\bigl(M ; X(\E_t')\bigr) \\
		&\Rightarrow\qquad & I\bigl(M ; X(\E_t') \given X(E_t)\bigr) &\overset{(d)}{>} I\bigl(M ; X(\E_t')\bigr),
	\end{alignat}
	where in (a), we simply added $I\bigl(M ; X(\E_t')\bigr)$ to both sides; in (b) and (c), we used the chain rule in two different ways; and in (d), we used the fact that $I\bigl(M ; X(E_t)\bigr) = 0$. So, by Property~\ref{ppty:cond-inc-info}, we have that $\F_M(E_t) = 1$. This proves the converse.
\end{IEEEproof}

\paragraph*{Remark} It should be noted that Definition~\ref{def:info-flow} only specifies \emph{whether or not} a given edge has $M$-information flow. It does not \emph{quantify} this flow. So Proposition~\ref{prop:uniqueness} demonstrates the uniqueness of our definition up to an unspecified information volume. If we require that the conditions in Property~\ref{ppty:uniqueness} hold, then any quantitative definition of information flow will go to zero at an edge if and only if the $M$-information flow carried by that edge is zero.


\section{Miscellaneous Derivations from Section~\ref{sec:canonical-examples}}

\subsection{Derivation of Expressions in the Second FFT Example from Section~\ref{sec:eg-fft}}
\label{app:fft-eg-2-derivation}

Here, we derive the expressions used in Figure~\ref{fig:fft-eg-2}. Recall that $Y_i = \omega^{-iM} / 4$, where $\omega = e^{-j2\pi/4} = -j$.
\begin{align}
	Y_{02} &= Y_0 + Y_2 = \frac{1}{4} + \frac{\omega^{-2M}}{4} = \frac{1}{4}(1 + \omega^{-2M}) \\
	Y_{13} &= Y_1 + Y_3 = \frac{\omega^{-M}}{4} + \frac{\omega^{-3M}}{4} = \frac{\omega^{-M}}{4}(1 + \omega^{-2M}) \\
	Y_{02}' &= Y_0 + \omega^2 Y_2 = \frac{1}{4} + (-1) \frac{\omega^{-2M}}{4} = \frac{1}{4}(1 - \omega^{-2M}) \\
	Y_{13}' &= Y_1 + \omega^2 Y_3 = \frac{\omega^{-M}}{4} + (-1) \frac{\omega^{-3M}}{4} = \frac{\omega^{-M}}{4}(1 - \omega^{-2M})
\end{align}
Next, we show that these intermediate values actually yield the expected values of $\widetilde Y$.
\begin{align}
	\widetilde Y_0 &= Y_{02} + Y_{13} = \frac{1}{4}(1 + \omega^{-2M} + \omega^{-M} + \omega^{-3M}) \\
	&= \begin{cases}
		\frac{1}{4}(1 + 1 + 1 + 1), & M = 0 \\
		\frac{1}{4}(1 + j + j^2 + j^3), & M = 1
	\end{cases} \\
	&= 1 - M \\
	\widetilde Y_1 &= Y_{02}' + \omega Y_{13}' = \frac{1}{4}(1 - \omega^{-2M} + \omega^{1 - M} - \omega^{1 - 3M}) \\
	&= \begin{cases}
		\frac{1}{4}(1 - 1 + \omega - \omega), & M = 0 \\
		\frac{1}{4}(1 - j^2 + 1 - j^2), & M = 1
	\end{cases} \\
	&= M \\
	\widetilde Y_2 &= Y_{02} + \omega^2 Y_{13} = \frac{1}{4}(1 + \omega^{-2M} + \omega^{2-M} + \omega^{2-3M}) \\
	&= \frac{1}{4}(1 + \omega^{-2M} - \omega^{-M} - \omega^{-3M}) \\
	&= \begin{cases}
		\frac{1}{4}(1 + 1 - 1 - 1), & M = 0 \\
		\frac{1}{4}(1 - 1 - \omega^{-1} + \omega^{-1}), & M = 1
	\end{cases} \\
	&= 0 \\
	\widetilde Y_3 &= Y_{02}' + \omega^3 Y_{13}'= \frac{1}{4}(1 - \omega^{-2M} + \omega^{3-M} - \omega^{3-3M}) \\
	&= \frac{1}{4}(1 - \omega^{-2M} + \omega^3(\omega^{-M} - \omega^{-3M})) \\
	&= \begin{cases}
		\frac{1}{4}(1 - 1 - \omega(1 - 1)), & M = 0 \\
		\frac{1}{4}(1 - (-1) - \omega(\omega^{-1} + \omega^{-1})), & M = 1
	\end{cases} \\
	&= 0
\end{align}

\subsection{Derivation of the Markov Chain Failure in Section~\ref{sec:eg-sk}}
\label{app:sk-markov-proof}

We wish to show that in the canonical example from Section~\ref{sec:eg-sk}, $M$---$[\what M_1, \; \what M_2]$---$(M - \what M_2)$ is not a valid Markov chain. Recall that $Z_1, Z_2, Z_3 \sim \text{i.i.d. } \mathcal N(0, \sigma^2)$ and $M \sim \mathcal N(0, 1)$. Let $h(\cdot)$ denote differential entropy. Then,
\begin{align}
	I(M ; M - \what M_2, \what M_2) > I(M ; \what M_3) &= h(\what M_3) - h(\what M_3 \,\vert\, M) \\
													&= \frac{1}{2} \log\biggl(2 \pi e \Bigl(1 + \frac{\sigma^2}{3}\Bigr)\biggr) - \frac{1}{2} \log\biggl(2 \pi e \Bigr(\frac{\sigma^2}{3}\Bigr)\biggr) \\
													&= \frac{1}{2} \log\Bigl(1 + \frac{3}{\sigma^2}\Bigr).  \label{eq:sk-markov-rhs}
\end{align}
Here, we have used the fact that if $Y \sim \mathcal N(0, \sigma^2)$ is a zero-mean multivariate Gaussian random variable with variance $\sigma^2$, then its differential entropy is given by~\cite[Thm.~8.4.1]{CoverThomas}
\begin{equation} \label{eq:diff-ent-gaussian}
	h(Y) = \frac{1}{2} \log(2 \pi e \sigma^2) \text{ nats}.
\end{equation}

Next, note that since $\what M_1 = M + Z_1$ and $\what M_2 = M + \frac{1}{2}(Z_1 + Z_2)$, $\what M_1$ has no extra information about $M$, given $\what M_2$. This is obvious when we think of $\what M_1$ as being $\what M_1 = \what M_2 + Z'$, where $Z' = \frac{1}{2}(Z_1 - Z_2)$, and it can be shown that $Z' \independent \what M_2$:
\begin{align}
	\mathbb E[\what M_2 Z'] &= \mathbb E\biggl[\Bigl(M + \frac{1}{2}(Z_1 + Z_2)\Bigr) Z'\biggr] \\
						   &= \mathbb E[M Z'] + \frac{1}{4} \mathbb E\bigl[(Z_1 + Z_2) (Z_1 - Z_2)\bigr] \\
						   &= 0 + \frac{1}{4} \mathbb E[Z_1^2 - Z_2^2] \\
						   &= \frac{1}{4} (\sigma^2 - \sigma^2) = 0.
\end{align}
Since all variables involved are zero-mean Gaussians, this naturally implies that $\what M_2 \independent Z'$. Thus, from our previous argument, $\what M_1$ has no extra information about $M$ when given $\what M_2$, or in other words, $M$---$\what M_2$---$\what M_1$ is a valid Markov chain. Therefore,
\begin{align}
	I(M ; \what M_1, \what M_2) &= I(M ; \what M_2) + I(M ; \what M_1 \,\vert\, M_2) \\
							  &= \frac{1}{2} \log\Bigl(1 + \frac{2}{\sigma^2}\Bigr) + 0,  \label{eq:sk-markov-lhs}
\end{align}
derived in the same way as~\eqref{eq:sk-markov-rhs}. From~\eqref{eq:sk-markov-rhs} and~\eqref{eq:sk-markov-lhs}, we can conclude that $I(M ; \what M_3) > I(M ; \what M_2)$, and therefore
\begin{align}
	I(M ; M - \what M_2, \what M_2) &> I(M ; \what M_1, \what M_2) \\
	I(M ; M - \what M_2, \what M_2, \what M_1) &> I(M ; \what M_1, \what M_2) \\
	I(M ; M - \what M_2, \what M_2, \what M_1) - I(M ; \what M_1, \what M_2) &> 0 \\
	I(M ; M - \what M_2 \,\vert\, \what M_1, \what M_2) &> 0.
\end{align}
Thus, the stated Markov chain, $M$---$[\what M_1, \; \what M_2]$---$(M - \what M_2)$, cannot hold.


\end{document}